# Comparative analysis of tungsten armor and a liquid-lithium film as first walls for the HiPER laser-fusion chamber: thermal response, vapor shielding and helium transmission to the steel

**Shohruxbek Bozorov** (ORCID: https://orcid.org/0009-0005-5165-093X),
Independent researcher, Chust, Uzbekistan
Corresponding author: Shohruxbek Bozorov, bshoxrux48@gmail.com

## Abstract

Tungsten armor and liquid-metal films, the usual sacrificial first walls of laser-fusion chambers, were modeled separately. Both go through one calculation for the three HiPER scenarios: NRL 154 MJ spectra scaled by yield, binned to keep each ion species' number and energy, a common deposition model, and a one-dimensional enthalpy solver with melting, evaporation and parameterized vapor shielding. In Demo (154 MJ, 6.5 m, 10 Hz), a 1 mm tungsten armor peaks 378 K below melting without neutron heating (275–319 K with it). A degraded near-surface conductivity can erase this margin, which also depends on the ion-spectra construction. The steel interface exceeds the 925 K limit, and the surface accumulates the $10^{18}$–$10^{19}$ $m^{-2}$ helium fluences associated with the onset of surface modification within 10–98 shots (not a lifetime). A lithium film keeps the steel below 925 K from 44 μm and below the 873 K compatibility limit from 56 μm, more with partial renewal. Burn ions then implant helium; keeping the steel below $10^{18}$ $m^{-2}$ for $10^6$ shots needs 184 μm, which binds in every scenario. The film evaporates $7.4 \times 10^{-4}$ kg $m^{-2}$ per shot (range 4.2–$8.4 \times 10^{-4}$ kg $m^{-2}$, full renewal). Vapor shielding, with a nearly un-ionized vapor (Saha estimate), changes this by −21 % to +5 % for the Saha-supported stopping factor $s = 1$ and −34 % to +8 % across $s = 1$–2. Both walls are thermally viable in Prototype and marginal for tungsten in Demo. The deciding issues are tungsten damage evolution, lithium film survival and chamber clearing.

## 1. Introduction

In the chamber of a laser-driven inertial fusion energy (IFE) plant without a protective gas, the first wall receives the x-rays and the ions of every target explosion directly. For the 154 MJ direct-drive target of the Naval Research Laboratory (NRL), the reference for dry-wall chamber studies, 2.1 MJ leave as x-rays and 41.6 MJ as burn and debris ions [1, 2]. At a chamber radius of 6.5 m almost all of this energy reaches the wall within about 40 μs and is mostly absorbed within ten micrometers of a tungsten surface. In a power plant the pulse repeats several times per second, about $3 \times 10^8$ times per full-power year at 10 Hz. The wall must survive surface temperature excursions of thousands of kelvin, implanted helium and sputtering, shot after shot, while the structure behind it must stay within its own temperature limit [1, 3].

Two sacrificial walls are usually considered for such chambers. A solid tungsten armor survives each pulse without melting if the chamber is large enough. But it accumulates damage: implanted helium, recrystallization and thermal fatigue [3, 4], and a thermal conductivity that irradiation lowers [5]. A thin liquid film, continuously replaced, does not itself accumulate damage. It suffers surface evaporation on every shot and must be renewed between shots [6]. As Section 5.3 shows, it also lets the most energetic ions through to the structure behind it. Each wall has its own weakness. Which of the two is the better choice depends on quantities that are rarely computed together: the melting margin, interface temperature and helium dose of the armor, and the minimum thickness, evaporated mass and renewal of the film and the helium dose in the steel behind it. These quantities have been computed for each wall separately, with different targets, deposition models and solvers. Differences between the walls therefore cannot be separated from differences between the calculations.

The tungsten armor of a dry chamber has been studied in detail. Raffray [1] described the threat spectra of the NRL target and derived design windows for the dry-wall chambers of the High Average Power Laser (HAPL) program. Blanchard and Martin [7] computed the thermomechanical response of the same armor. For the European High Power laser Energy Research facility (HiPER), Garoz et al. [8] computed the temperature, plastic strain and crack growth of a 1 mm tungsten first wall in three scenarios, which González-Arrabal et al. [3] reviewed.

Liquid walls were studied mostly with lead, lead–lithium or FLiBe, often for heavy-ion or indirect-drive targets [6, 9–15]. Few studies have applied one code to both wall types. The main precedent is the ARIES-IFE study of operational windows [10], which used the radiation-hydrodynamics code BUCKY for a graphite wall behind a xenon buffer gas and for a lead film. Hassanein and Morozov [16] used HEIGHTS-IFE for bare tungsten and carbon walls and for a composite Li/Pb film on carbon composite. For the film they computed only the deposition (x-rays, and in the preprint ion ranges in lithium [17]). Appendix C compares these studies with this one.

This work compares a 1 mm tungsten armor and a liquid-lithium film, both on 1 cm of steel, as the first wall of the evacuated HiPER chamber in its three scenarios. Both walls get the NRL 154 MJ target spectra scaled by yield as their source term, identical photon and ion deposition models and a single one-dimensional enthalpy solver. Over hundreds of shots the solver follows the tungsten wall and the steel behind the film to a periodic state. Both walls are judged by the same indicators, including the helium and displacement damage in the steel. The solver is benchmarked against the Blanchard–Martin calculation on its own inputs, and the tungsten results against the published HiPER temperatures. A reduced vapor-shielding model, coupled in time to the evaporation, brackets the vapor's effect with two parameters that radiation hydrodynamics would otherwise supply: the stopping power of the vapor and the fraction of the stopped energy re-radiated to the wall.

Neither of these precedents treated a lithium film as a wall: the ARIES-IFE wetted-wall analysis used lead and FLiBe films, mainly for a heavy-ion indirect-drive target, and Hassanein and Morozov similarly computed the thermal response only of the bare walls. To the author's knowledge, no published study has computed the thermal response, melting and evaporation of a tungsten armor and a lithium film within one shared calculation. Nor has a liquid-wall analysis of the HiPER scenarios been reported. Any difference found is then a difference between the walls, at the cost of physics that neither wall model contains: stress, damage evolution and the hydrodynamics of the vapor. The code is released with the paper. Section 2 describes the model, Section 3 its verification and benchmarks, Sections 4–6 the results for tungsten, lithium and vapor shielding, and Section 7 discusses what the two walls require at the Demo conditions.

## 2. Model

### 2.1 Scenarios and wall configurations

The three HiPER scenarios differ in target yield, chamber radius and repetition rate (Table 1). Symbols are listed in Appendix A. Supplementary Section S4 separates the literature inputs and choices from the computed quantities and describes the walls in full. In every scenario the chamber is evacuated and the wall is a sphere of radius $R$ irradiated by an isotropic point source at normal incidence.

The tungsten wall is 1 mm of W on 1 cm of steel, as in Garoz et al. [8]. The steel has the ferritic-steel properties of Blanchard and Martin [7]. They substitute for the ODS-RAFM steel of that configuration, whose property set was not available. The steel is judged against the 925 K operating limit used for ODS-RAFM. Its conductivity is the constant 33 W m$^{-1}$ K$^{-1}$ used by Blanchard and Martin, who take it from ferritic-steel data [7]. A lower value, 29 W m$^{-1}$ K$^{-1}$, raises the mean Demo interface temperature from 930 to 965 K and the quasi-steady surface peak from 3317 to 3346 K (Supplementary Section S6). In Prototype and Demo its back face is cooled by lead–lithium at $T_c = 600$ K with $h = 10$ kW m$^{-2}$ K$^{-1}$ [8]. The absorbed heat leaves only through the steel. The Experimental scenario is treated as single shots on a wall with an adiabatic back face.

The lithium wall is a film of thickness $\delta$ on the same 1 cm of steel, computed for thirteen thicknesses from 1 μm to 1 mm. A single-shot calculation starts from a fresh film at $T_{Li} = 600$ K on steel at the same temperature. The repeated-shot calculations of Section 5.3 replace the film at full thickness and 600 K before every shot and carry the steel's temperature from shot to shot. Film, steel and coolant together at 500 and 800 K are run as a sensitivity case (the film-and-coolant temperature).

**Table 1.** HiPER scenarios and wall conditions [3, 8].

| | Experimental | Prototype | Demo |
|---|---|---|---|
| Target yield $Y$ [3, 8] | 20 MJ | 50 MJ | 154 MJ |
| Chamber radius $R$ [3, 8] | 5 m | 6.5 m | 6.5 m |
| Repetition rate [3, 8] | single shots (this work) | 1 Hz | 10 Hz |
| Initial wall temperature (W case) [8] | 300 K | 600 K | 600 K |
| Back face of the steel | adiabatic (this work) | convection, $h$ = 10 kW m$^{-2}$ K$^{-1}$, coolant at 600 K [8] | convection, $h$ = 10 kW m$^{-2}$ K$^{-1}$, coolant at 600 K [8] |
| Steel operating limit [8] | 925 K | 925 K | 925 K |

### 2.2 Thermal model

The wall is a one-dimensional slab described in a Lagrangian mass coordinate. Let $\mu$ be the areal mass between a fixed material reference (the initial surface) and a material element, so that $d\mu = \rho\, dx$. Each layer then obeys

$$\partial H/\partial t|_{\mu} = \partial/\partial\mu\,(\,\rho\,k\,\partial T/\partial\mu\,) + q_v/\rho, \quad (1)$$

where $H(T)$ is the specific enthalpy, ρ the density, $k$ the thermal conductivity and $q_v$ the volumetric heating by the x-rays and ions. The plasma-facing surface sits at $\mu = M_e(t)$, the areal mass evaporated so far, which grows as $dM_e/dt = j$. The receding surface therefore needs no mesh motion, and melting enters through the latent heat in $H(T)$.

The conductance to the chamber is zero at the plasma-facing surface, and no radiative heat is exchanged, following the isothermal-chamber argument of Blanchard and Martin [7]. The surface instead loses energy only by evaporation, through the term $j\,L_v$:

$$k\,\partial T/\partial x|_{\mathrm{surface}} = j\,L_v,\; j = \alpha\,p_{\mathrm{sat}}(T_0)\,\sqrt{(\,M\,/\,(2\pi\,R_g\,T_0)\,)}, \quad (2)$$

with $x$ measured into the wall: the Hertz–Knudsen–Langmuir flux into vacuum with an evaporation coefficient $\alpha = 1$ (0.1 as a sensitivity case) and no return flux. $\alpha = 1$ is an upper bound. The steel back face is cooled by convection, $-k\,\partial T/\partial x = h\,(T - T_c)$, and the W/steel and Li/steel interfaces are in perfect thermal contact.

The slab is divided into cells of fixed areal mass, graded from 1 nm at the tungsten surface (10 nm for lithium). Each time step is solved by backward Euler with Newton iteration under a step control that resolves the x-ray pulse and the ion arrival. Repeated-shot runs continue until the pre-shot temperature of the plasma-facing surface (or, for the lithium film, the steel front face behind it) rises by less than 0.05 K per shot over the last 10 % of the shots. The state then reached is called quasi-steady below, because a slow drift remains. Supplementary Section S1 gives the grids, the step control and the verification tests in full.

### 2.3 Source term

The source term is that of the NRL 154 MJ direct-drive target of the ARIES-IFE study: 2.1 MJ of x-rays, 19.5 MJ of burn ions and 22.1 MJ of debris ions, while the 109 MJ of neutrons deposit most of their energy deep in the blanket [1, 3]. The ion spectra are the ARIES-IFE tables for burn and debris ions [2]. They are binned so that every species keeps both its number of ions and its energy in the ARIES accounting: helium at one alpha per D–T fusion, $5.46\times10^{19}$ for 154 MJ at 17.6 MeV, and every other species at its tabulated number (Supplementary Section S2), with the raw tables as a sensitivity case. The x-ray spectrum was digitized from Raffray's Fig. 2 [1]. Following Raffray's practice for other yields [1], all energies are scaled by $Y/154$ MJ for the 20 and 50 MJ scenarios. This is an estimate—the HiPER group's own target spectra are not published.

Neutron heating of the armor and the steel, which Garoz et al. neglected [8], is left out of the baseline and added as a separate case (Sections 4.2 and 5.3). In that case tungsten receives the volumetric heating that Sharafat et al. give for the same target and chamber, 800 MW cm$^{-3}$ for 5 ns [18], or 4 J cm$^{-3}$ per shot at Demo conditions, and the steel receives 0.5 or 1 times that value per unit volume. No sourced value for the steel heating per unit volume at HiPER conditions was found, so the two cases are reported as a bracket.

The x-rays arrive at $R/c$ as a square pulse of 20 ns. Every ion component flies freely from the target to the wall, so its power on the wall follows from its spectrum and its time of flight. At 6.5 m the first ions arrive at 0.105 μs and the last at 83.5 μs. Table 2 gives the resulting energy and power densities on the wall, and Supplementary Section S2 the binning, the particle count of every species and the neutron source in full.

**Table 2.** Energy and power density on the wall per shot, and averaged over one second, from the source term of this work (normal incidence; Experimental: one shot in that second). Experimental, Prototype and Demo are the HiPER scenarios defined in [3, 8] (Table 1).

| | Experimental | Prototype | Demo |
|---|---|---|---|
| X-ray energy per shot | 868 J m$^{-2}$ | 1284 J m$^{-2}$ | 3955 J m$^{-2}$ |
| Ion energy per shot | $1.72\times10^4$ J m$^{-2}$ | $2.55\times10^4$ J m$^{-2}$ | $7.84\times10^4$ J m$^{-2}$ |
| X-ray power density during the 20 ns pulse | $4.3\times10^4$ MW m$^{-2}$ | $6.4\times10^4$ MW m$^{-2}$ | $1.98\times10^5$ MW m$^{-2}$ |
| Peak ion power density (time after ignition) | $1.2\times10^5$ MW m$^{-2}$ (0.38 μs) | $1.4\times10^5$ MW m$^{-2}$ (0.50 μs) | $4.29\times10^5$ MW m$^{-2}$ (0.50 μs) |
| Mean ion power density over the 10–90 % arrival window | $8.8\times10^3$ MW m$^{-2}$ (0.21–1.78 μs) | $9.96\times10^3$ MW m$^{-2}$ (0.27–2.32 μs) | $3.07\times10^4$ MW m$^{-2}$ (0.27–2.32 μs) |
| Power density averaged over 1 s | 0.018 MW m$^{-2}$ | 0.027 MW m$^{-2}$ | 0.82 MW m$^{-2}$ |

### 2.4 Energy deposition

Photons are attenuated by the Beer–Lambert law in each of 100 energy bins, with the photoelectric plus incoherent-scattering mass attenuation coefficients of NIST XCOM above 1 keV [19] and the photoabsorption coefficients of Henke et al. below [20].

Ion ranges come from SRIM-2012 tables of projected range for every ion in W, Li and Fe [21], corrected for keV hydrogen isotopes and helium in tungsten with TRIM runs made for this study. The energy is placed at depth so that the cell deposits sum to the incident energy. Depositing the keV hydrogen isotopes and helium in tungsten with TRIM depth profiles, and the faster hydrogen isotopes with an estimated depth stretch, instead lowers the single-shot peak temperatures by 0.3–0.6 % (Demo 3027 → 3010 K, Supplementary Section S2). The reported tungsten temperatures are therefore slightly high, in the conservative direction.

### 2.5 Reduced vapor-shielding model

The vapor matters for the ions. The x-rays arrive within 42 ns, before there is vapor, but the ions arrive over 0.1–84 μs while the lithium surface evaporates, so the vapor made by the first debris ions shields the later ones. Because the vapor moves only centimeters during the ion pulse, the column crossed by an ion arriving at time $t$ is the mass evaporated so far in the shot, $\sigma_v(t) = M_e(t)$. The model represents it as one cell of the wall's own front material, of areal mass $s\ \sigma_v$, in front of the mesh. Every ion component is deposited through it with the production depth model. Here $s = 1$ is cold, neutral vapor, $s = 2$ an assumed enhancement by ionization, and $s = 5$ and 10 test larger stopping. A fraction $f_{rr}$ of the energy stopped in the vapor is re-radiated to the surface cell, at once or relaxing from a store with time constant $\tau_{rr}$,

$$dE_{st}/dt = f_{rr}\, P_v(t) - E_{st}/\tau_{rr},\ q_{rr}(t) = E_{st}/\tau_{rr},\ (3)$$

where $P_v$ is the power stopped in the vapor, and the rest leaves the wall's energy budget for the shot. The calculations use $f_{rr} = 0$, 0.5 and 1 with immediate return and $s = 1$ and 2, one xenon-like case with $f_{rr} = 1$ and $\tau_{rr} = 300$ μs, and $f_{rr} = 0$, $s = 1$ in the repeated-shot runs. The model neglects the hydrodynamics, ionization balance and radiation transport of the vapor, which are the content of radiation-hydrodynamics codes [22, 23]. It is therefore a parameterized screening calculation that brackets the effect of the vapor on the wall (Supplementary Section S2).

### 2.6 Material properties

Tungsten follows the recommended analytical expressions of Tolias [24] for the solid and the liquid, with $T_m = 3695$ K. Its vapor pressure and heat of vaporization are those tabulated by Raffray [1]. Lithium is liquid throughout, with the properties compiled by Davison [25], whose vapor-pressure correlation is extrapolated above 1800 K. The steel has the properties of Blanchard and Martin [7], and Appendix B lists every property with its source and range.

### 2.7 Derived quantities and indicators

The minimum lithium film is the thickness at which the peak steel temperature crosses 925 K, or the 873 K compatibility limit, interpolated linearly in temperature and logarithmically in $\delta$ between the computed thicknesses. A third, helium-exclusion criterion is the film beyond which the helium reaching the steel, all of it retained, stays below $10^{18}$ m$^{-2}$ for $10^6$ shots. A second, stricter exposure level is one full-power year ($3.2\times10^8$ shots at 10 Hz, $3.2\times10^7$ at 1 Hz). The helium and displacements in the steel are computed from the same ion source term and range data (Supplementary Section S3). Supplementary Section S3 defines the thick-film evaporated mass, the whole-chamber totals, the energy partition of a shot and the flow speed for film replacement.

Two damage indicators are reported for tungsten, and both mark the onset of surface modification, not a lifetime. The first is the number of shots after which the retained helium reaches the pulsed-damage fluence of $10^{18}$–$10^{19}$ m$^{-2}$ that González-Arrabal et al. [3] report, counting every implanted helium atom (a lower bound on the shots, with a secondary estimate from a reaction–diffusion model in Supplementary Section S7). The second is the number of cycles allowed by the surface-load design rule of Mitteau et al. [26] at a heat-flux factor of the ion load.

Sputtering follows the Eckstein–Preuss fit formula [27, 28]. The mean W/steel interface temperature at periodic steady state follows exactly, in this model, from the heat balance $\langle T_{if}\rangle = T_c + \langle q\rangle\ (1/h + L_s/k_s)$, where $\langle q\rangle$ is the repetition rate times the energy absorbed per shot and $L_s/k_s$ the thermal resistance of the steel. Supplementary Section S6 gives the indicators in full.

## 3. Verification and benchmarks

The numerics are checked first. The solver reproduces problems with known solutions (a constant surface flux on a semi-infinite slab, an instantaneous exponential source and the two-phase Stefan problem) within 0.065–0.28 %. It conserves energy to a relative residual of $1\times10^{-14}$ in a synthetic case with melting, evaporation, rezoning, two layers and a coolant, and it shows the expected orders of convergence. These constant-property tests check the numerics only (Supplementary Table S1, Supplementary Fig. S1, Supplementary Section S1).

In the production runs the energy ledger closes to below $10^{-11}$ in single shots and to about $10^{-7}$ over the 600-shot Demo run. Halving or doubling the cells and the time-step limits changes no result by more than 1.2 K or 0.8 % (Supplementary Table S2). The paper quotes the production values (Supplementary Section S1 gives the extrapolated ones).

The vapor-shielding option is checked by tests of its limiting cases, coupling and energy bookkeeping, not of the stopping power of the vapor itself (Supplementary Sections S1 and S3).

The source term, the deposition and the whole calculation were checked separately against published values (Table 3, full table in Supplementary Section S3). No measurement of the thermal response of a wall under these loads exists, so these are comparisons with other calculations and with tabulated data, not an experimental validation. The energy fluences agree with those of Garoz et al. [8] within their rounding for 31 of 33 per-species, group and x-ray values. The other two, the debris totals of Prototype and Demo, are 4.1 % and 4.2 % higher, and the total Demo ion fluence is 1.8 % higher (Supplementary Section S3). That comparison tests the bookkeeping rather than the spectral shapes.

At every energy of the NIST tables from 0.01 to 10 MeV [29], the SRIM ranges of protons and alphas in tungsten and, above 0.1 MeV, their electronic stopping powers agree within 7.3 % and 5.1 %. The depth model follows TRIM except for keV deuterons and helium in tungsten, which it places too close to the surface (Section 2.4), and for the 90 %-energy depth of 15 MeV gold, 0.85 of TRIM's (Supplementary Fig. S2). For lithium, the ranges agree within 15 % with those of Hassanein and Morozov [17].

On the inputs of Blanchard and Martin [7] the solver reproduces the rises of their finite-element peak temperatures above the 400 °C initial temperature within 4.4 % at the surface and 10 μm and 9.9 % at 25 μm. With this study's spectra and deposition, the peak rises agree with those of Raffray [1] within 3 % and the peaks with those of Garoz et al. [8] within 3 %.

**Table 3.** Benchmark summary. The Blanchard–Martin and Raffray rows compare temperature rises above the initial temperature of each calculation (400 °C and 500 °C), with the peaks in °C as in the original papers; fluences in J cm$^{-2}$, as in Garoz et al.

| Check | Reference | This work | Ratio |
|---|---|---|---|
| Total ion fluence at the wall, Exp. / Proto. / Demo | 1.7 / 2.5 / 7.7 J cm$^{-2}$ [8] | 1.721 / 2.545 / 7.840 J cm$^{-2}$ | within rounding / within rounding / 1.018 |
| Peak W surface temperature, Experimental, first shot | 875 K [8] | 868 K | 0.992 |
| Peak W surface temperature, Prototype, last of 100 shots | 1410 K [8] | 1381 K | 0.979 |
| Peak W surface temperature, Demo, last of 600 shots | 3400 K [8] | 3317 K | 0.976 |
| Blanchard–Martin benchmark: surface / 10 μm / 25 μm, peak (ratio of rises above 400 °C) | 2554 / 1393 / 849 °C [7] | 2460 / 1437 / 893 °C | 0.956 / 1.044 / 1.099 |
| Raffray, 3 mm W slab: surface / 10 μm, peak (ratio of rises above 500 °C) | ~2915 / ~1525 °C [1] | 2895 / 1549 °C | 0.992 / 1.023 |
| SRIM vs NIST in W: projected range (0.01–10 MeV) / electronic stopping (0.1–10 MeV) | NIST PSTAR/ ASTAR [29] | SRIM-2012 [21] | 0.927–1.041 / 0.958–1.051 |
| Depth model vs TRIM, 50 %-energy depth: Li targets and MeV ions in W / keV d and He in W | TRIM [21] | model | 0.935–1.005 / 0.69–0.76 |

## 4. Results: tungsten armor

Table 4 collects the tungsten results, whose temperatures, erosion rates and displacement doses are outputs of the model. The helium count divides a computed fluence by a published threshold, and the design-rule count uses only a heat-flux factor of the ion load.

### *4.1 Single shot*

The surface temperature follows the same history in all three scenarios (Figure 1), with the peak set by the debris ions 1.6–2.1 μs after ignition. With the conductivity of undamaged tungsten no scenario melts. The Demo single-shot peak of 3027 K lies 668 K below the 3695 K melting point. The Prototype peak, 1371 K, falls inside the 1300–1800 K recrystallization range given by González-Arrabal et al. [3], but below the 1523 K one-hour onset of a pure-tungsten plate [30] (Table B.2).

### *4.2 Repeated shots and the interface with the steel*

At 10 Hz the Demo wall settles into a quasi-steady state. Before every shot its surface returns to 921 K, rising by only 0.035 K per shot over the last 60 of 600 shots (Figure 2); the Prototype surface returns to 609 K. The wall does not ratchet, but it does settle well above its 600 K initial temperature. The quasi-steady Demo shot peaks at 3317 K, 378 K below melting.

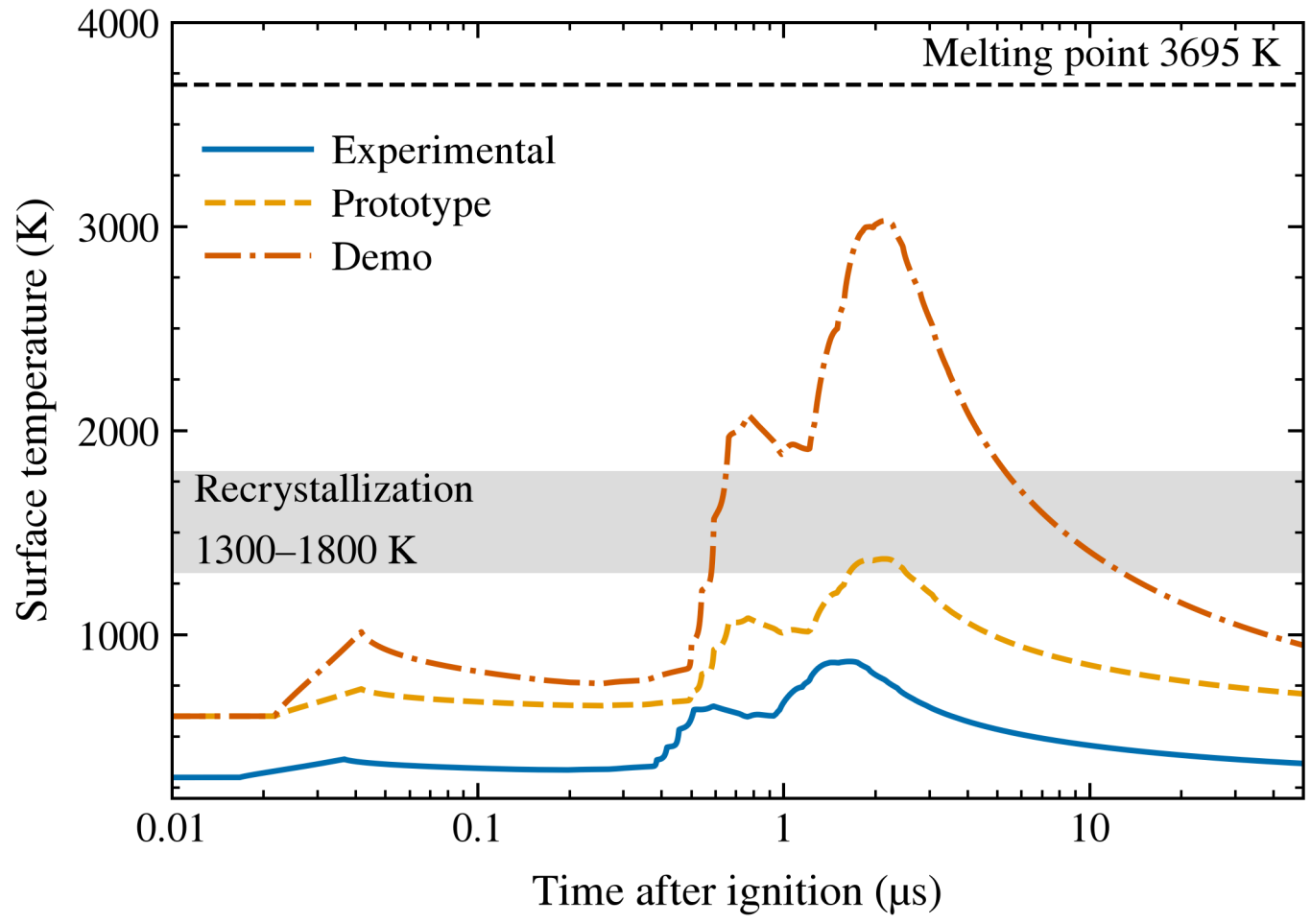

**Figure 1.** Tungsten surface temperature during a single shot in the three scenarios (Experimental from 300 K, Prototype and Demo from 600 K). Dashed line: melting point, 3695 K [24]; gray band: the recrystallization range, 1300–1800 K [3].

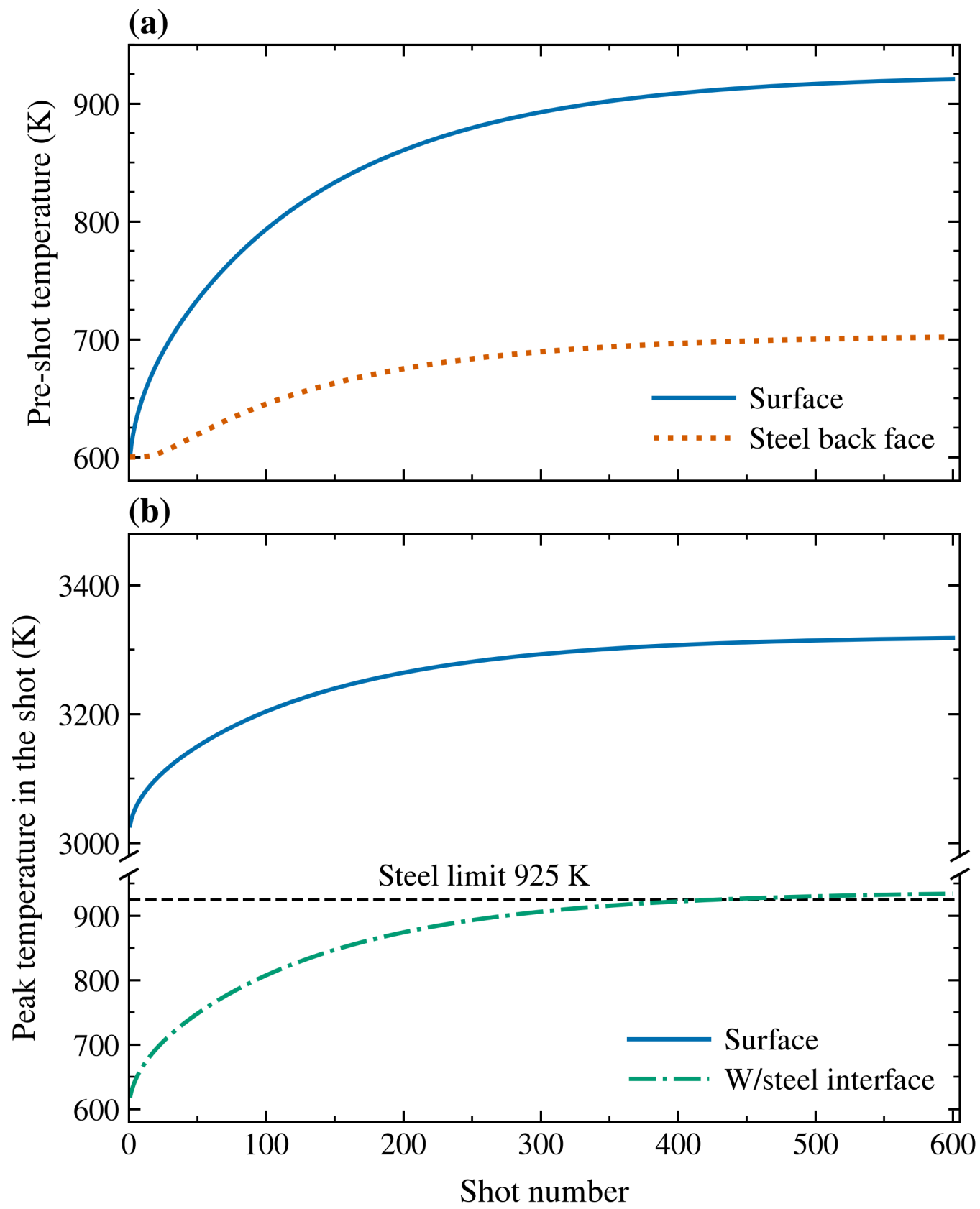

**Figure 2.** Demo at 10 Hz, 600 shots from a uniform 600 K. (a) Surface and steel back-face temperatures just before each shot. (b) Peak surface and peak W/steel interface temperatures in each shot; dashed line, the 925 K steel limit [8].

The interface with the steel does not share the surface's margin. Its Demo peak first exceeds the 925 K limit at shot 433 and reaches 934 K at shot 600, still rising. An exponential fit to shots 201–600 extrapolates to 938 K. The excess is set by the cooling path rather than by the armor (Supplementary Section S6).

**Table 4.** Tungsten armor, 1 mm W on 1 cm steel, without vapor shielding (which changes no tungsten result; Section 6). Experimental, Prototype and Demo are the HiPER scenarios defined in [3, 8] (Table 1).

| Quantity | Experimental | Prototype | Demo |
|---|---|---|---|
| Peak surface temperature, single shot from the initial temperature | 868 K | 1371 K | 3027 K |
| Pre-shot surface temperature at the end of the multi-shot run | single shots | 609 K (100 shots) | 921 K (600 shots) |
| Peak surface temperature, last shot | – | 1381 K | 3317 K |
| Peak W/steel interface temperature over the run | – | 615 K | 934 K |
| Mean interface temperature at periodic steady state (heat balance) | – | 611 K | 930 K |
| With neutron heating, steel at 0.5 / 1 times the W value (Section 4.2): pre-shot surface temperature | – | 612 / 613 K | 986 / 1036 K |
| … peak surface temperature, last shot | – | 1383 / 1385 K | 3376 / 3420 K |
| … peak W/steel interface temperature | – | 617 / 619 K | 999 / 1050 K |
| … mean interface temperature (heat balance) | – | 613 / 615 K | 997 / 1047 K |
| Evaporation, first shot | below $10^{-20}$ nm | below $10^{-20}$ nm | $3.4\times10^{-7}$ nm |
| Sputtering per shot (with C and Au proxies) | $2.4\times10^{-5}$ nm ($3.8\times10^{-5}$) | $3.6\times10^{-5}$ nm ($5.6\times10^{-5}$) | $1.1\times10^{-4}$ nm ($1.7\times10^{-4}$) |
| Shots for the retained He to reach $10^{18}$ / $10^{19}$ m$^{-2}$ [3], all helium retained (lower bound): burn helium only | 83 / 825 | 56 / 558 | 18 / 181 |
| … burn and debris helium (headline count) | 45 / 449 | 30 / 303 | 10 / 98 |
| … burn and debris helium, with the helium-retention model (secondary estimate, Supplementary Section S7; displacements counted along the ion path in Prototype and Demo, over the projected depth in Experimental) | 45 / 452 | 31 / 308 | 11 / 125 |
| Displacements per shot, mean over the first 5 µm (ions) | $8.1\times10^{-6}$ dpa | $1.2\times10^{-5}$ dpa | $3.7\times10^{-5}$ dpa |
| Helium and ion displacements in the steel | none: 1 mm of W stops every ion | none | none |
| Time-mean heat flux through the steel (heat balance) | single shots | 0.027 MW m$^{-2}$ | 0.82 MW m$^{-2}$ |
| Cycles allowed by the surface-load design rule [26], at the heat-flux factor of González-Arrabal et al. [3] / of this work (Section 4.4) | $3.0\times10^{4}$ / $9.3\times10^{3}$ | $1.0\times10^{3}$ / $1.14\times10^{3}$ | 9.5 / 9.8 |

Neutron heating raises the interface further (Section 2.3). Continued from the 600-shot state with the neutron heat added, the Demo wall settles again after 390 and 490 more shots for steel heating of half and all of the tungsten value per unit volume. Before each shot the surface is then at 986 K and 1036 K. The shot peaks at 3376 K and 3420 K, 319 K and 275 K below melting. The interface peaks at 999 K and 1050 K, 74 K and 125 K above the steel limit. Because the steel factor is an assumption, the two cases bracket the effect rather than predict it.

The sensitivity cases of Sections 4.3 and 4.5 were computed without neutron heating, and with it they would start from a narrower melting margin. Supplementary Section S6 gives the heat balance with neutron heating, the cooling that would restore 925 K and the Prototype values.

### 4.3 Sensitivity of the Demo peak

Two inputs control the Demo single-shot peak most (Figure 3). Using the raw ion tables instead of the production spectra, which keep the number and energy of every species, raises it by 526 K. The raw tables carry 13 % more ion energy than the ARIES accounting, 47.2 instead of 41.6 MJ (19 % more in the debris ions), and 1.18 alphas ($^4$He) per fusion instead of one (Supplementary Section S2). No run separates the two effects. A 15 % lower tungsten conductivity raises the peak by 180 K (a 15 % higher one lowers it by 148 K). The other inputs matter much less. From the quasi-steady state the raw tables would, by extrapolation, bring the peak above the melting point. The Demo melting margin therefore depends on how the ion spectra are constructed and on the conductivity (Supplementary Section S6).

### 4.4 Erosion and damage indicators

With the conductivity of undamaged tungsten, erosion is negligible. Evaporation removes $6.9\times10^{-6}$ nm per shot in the quasi-steady state and sputtering up to $1.7\times10^{-4}$ nm, so 1 mm lasts 18–27 full-power years at 10 Hz (Supplementary Section S6). For 16–24.5 full-power years (evaporation included), the armor retains more than the 100 µm that Garoz et al. found sufficient to protect the steel [8]. A degraded conductivity shortens this to 0.9–3.8 years (Section 4.5 and Supplementary Section S6).

The helium and design-rule counts mark the onset of surface modification. They are not lives, and the life of the armor is not computed here. The headline count of Table 4 is this full-retention count of burn and debris helium. If every implanted helium atom stays, the retained helium in the Demo armor reaches $10^{18}$ m$^{-2}$ after 10 shots and $10^{19}$ m$^{-2}$ after 98 shots, in Prototype after 30 and 303 shots and in Experimental after 45 and 449. González-Arrabal et al. give about $10^3$ shots to $10^{19}$ m$^{-2}$ for Experimental [3], which matches the burn helium alone here (825 shots). The debris helium nearly doubles the fluence.

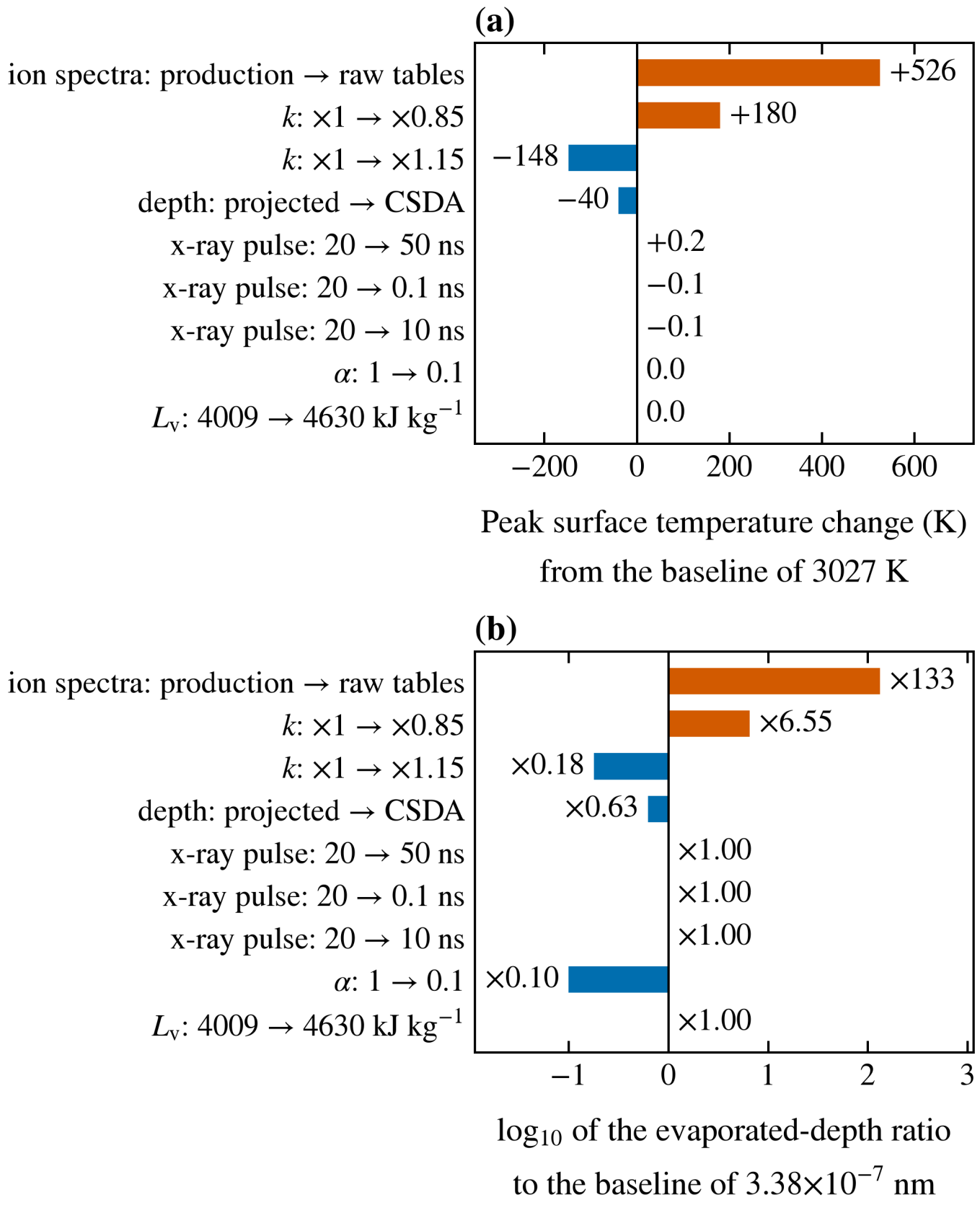

**Figure 3.** Sensitivity of the Demo single shot from 600 K, one input changed at a time: change of the peak surface temperature from the baseline 3027 K (a) and $\log_{10}$ of the ratio of the evaporated depth to the baseline $3.38 \times 10^{-7}$ nm (b), bars in the same order. Ion spectra: the ARIES tables as read, with 1.18 alphas ($^4$He) per fusion instead of one (18 % more helium at the wall) and 47.2 instead of 41.6 MJ of ions, in place of the production spectra (Supplementary Section S2). Depth: the CSDA path length in place of the projected range (Supplementary Fig. S2). $k$: tungsten conductivity 15 % lower and 15 % higher. X-ray pulse: 0.1, 10 or 50 ns in place of 20 ns. $\alpha$: evaporation coefficient 0.1 in place of 1. $L_V$: heat of vaporization 4630 kJ kg$^{-1}$, the Clausius–Clapeyron slope of the vapor-pressure law of Raffray [1], in place of his tabulated 4009 kJ kg$^{-1}$.

These counts are lower bounds, because any release of helium can only delay the threshold. The retention model of Supplementary Section S7, which probably overstates the release, gives 11 and 125 Demo shots (31 and 308 in Prototype, 45 and 452 in Experimental). The threshold itself is uncertain, and helium was not confirmed as the cause of the damage in the pulsed experiments behind it [4], so the count could be several times larger (Supplementary Section S6). The modified layer is shallow—95 % of the Demo helium stops within 5 µm, one twentieth of the 100 µm of tungsten that Garoz et al. found sufficient, with a crack-growth analysis, to protect the steel. For the same armor they estimated a fatigue life of about $10^6$ cycles [8].

The surface-load design rule of Mitteau et al. [26] allows 9.5 cycles at the Demo heat-flux factor of 44.4 MW m$^{-2}$ s$^{1/2}$ printed by González-Arrabal et al. [3], $1.0 \times 10^3$ in Prototype and $3.0 \times 10^4$ in Experimental. Similar factors follow from the ion power history of this work: 43.7, 14.2 and 10.9 MW m$^{-2}$ s$^{1/2}$ from the absorbed energy over the 10–90 % arrival window divided by the square root of the window, and 9.8, $1.14 \times 10^3$ and $9.3 \times 10^3$ cycles. Using the peak absorbed power times the square root of the window gives 612, 199 and 153 MW m$^{-2}$ s$^{1/2}$ and 1.2–1.9 cycles (Supplementary Section S6). The rule is stated for surface loads, and the data behind it differ from these loads in depth and duration (Supplementary Section S6). The counts are therefore order-of-magnitude indicators of surface damage, not a fatigue life. That the Demo count is similar to the helium count is a coincidence of unrelated inputs.

### *4.5 Reduced conductivity and the near-surface layer*

The melting margin depends on the conductivity of the first micrometers, which irradiation and surface preparation can lower. With the conductivity of solid tungsten multiplied by a factor $\varphi$ = 0.5 and molten tungsten left undamaged, the Demo surface melts on every shot of a 550-shot run, 0.31 μm deep in the first shot and 0.77 μm in the last (Supplementary Table S12). It also evaporates enough to consume 1 mm in about 5 full-power years. Reductions of this size have been measured at room temperature, after helium implantation and self-ion damage [5, 31], but whether they occur in the HiPER wall is not established (Supplementary Section S6).

The surface itself need not have bulk properties. Polishing can leave a defect-rich layer near the surface: about 74 nm deep in electropolished tungsten, and 0.23 μm deep, with less than half the conductivity of the material below it, in a mechanically polished W–Cr–Hf alloy [32]. Figure 4 shows the peak for a surface layer of thickness $d$ = 1, 3 or 10 μm whose solid conductivity is a hypothetical fraction $\varphi$ of the bulk value. From the quasi-steady state a 10 μm layer at $\varphi$ = 0.75 brings the peak to 3622 K, 73 K below the melting point, and at $\varphi$ = 0.25 the surface reaches 4102 K and melts 1.8 μm per shot. The layer changes neither the interface nor the temperature before the next shot; the top ten micrometers set the peak. A measured effusivity of the as-installed surface, for example by pulsed photothermal radiometry [33, 34], would decide between these cases (Supplementary Section S6).

In pure tungsten damaged at room temperature by 5 MeV gold ions to about 1–100 dpa, this method gave layers 0.5–0.9 μm thick with an apparent conductivity 0.30–0.69 times that of the unirradiated sample [32]. The factors lie inside the range of Fig. 4, and the layers are thinner than its thinnest case, 1 μm, which from the quasi-steady state gives Demo peaks of 3376–3713 K for $\varphi$ = 0.75–0.25. The damage was made by gold ions, not by fusion ions, so these values do not fix $\varphi$ or $d$ for the HiPER wall.

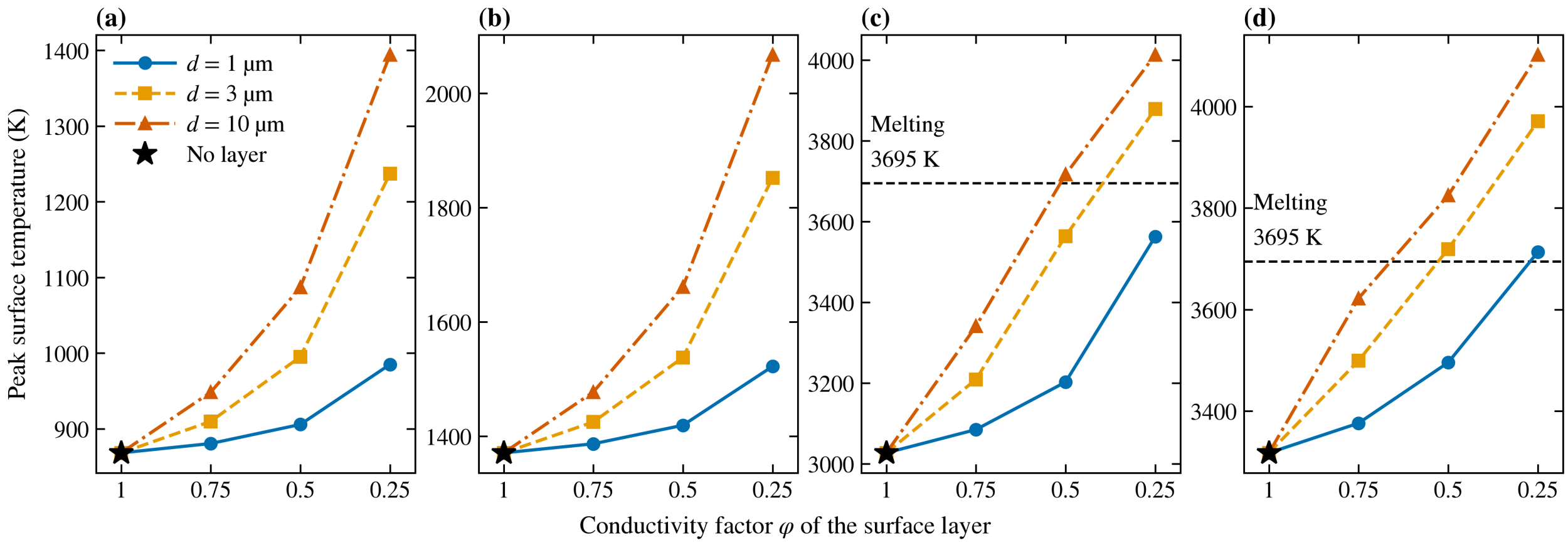

**Figure 4.** Peak tungsten surface temperature in a single shot against the conductivity factor $\varphi$ of a surface layer of thickness $d$ = 1, 3 and 10 μm (density and specific heat at bulk values). (a) Experimental from 300 K, (b) Prototype from 600 K, (c) Demo from 600 K and (d) Demo from the quasi-steady 921 K state. Star: no layer; dashed line: melting point, 3695 K [24]. The factors are hypothetical.

### *4.6 Tungsten grades*

Potassium-doped tungsten (W–K) and $La_2O_3$-strengthened tungsten (ODS W–$La_2O_3$) enter this model only through their thermal conductivity. Rolled W–K plate conducts like pure tungsten within 1 % [35], so every result above applies to it unchanged. For ODS W–$La_2O_3$, whose $k(T)$ was not found, 0.8 times the pure-tungsten curve gives a quasi-steady Demo peak of 3557 K, 138 K below melting. A thermal-exposure indicator of recrystallization reaches its one-hour value after about $2.2\times10^3$ Demo shots (3.7 min) for pure tungsten, $1.1\times10^4$ (19 min) for W–K and $7.4\times10^4$ (2.1 h) for ODS W–$La_2O_3$, and in Prototype only after more than a thousand years (Supplementary Section S6). These counts depend strongly on the uncertain onset temperature, and the grades change neither the interface temperature nor the helium count. Their real differences, in surface damage and ductility, need the thermomechanics that this model lacks.

## 5. Results: lithium film

### *5.1 What the film absorbs*

Lithium is nearly transparent to the x-rays. A 1 mm film absorbs 91 % of the threat energy and 96 % of the ion energy, but only 9 % of the x-ray energy (Figure 5), whereas 1 mm of tungsten stops both. The film thickness needed to protect the steel is therefore set by the stopping of the most penetrating ions—the MeV burn ions. A thicker film barely reduces the x-ray load on the steel behind it.

### 5.2 Minimum film thickness and evaporation in a single shot

Table 5 gives the lithium results. In a single Demo shot on steel at 600 K, the steel stays below 925 K for films of 30.1 µm and more. The limit is crossed between the computed thicknesses of 17.8 µm, at which the steel reaches 1116 K, and 31.6 µm, at which it reaches 907 K (Figure 6). Across the sensitivity cases the Demo minimum ranges from 24 to 79 µm, mostly through the film-and-coolant temperature (Supplementary Section S3 and Fig. S4). In Prototype the minimum is 11.9 µm (8.2–34 µm, the upper end at $T_{\mathrm{Li}}$ = 800 K), and in Experimental 7.2 µm.

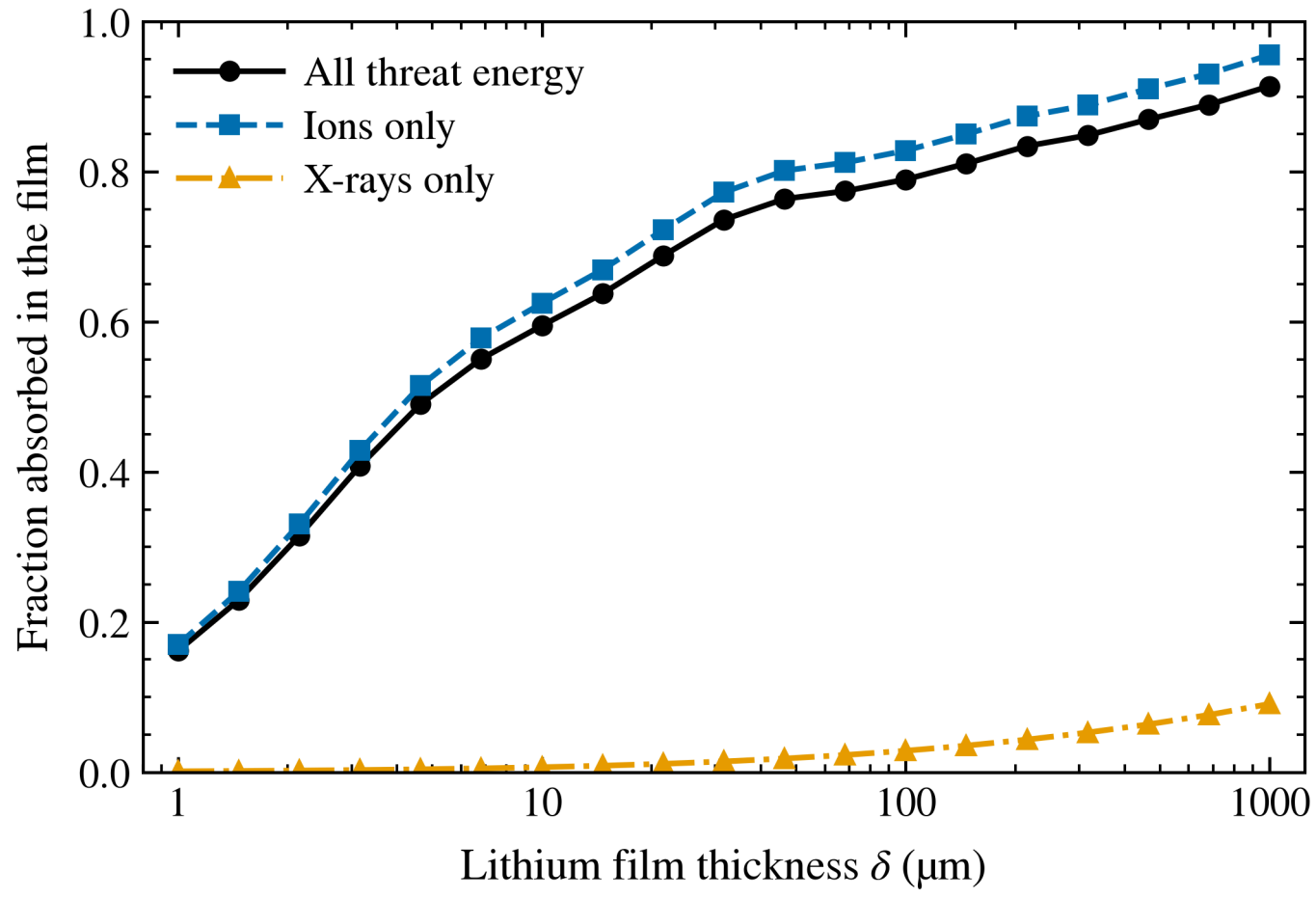

**Figure 5.** Fraction of the incident threat energy absorbed in a lithium film on steel against film thickness, for all threat energy and for the ions and the x-rays alone, each as a fraction of its own incident energy (static deposition, no vapor shielding).

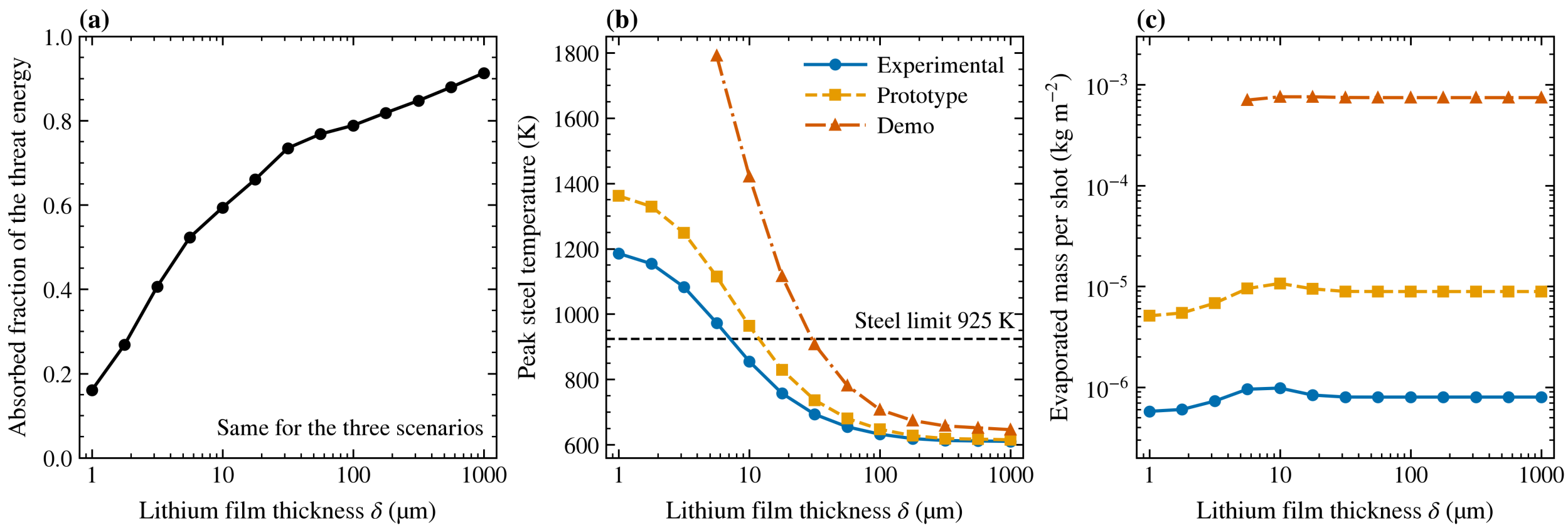

**Figure 6.** Lithium film on 1 cm steel, single shot at $T_{\mathrm{Li}}$ = 600 K, against film thickness: absorbed fraction of the threat energy (a), peak steel temperature with the 925 K limit [8] dashed (b), and lithium evaporated per shot (c). Demo films of 3.2 µm and less are missing from (b) and (c) because the steel passed the 1800 K limit of its property table.

Beyond a few tens of micrometers the evaporated mass no longer depends on the film thickness. A Demo shot then evaporates $7.4 \times 10^{-4}$ kg m$^{-2}$ of lithium, 0.39 kg over the whole chamber if the film covers the whole sphere. This is a surface-evaporation estimate. It rests on a vapor-pressure law extrapolated above its 800–1800 K data, on an estimated heat of vaporization and on the evaporation coefficient. An energy-balance estimate made independently of the solver is consistent with it (Supplementary Section S3).

With the film fully renewed every shot, the plausible range of the evaporated mass is $4.2$–$8.4 \times 10^{-4}$ kg m$^{-2}$, with the evaporation coefficient $\alpha$ = 0.1 giving the low end; partial renewal gives up to $9.9 \times 10^{-4}$ kg m$^{-2}$ (Section 5.4). Vapor shielding moves the mass by −34 % to +8 % across $s$ = 1–2 (−21 % to +5 % for the Saha-supported $s$ = 1; Section 6), and neutron heating of the steel raises it to $8.4 \times 10^{-4}$ kg m$^{-2}$ at the minimum film (Section 5.3). The film-and-coolant temperature, a design choice rather than an uncertainty, gives up to $9.5 \times 10^{-4}$ kg m$^{-2}$ at 800 K. Splashing, bubble bursting, droplet ejection and the volumetric or explosive boiling of the superheated layer [11] are not modeled, and they can remove more liquid than evaporation [16]. In Prototype the thick film evaporates $8.9 \times 10^{-6}$ kg m$^{-2}$ per shot, two orders of magnitude less, because its surface peaks at 1445 K.

The model's own vapor-pressure law gives the load on the Demo film, with a saturation pressure of 2.4 MPa at the 2218 K surface peak. The recoil of the evaporating vapor, 0.55 times that (computed here from the sonic-point state of Anisimov [37]), is 1.33 MPa, with an impulse of 1.14 Pa s over the debris-ion arrival. The liquid below the surface is hotter, 2483 K at 2.3 μm depth, 411 K above the boiling point at the recoil pressure and only 119 K below the lowest limit of superheat, the van der Waals spinodal at 27/32 of the lowest critical temperature in Zaghloul and Hassanein [38] (computed here). Volumetric boiling of this layer cannot be excluded. No cavitation threshold of liquid lithium was found in a source that could be read in full. The film model is static, so these numbers describe the load on the film, not its response (Supplementary Section S3).

**Table 5.** Lithium film on 1 cm steel at $T_{\mathrm{Li}} = 600$ K, without vapor shielding or neutron heating unless stated (Section 6 gives the shielded single-shot values). Single shots unless stated otherwise; repeated-shot minima are interpolated between computed films, those of Demo including direct runs near each minimum (45 and 56 μm without neutron heating or vapor shielding). The rows on the helium, the displacements and the viscous stopping behind the 925 K minimum film use the repeated-shot minima (Experimental: single shot). Helium in the steel counted with all of it retained. Experimental, Prototype and Demo are the HiPER scenarios defined in [3, 8] (Table 1).

| Quantity | Experimental | Prototype | Demo |
|---|---|---|---|
| 925 K [8] minimum film, single shot (range over sensitivity cases) | 7.2 μm | 11.9 μm (8.2–34 μm; 34 μm at $T_{\mathrm{Li}}$ = 800 K) | 30.1 μm (24–79 μm; 79 μm at $T_{\mathrm{Li}}$ = 800 K) |
| … repeated shots to the quasi-steady state | single shots | 12.1 μm | 44 μm |
| … with steel neutron heating at 0.5 / 1 times the W value (44 μm without; minima with direct runs) | – | – | 48 / 53 μm |
| … with vapor shielding, $f_{\mathrm{rr}}$ = 0, $s$ = 1 (Section 6) | – | – | 42 μm |
| 873 K (Li/steel compatibility [1]) minimum film, single / repeated shots | – | 14.8 / 15.1 μm | 37 / 56 μm |
| … repeated, with steel neutron heating at 0.5 / 1 (56 μm without; minima with direct runs) | – | – | 63 / 83 μm |
| Helium-exclusion film: steel below $10^{18}$ / $10^{19}$ $m^{-2}$ of He [3] after $10^6$ shots | 165 / 133 μm | 174 / 140 μm | 184 / 156 μm |
| … after one full-power year ($3.2\times10^7$ shots at 1 Hz, $3.2\times10^8$ at 10 Hz) | – | 205 / 184 μm | 236 / 205 μm |
| He reaching the steel per shot behind the 925 K minimum film (repeated shots; Experimental single shot) | $1.4\times10^{16}$ $m^{-2}$ | $1.6\times10^{16}$ $m^{-2}$ | $4.8\times10^{15}$ $m^{-2}$ |
| … shots to $10^{18}$ / $10^{19}$ $m^{-2}$ (W surface, Table 4) | 73 / 726 (45 / 449) | 61 / 608 (30 / 303) | 208 / 2084 (10 / 98) |
| … ion dpa per shot, first 5 μm of steel (W surface) | $2.9\times10^{-6}$ dpa ($8.1\times10^{-6}$) | $2.0\times10^{-6}$ dpa ($1.2\times10^{-5}$) | $6.3\times10^{-7}$ dpa ($3.7\times10^{-5}$) |
| Li evaporated per shot, thick film | $8.0\times10^{-7}$ kg $m^{-2}$ | $8.9\times10^{-6}$ kg $m^{-2}$ | $7.4\times10^{-4}$ kg $m^{-2}$ |
| … over the whole chamber | $2.5\times10^{-4}$ kg | $4.7\times10^{-3}$ kg | 0.39 kg |
| … range over the single-shot sensitivity cases | – | $9.8\times10^{-7}$–$4.6\times10^{-5}$ kg $m^{-2}$ ($4.6\times10^{-5}$ at $T_{\mathrm{Li}}$ = 800 K) | $4.2\times10^{-4}$–$9.5\times10^{-4}$ kg $m^{-2}$ ($9.5\times10^{-4}$ at $T_{\mathrm{Li}}$ = 800 K, outside the plausible range; Section 5.2) |
| Flow speed for full replacement every shot from one pole, $\pi R f$ ($2Rf$) | – | 20 m $s^{-1}$ (13 m $s^{-1}$) | 204 m $s^{-1}$ (130 m $s^{-1}$) |
| Viscous stopping distance of a free film of the 925 K minimum (repeated shots), at 15 m $s^{-1}$ [36] / $\pi R f$ | – | 0.7 / 1.0 mm | 0.9 / 13 cm |

### 5.3 Repeated shots and the steel behind the film

The single-shot results assume steel that keeps no heat from earlier shots, but in repeated operation it accumulates heat as the tungsten wall does. At 10 Hz the single-shot minimum fails (Figure 7). With a 31.6 μm film the steel exceeds 925 K from the ninth shot and settles at a 1006 K peak, its pre-shot front face rising from 600 to 783 K. A direct run at 45 μm settles at 917 K, so the Demo minimum for repeated operation is 44 μm, about 1.5 times the single-shot value. At 1 Hz the steel cools back to 609 K before every shot, and the Prototype minimum rises only from 11.9 to 12.1 μm.

The 925 K limit is a structural one. Applied to the peak, the separate compatibility limit of about 600 °C (873 K) for ferritic steel and lithium coolant [1] raises the minimum film to 37 μm in a single shot and 56 μm in repeated operation (14.8 and 15.1 μm in Prototype). Both limits are crossed only at the peak of each shot. Averaged over a cycle, the steel front face is only 6–10 K above its pre-shot temperature (Demo, 17.8–100 μm), at 777 K behind 45 μm. Behind 31.6 μm it spends 0.10 ms of each 100 ms above 925 K and 0.32 ms above 873

K (Supplementary Table S6).

The steel also receives the ions that pass the film. The 3.5 MeV burn alphas have a range of 50 μm in lithium at 600 K (60 μm at 4.0 MeV, the top of their bin), but the high-energy tail of the burn helium reaches the steel through up to 0.68 mm. Behind the 44 μm minimum, $4.8 \times 10^{15}$ He $m^{-2}$ reach the steel per shot (Supplementary Fig. S3). If the steel keeps all of it, it reaches $10^{18}$ and $10^{19}$ $m^{-2}$ after 208 and 2084 shots, against 10 and 98 for the tungsten surface. Its first 5 μm receive $6 \times 10^{-7}$ dpa per shot, about a sixtieth of the tungsten-surface dose (Table 5).

Helium retention gives a third criterion. Keeping the steel below $10^{18}$ $m^{-2}$ for $10^6$ shots needs 184 μm in Demo, 174 μm in Prototype and 165 μm in Experimental. A full-power year needs 236 μm in Demo and 205 μm in Prototype; this criterion binds in every scenario. Both films are set by the upper part of single bins of the burn spectrum and depend on how its energy is matched to the accounting (Supplementary Section S3). Each of these films sits at a step created by the spectrum binning, because a sub-bin of the spectrum either reaches the steel or not, and is given as the next whole micrometer above the step (183.2 μm for $10^6$ Demo shots). A direct run at 183 μm settles at a 754 K steel peak.

Per shot the steel face swings by 147 K behind the 45 μm film, equivalent to a square pulse with a heat-flux factor of 1.9 MW $m^{-2}$ $s^{1/2}$, on top of 27 MW $m^{-2}$ $s^{1/2}$ in the 20 ns x-ray flash that the film lets through. No thermal-fatigue rule for pulsed loads on steel was found, so these are indicators only (Supplementary Table S6).

Neutron heating of the steel (Section 2.3), which the film does not receive, raises the repeated-shot minimum to 48 μm with half the tungsten heating per unit volume and 53 μm with the full value, and the 873 K film to 63 and 83 μm. They are interpolated between the runs at 31.6, 56.2 and 100 μm and direct runs near each minimum (at 45 and 56 μm without neutron heating). The two cases bracket the effect (Section 2.3).

Of the energy that a Demo shot deposits in a 45 μm film and its steel, 27 % leaves with the vapor (21 % as latent and 6 % as sensible heat), 18 % with the replaced film, whose enthalpy above 600 K the model removes before every shot, and 53 % passes through the steel to the coolant, a time-mean 0.43 MW $m^{-2}$ against 0.82 MW $m^{-2}$ behind tungsten (Supplementary Table S6). If the latent heat of the vapor returned to the film between shots, as a bounding case for vapor condensing on the wall, the steel behind 45 μm would settle at 944 K instead of 917 K. The minimum films would rise to 50 μm (925 K) and 69 μm (873 K), still far below the helium-exclusion film (Supplementary Section S3).

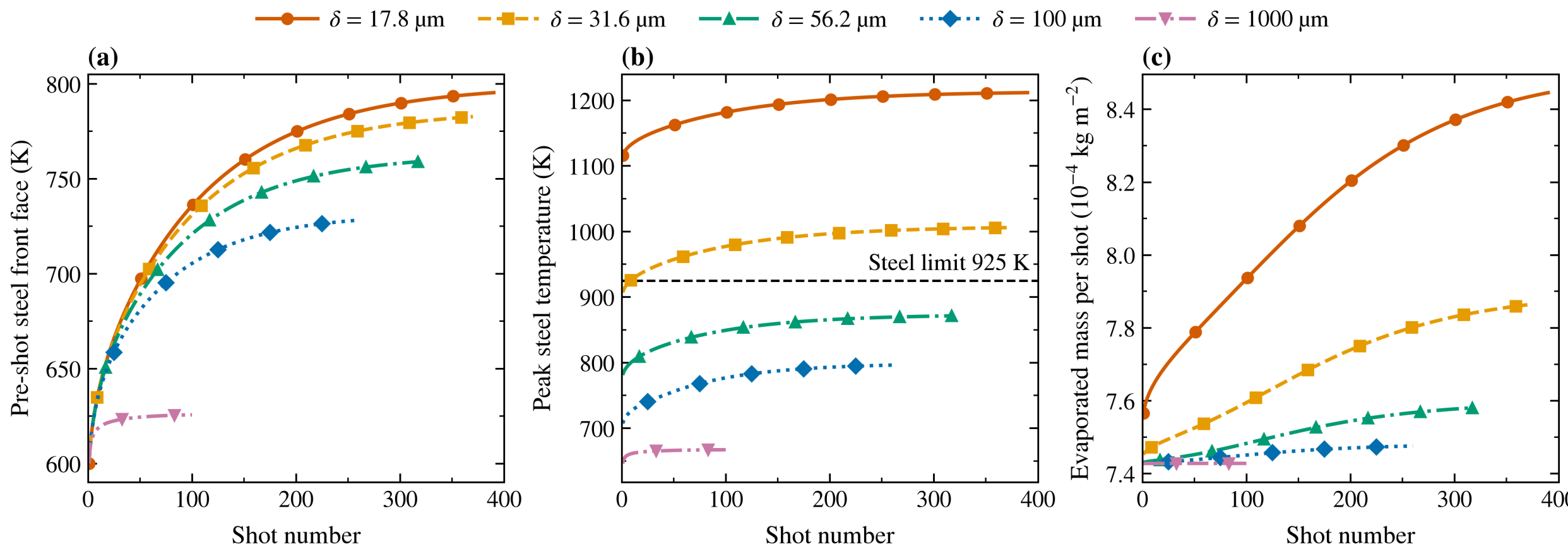

**Figure 7.** Demo, lithium film on 1 cm steel at 10 Hz, one curve per film thickness: steel front-face temperature before each shot (a), peak steel temperature in each shot with the 925 K limit [8] dashed (b), and lithium evaporated per shot (c). The film is replaced at full thickness and 600 K before every shot; the steel is carried forward.

The evaporated mass per shot rises as the steel warms, by 6 % at 31.6 μm and by less in thicker films, so the single-shot value of $7.4 \times 10^{-4}$ kg $m^{-2}$ holds to within 6 % for any film that passes. With neutron heating, a shot at the minimum films of 48 and 53 μm evaporates $7.9 \times 10^{-4}$ and $8.4 \times 10^{-4}$ kg $m^{-2}$, 7 % and 13 % more (0.42 and 0.45 kg over the chamber). The steel is heated mainly by the ions. The film replacement acts as a heat sink on the steel front face whose effect relative to a real flowing film is unknown (Supplementary Section S3).

### *5.4 Film renewal*

Replacing the whole film once per shot from inlets at one pole would require the lithium to sweep the wall within one period, at $\pi R f = 204$ m $s^{-1}$ in Demo and 20 m $s^{-1}$ in Prototype, against the 15 m $s^{-1}$ of demonstrated free-surface lithium flows [36]. This arithmetic is only a reference point, because a film this thin cannot travel that far.

Without a driving force, wall friction stops a free film of thickness $\delta$ launched at speed $U$ within about $U\delta^2/$

(3ν), where ν is the kinematic viscosity (Supplementary Section S3). With the lithium viscosity at 600 K [25], a film of the 44 μm minimum stops within 0.9 cm at 15 m $s^{-1}$ and within 13 cm even at 204 m $s^{-1}$. A 184 μm helium-exclusion film stops within 17 cm and 2.3 m at the same speeds. Renewal is therefore a problem of injection spacing. The liquid has to enter the wall every few centimeters, through distributed or porous injection, rather than sweep the chamber from a few inlets. Whether such a wall keeps the film at its minimum thickness everywhere is not addressed here.

If only part of the film is renewed each shot, the retained lithium keeps its heat and starts the next shot at 720–791 K. Renewing 30 % per shot raises the Demo minimum films to 59 μm (925 K) and 86 μm (873 K), renewing 3 % to 75 μm and beyond 100 μm, against 45 and 56 μm interpolated in the same way with full renewal. At the same films the evaporation rises by 17–19 % with 30 % renewal and by 25–32 % with 3 % renewal (Supplementary Section S3). Intermediate fractions and Prototype were not computed.

## 6. Results: vapor shielding

Vapor shielding matters only for lithium in Demo (Figure 8 and Supplementary Table S4). Tungsten evaporates too little. Even in a strongly degraded case its vapor would stop only $3–7\times10^{-6}$ of the ion energy (Supplementary Section S2). In Prototype the column stops 0.08–0.17 % of the ion energy and changes the evaporated mass and the minimum film by 1 % or less. In Experimental the change is below 0.1 %.

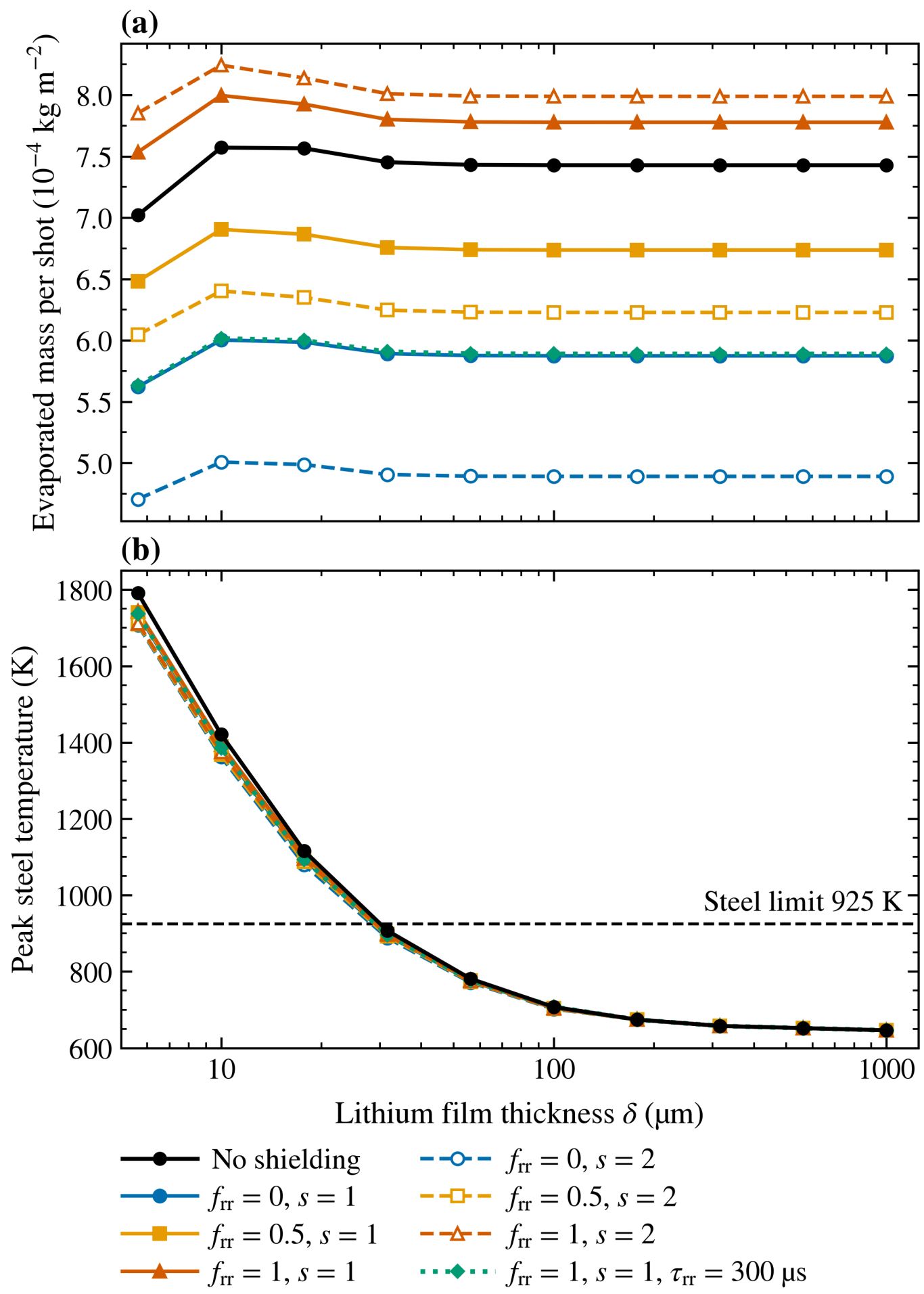

**Figure 8.** Demo, lithium film on 1 cm steel, single shot at $T_{Li}$ = 600 K, with the reduced vapor-shielding model: lithium evaporated per shot (a) and peak steel temperature (b) against film thickness, for the re-radiated fractions $f_{rr}$ = 0, 0.5 and 1 with cold ($s$ = 1, solid) and enhanced ($s$ = 2, dashed) stopping, and for $f_{rr}$ = 1, $s$ = 1 with the return spread over 300 μs (dotted, on top of $f_{rr}$ = 0, $s$ = 1). Black: no shielding; dashed in (b): the 925 K steel limit [8].

Coupled in time, the vapor stops much less than the static bound suggests. A column holding the whole per-shot mass from the start would stop 24 % of the ion energy, but the column grows during the shot. Cold vapor stops 7.5–8.8 % of the ion energy, and vapor with twice that stopping power 12–15 %.

The fate of the stopped energy decides the sign of the effect. If none of it returns to the wall ($f_{rr}$ = 0), the thick-film evaporation falls by 21 % with cold vapor and by 34 % with $s$ = 2. If all of it returns at once ($f_{rr}$ = 1), it rises by 5 % and 8 %. A return spread over 300 μs acts like $f_{rr}$ = 0, and the literature on lithium vapor points to the low end of $f_{rr}$ [39]. An equilibrium Saha estimate gives the cold vapor an ionized fraction of only 0.8–3

%. This makes $s = 1$, and with it the range of $-21$ % to $+5$ %, the closer of the two cases (Supplementary Section S2). Stopping factors of 5 and 10 test the sensitivity of the model rather than a likely state of the vapor (Supplementary Table S4).

The steel hardly notices the vapor, because the column stops mainly debris ions, which deposit their energy in the first micrometers of the film. Shielding lowers the single-shot minimum film from 30.1 μm to 28.4–29.5 μm, 2–6 % less. In repeated operation, with $f_{rr} = 0$ and $s = 1$ and the column restarting every shot, it lowers the minimum, with direct runs at 43 and 54 μm, from 44 to 42 μm (4.5 %). The film for the 873 K compatibility limit falls from 56 to 54 μm.

## 7. Discussion

### 7.1 The two walls side by side

The walls are compared here by the same indicators: temperatures, melting, erosion, and the helium and displacement damage in the plasma-facing material and in the steel behind it. In the Experimental and Prototype scenarios both pass every thermal test of this model. The tungsten surface, from 300 K in Experimental and 600 K in Prototype, stays far from melting and erodes negligibly (Section 4). A lithium film at 600 K keeps the steel below 925 K from 7.2 μm in Experimental and below 925 and 873 K from 12.1 and 15.1 μm in Prototype. A thick film evaporates only 0.3–5 g per shot over the whole chamber (Section 5).

Both walls collect helium. The tungsten surface reaches $10^{18}$–$10^{19}$ m$^{-2}$ within 30–303 shots in Prototype and 45–449 in Experimental, the onset of surface modification. The lithium is renewed with the film, but the burn helium passes it. Behind a film of the thermal minimum the steel reaches the same fluences within 61–608 and 73–726 shots, and only films of 174 and 165 μm keep it below $10^{18}$ m$^{-2}$ for $10^6$ shots (Section 5.3).

In Demo the walls differ in kind. The quasi-steady tungsten surface peaks 378 K below melting, 275–319 K with neutron heating. That margin depends on how the ion spectra are constructed. The raw tables, with 13 % more ion energy and 18 % more helium than the production spectra, raise the single-shot peak by 526 K, and comparing that rise with the quasi-steady margin is an extrapolation (Supplementary Section S6). It survives a 15 % lower conductivity, which raises the peak by 180 K, but a degraded near-surface layer can erase it. Melting can be neither excluded nor confirmed (Sections 4.2–4.5). The interface exceeds the steel limit, slightly without neutron heating and clearly with it. The helium and design-rule counts, 10–98 shots and about ten cycles, mark the onset of surface modification in the first 5 μm. That depth is one twentieth of the 100 μm of tungsten that Garoz et al. found sufficient to protect the steel, in an armor whose fatigue life they estimated at about $10^6$ cycles [8]. Its life is not computed here.

The lithium film must be 44 μm thick to keep the steel below 925 K, 56 μm for the 873 K limit and 184 μm to keep the helium in the steel below $10^{18}$ m$^{-2}$ for $10^6$ shots, the criterion that binds. If the latent heat of the vapor returned to the film, the two thermal minima would rise to 50 and 69 μm, and if only 30 % of the film were renewed per shot, to 59 and 86 μm. A quarter of the absorbed energy leaves with the vapor. The 45 μm film evaporates 0.41 kg per shot, 4.1 kg s$^{-1}$ at 10 Hz that the chamber must clear (3.9 kg s$^{-1}$ for a thick film). Renewal needs injection every few centimeters (Sections 5.2–5.4). Tritium bred in the film would have to be extracted from it and from the cleared vapor, which this study does not address.

The comparison is therefore not a ranking. The tungsten margin to melting rests on poorly known surface properties and on the construction of the ion spectra, and the consequence of its early surface modification is not known. Whether a lithium wall is feasible depends on whether the film survives the shot, under a 1.33 MPa recoil pressure with a superheated layer below its surface (Section 5.2), and on whether the chamber can be cleared between shots. This study assesses neither, so it gives the film thickness and the evaporated mass but no verdict on feasibility. Neither wall model contains what would decide between them: the stress history and damage evolution of tungsten and the hydrodynamics of the film and its vapor.

### 7.2 What vapor shielding changes

Shielding is asymmetric between the walls. Tungsten evaporates too little for its vapor to matter, so every tungsten result, including the melting margin, is unaffected. For lithium the vapor made by the debris ions stops only a small part of the later ion energy, because the column grows during the shot. Whether it lowers or raises the evaporation depends mostly on how much of the stopped energy is radiated back to the wall during the shot (Section 6). An equilibrium Saha estimate gives the cold vapor an ionized fraction of only 0.8–3 %. This favors its stopping ($s = 1$) and with it the range of $-21$ % to $+5$ % on the evaporated mass (Supplementary Section S2). Even with the smallest re-radiated fraction the chamber must clear about 0.3 kg of vapor per shot, and the minimum film moves by only a few percent. No conclusion of Section 7.1 changes, though fixing the two parameters needs radiation hydrodynamics [22, 23]. Liquid-metal experiments under very different conditions show the same direction of the effect (Supplementary Section S3).

### *7.3 Tungsten grades*

Neither potassium doping nor an oxide dispersion changes the Demo indicators that limit tungsten, the interface temperature and the helium count, which do not depend on the grade in this model. Apart from the melting margin, which the conductivity assumed for the oxide grade, 0.8 times that of pure tungsten, cuts from 378 to 138 K (Section 4.6), the grades move only the recrystallization indicator, by an amount comparable with the uncertainty of the onset temperature. Their possible advantages need a stress calculation (Supplementary Section S6).

### *7.4 Limitations*

The largest gap is the absence of thermomechanics. The model computes no stress or strain, so the tungsten damage indicators of Section 4.4 rest on a published helium threshold and an empirical design rule, not on a computed damage state. No life is computed, though a thermomechanical calculation driven by the temperature histories computed here would give one. The second gap is material damage under irradiation. The helium threshold is uncertain by more than an order of magnitude (Section 4.4 and Supplementary Section S6) and is taken over for the steel behind the film. Displacement damage enters only the helium model and the steel indicators (Supplementary Section S7.4 and Section 5.3). Its effect on the conductivity is treated only as a sensitivity.

The third gap is the system around the lithium film: chamber clearing, recondensation, film stability under the recoil pressure, boiling of the superheated layer, waves and droplet ejection. This study leaves it out entirely. The evaporated masses of Sections 5 and 6 are surface-evaporation estimates. Splashing and boiling can remove far more mass than surface evaporation [16], and the same holds for the melt layer of a damaged tungsten surface (Section 4.5).

The displacement damage computed for the ions points to a possibility, not a result. After about $3 \times 10^3$ Demo shots, 5 minutes at 10 Hz, the ions alone could bring the first 5 μm to 0.1 dpa, the dose from which the measured reduction of tungsten's conductivity saturates at about half (Supplementary Section S6). The neutrons alone would need about 14 hours at the dose per shot used in the helium model, $2 \times 10^{-7}$ dpa, which this work derives from the $10^{-6}$ dpa $s^{-1}$ of Sharafat et al., Section 3.2 (material not stated), at the 5 Hz HAPL rate of their Section 4.5 [18], or 2.5–8 days at the 4.5–14.5 dpa per full-power year that Gilbert et al. computed for tungsten first-wall armor in a magnetic-fusion DEMO [40]. If the conductivity then fell as measured, the Demo surface could melt on every shot (Section 4.5). It may not. The reductions were measured at room temperature under steady irradiation, whereas the Demo surface stays near 921 K between shots and passes 2000 K in every shot, which may repair much of the damage [41]. The dose is an average counted with a standard displacement model, not measured (Supplementary Section S6). An irradiation experiment under pulsed heating could settle it.

Several simplifications act in a known direction. The slightly too shallow deposition of keV ions in tungsten is conservative, and so is the neglect of surface radiation if the chamber interior is not isothermal (it has no effect under the isothermal-chamber argument of Section 2.2); leaving neutron heating out of the baseline is not. The source term is the ARIES spectrum scaled by yield, not the HiPER group's own. How its tables are made into spectra is a first-order input (the raw tables raise the Demo single-shot peak by 526 K). Supplementary Section S5 lists each limitation with the results it affects and its direction.

### *7.5 What each wall would need at Demo conditions*

Moving the coolant closer to the surface removes the interface problem of the armor. With the coolant 3.5 mm behind the surface instead of 11 mm, the Demo interface peaks at 752 K, with a mean of 744 K (758–765 K with neutron heating), and a lithium film on 3.5 mm of steel needs 37 μm instead of 44 μm, both with direct runs (Supplementary Section S6). Whether 2.5–3.5 mm of steel can carry the coolant pressure and the thermal stresses is not addressed.

Each wall has its own conditions, and for the tungsten armor three must hold. First, the interface must stay below the steel limit, which needs a modest improvement of the cooling path without neutron heating and the coolant closer to the surface with it. Second, the near-surface conductivity must stay close to that of undamaged tungsten, which needs a measured effusivity of the as-installed and of the irradiated surface. Third, the surface modification that sets in with $10^{18}$–$10^{19}$ $m^{-2}$ of helium must not compromise the function of the armor.

The lithium film must be at least 44 μm thick at 10 Hz, 48–53 μm with neutron heating, 56 μm for the 873 K limit and about 184 μm to keep the helium in the steel below $10^{18}$ $m^{-2}$ for $10^6$ shots. It must be renewed by injection a few centimeters apart, and the chamber must clear 0.22–0.45 kg of lithium vapor between shots (full renewal). Whether the film survives the shot and whether the chamber can be cleared are not assessed here (Supplementary Section S6).

## 8. Conclusions

A 1 mm tungsten armor and a renewable lithium film were compared as the first wall of the evacuated HiPER chamber in a single calculation and judged by the same indicators. In the Experimental and Prototype scenarios both walls meet their thermal limits; the lithium film does so at thicknesses of 7.2–15.1 μm and above. Both collect helium. The tungsten surface reaches the $10^{18}$–$10^{19}$ $m^{-2}$ associated with surface modification within 30 to 449 shots. Behind a film of the thermal minimum, the steel does so within 61 to 726 shots. Films of 165 to 174 μm keep the steel below $10^{18}$ $m^{-2}$ for $10^6$ shots.

In Demo, the quasi-steady tungsten surface peaks at 3317 K, 378 K below melting and 275 to 319 K with neutron heating. A 15 % lower conductivity would not close this margin, but a conductivity degraded as measured after irradiation would. The margin also depends on how the ion spectra are constructed: the raw ion tables, with 13 % more ion energy and 18 % more helium than the production spectra, raise the single-shot peak by 526 K, enough to bring the surface to melting by extrapolation (not in a computed run). The W/steel interface exceeds the 925 K limit slightly without neutron heating and by 74 to 125 K with it, by an amount that the cooling path controls. The helium and design-rule counts, 10 to 98 shots and about ten cycles, mark the onset of surface modification, not a life. Garoz et al. estimated a fatigue life of about $10^6$ cycles for this armor. This study computes none.

The lithium film must be 44 μm thick to keep the steel below 925 K at 10 Hz, 48 to 53 μm with neutron heating and 56 μm for the 873 K compatibility limit. Keeping the helium in the steel below $10^{18}$ $m^{-2}$ for $10^6$ shots, the binding criterion, needs 184 μm. The thermal minima rise to 50 and 69 μm if the latent heat of the vapor returns to the film, and to 59 and 86 μm if only 30 % of the film is renewed per shot. A thick film evaporates $7.4 \times 10^{-4}$ kg $m^{-2}$ per shot, 0.39 kg over the chamber, a surface-evaporation estimate with a plausible range of $4.2–8.4 \times 10^{-4}$ kg $m^{-2}$ for a fully renewed film. Splashing, bubble bursting, droplet ejection and volumetric boiling are not modeled, and renewing a film this thin needs injection every few centimeters.

Whether a lithium wall is feasible depends on whether the film survives the shot and whether the chamber can be cleared between shots, which this study does not assess. It gives the film thickness and the evaporated mass, but no feasibility verdict.

Vapor shielding, treated as a parameterized screening calculation coupled in time to the evaporation, changes the Demo lithium evaporation by $-21$ % to $+5$ % for the cold vapor that an equilibrium Saha estimate supports. With doubled stopping, the change is $-34$ % to $+8$ %. It moves the minimum film by 6 % or less and leaves every tungsten result unchanged.

## Declaration of generative AI and AI-assisted technologies in the manuscript preparation process

During the preparation of this work the author used Claude Code (Anthropic) for three tasks: to check the values in the data, benchmark and convergence tables of the early Supplementary sections against the saved simulation outputs; to review the simulation runs, including their input settings, numerical convergence and consistency between related cases; and to help find relevant published literature, all of which the author read and verified before citing. After using this tool, the author reviewed and edited the content as needed and takes full responsibility for the content of the publication.

## CRediT authorship contribution statement

Shohruxbek Bozorov: Conceptualization, Methodology, Software, Validation, Formal analysis, Investigation, Data curation, Writing – original draft, Writing – review and editing, Visualization.

## Declaration of competing interest

The author declares that he has no known competing financial interests or personal relationships that could have appeared to influence the work reported in this paper.

## Funding

This research did not receive any specific grant from funding agencies in the public, commercial, or not-for-profit sectors.

## Code and data availability

The code (the Python package `ife_wall` with its run scripts, tests and validation scripts) and the outputs behind every table and figure are archived at Zenodo (https://doi.org/10.5281/zenodo.23577155) and developed at https://github.com/Shohruxbek-cyber/first-wall-comparison, under the MIT licence. The commands that regenerate each table and figure are listed in Supplementary Section S8. Third-party input data (NIST stopping-power and attenuation tables, CXRO photoabsorption data, and SRIM-2012 stopping and range tables) are not redistributed with this work. The repository provides scripts and checksums to retrieve or regenerate them from their original sources.

## Acknowledgements

The author is deeply grateful to Professor Jake Blanchard (University of Wisconsin–Madison) for his guidance throughout this work, to Professor Hiroaki Abe (The University of Tokyo) for his careful review of the manuscript, to Dr. Gerald Pintsuk (Forschungszentrum Jülich) and Dr. Jiří Matějíček (Institute of Plasma Physics of the Czech Academy of Sciences) for valuable discussions and expert advice, to T. Hallam, R. Mitteau, R. González-Arrabal, W. Meier and Z. Zhou for discussions and suggestions, and to Min Ki Jung for his support with the arXiv submission. Any remaining errors are the author's own.

## Supplementary material

The supplementary material contains the thermal model, numerical scheme and verification (S1); source-term, deposition and vapor-shielding details (S2); benchmark detail, further lithium-film results and helium in the steel (S3); the literature inputs (S4); the full limitations table (S5); further tungsten results, damage indicators and tungsten grades (S6); the helium-retention model (S7); and the reproduction commands (S8).

# Appendix A. Nomenclature

**Table A.1.** Symbols. SI units unless stated; temperatures in kelvin except in the Blanchard–Martin and Raffray comparisons, which keep the °C of the original papers. Some letters have a second meaning in the supplementary material, defined where it is used: $\mu$, the dynamic viscosity of lithium (Section S3); $N$, the index of the last cell (Section S1) and the number of ions in $dN/dE$ (Section S2); $n$, the time level (Section S1), a number density (Section S3) and the shot number (Section S7); $G$, a diffusive conductance; $\nu$, the attempt frequency of detrapping; $c$, the recombination coefficient (these three in Section S7, Table S14); $f(y)$, the Thomas–Fermi scattering function (Section S7.4).

| Symbol | Meaning | Unit |
|---|---|---|
| $x$ | depth below the current plasma-facing surface | m |
| $\mu$ | areal mass from a fixed material reference (the initial surface), the Lagrangian coordinate of Eq. (1), $d\mu = \rho\, dx$ | kg m$^{-2}$ |
| $\xi$ | areal mass between the current surface and depth $x$, $\mu - M_e$ | kg m$^{-2}$ |
| $M_e$ | areal mass evaporated so far; the surface is at $\mu = M_e$ | kg m$^{-2}$ |
| $m_i$, $\Delta x_i$ | areal mass and thickness of cell $i$ ($\Delta x_i = m_i/\rho(T_i)$) | kg m$^{-2}$, m |
| $T$, $T_0$, $T_{surf}$, $T_c$ | temperature; surface-cell temperature; extrapolated surface temperature; coolant temperature | K |
| $H$ | specific enthalpy (reference 298.15 K for W and steel; the liquid at 453.7 K for Li) | J kg$^{-1}$ |
| $\rho$, $c_p$, $k$ | density, specific heat, thermal conductivity | kg m$^{-3}$, J kg$^{-1}$ K$^{-1}$, W m$^{-1}$ K$^{-1}$ |
| $\varphi$, $d$ | factor applied to the conductivity of solid tungsten, uniformly or in a surface layer of thickness $d$ | –, m |
| $T_m$, $L_f$, $\Delta T_m$ | melting point, latent heat of fusion, width of the melting interval (1 K) | K, J kg$^{-1}$, K |
| $p_{sat}$, $L_v$, $M$ | saturated vapor pressure, latent heat of vaporization, molar mass | Pa, J kg$^{-1}$, kg mol$^{-1}$ |
| $\alpha$ | evaporation coefficient of the Hertz–Knudsen–Langmuir flux | – |
| $j$ | evaporative mass flux | kg m$^{-2}$ s$^{-1}$ |
| $R_g$ | molar gas constant, 8.314462618 | J mol$^{-1}$ K$^{-1}$ |
| $h$ | coolant heat-transfer coefficient | W m$^{-2}$ K$^{-1}$ |
| $\langle T_{if}\rangle$, $\langle q\rangle$ | time-mean W/steel interface temperature at periodic steady state; time-mean absorbed heat flux, the repetition rate times the energy absorbed per shot | K, W m$^{-2}$ |
| $L_s$, $k_s$ | thickness and thermal conductivity of the steel | m, W m$^{-1}$ K$^{-1}$ |
| $G_{i+½}$ | thermal conductance between cells $i$ and $i + 1$ | W m$^{-2}$ K$^{-1}$ |
| $S_i$ | power deposited in cell $i$ | W m$^{-2}$ |
| $q_v$ | volumetric heating (x-rays and ions; neutrons in the separate case of Section 4.2) | W m$^{-3}$ |
| $Y$, $R$, $f$ | target yield, chamber radius, repetition rate | J, m, Hz |
| $c$ | speed of light (the x-rays reach the wall at $R/c$) | m s$^{-1}$ |
| $\kappa$ | photon mass attenuation coefficient | m$^2$ kg$^{-1}$ |
| $R_p$, $f_{TRIM}$ | projected range (areal mass); ratio of TRIM mean range to table projected range | kg m$^{-2}$, – |
| $\eta_E$, $\eta_N$ | backscattered energy and number fractions (TRIM) | – |
| $\sigma_v$ | areal mass of the vapor column in front of the wall | kg m$^{-2}$ |
| $s$ | stopping power of the vapor relative to cold vapor (ionization) | – |
| $f_{rr}$, $\tau_{rr}$ | fraction of the energy stopped in the vapor returned to the wall; time constant of the return | –, s |
| $E_{st}$, $P_v$ | energy stored in the vapor awaiting return; power stopped in the vapor | J m$^{-2}$, W m$^{-2}$ |
| $q_{rr}$ | heat flux returned to the wall from the vapor store, Eq. (3) | W m$^{-2}$ |
| $F_{HF}$ | heat-flux factor, absorbed power density × (pulse duration)$^{½}$ | MW m$^{-2}$ s$^{½}$ |
| $N$ | number of allowed load cycles of the design rule [26] | – |
| $\delta$, $T_{Li}$ | lithium film thickness; initial temperature of film, steel and coolant in the Li cases | m, K |
| $U$, $\nu$ | mean speed of a lithium film; kinematic viscosity of lithium | m s$^{-1}$, m$^2$ s$^{-1}$ |
| fpy | full-power year, $3.156\times10^7$ s at the design repetition rate | s |

## Appendix B. Material properties

**Table B.1.** Material properties, their sources, the range over which the source gives them, and what the code does outside that range. ESTIMATED marks a property without a source.

| Material | Property | Expression or value | Source | Range given by the source | Outside the range |
|---|---|---|---|---|---|
| W | $T_m$ | 3695 K | [24] | – | – |
| | $L_f$ | 52.3 kJ mol⁻¹ (284.5 kJ kg⁻¹), spread over 1 K | [24] | – | – |
| | $\rho$, solid | $19.25 - 2.66207\times10^{-4}\,\Delta T - 3.0595\times10^{-9}\,\Delta T^2 - 9.5185\times10^{-12}\,\Delta T^3$ g cm⁻³, $\Delta T = T - 293.15$ K | [24] | 300–3695 K | evaluated as written |
| | $\rho$, liquid | $16.267 - 7.679\times10^{-4}\,(T - T_m) - 8.091\times10^{-8}\,(T - T_m)^2$ g cm⁻³ | [24] | 3695–6000 K | no run reached 6000 K |
| | $c_p$, solid | $21.868372 + 8.068661\times10^{-3}\,T - 3.756196\times10^{-6}\,T^2 + 1.075862\times10^{-9}\,T^3 + 1.406637\times10^{4}/T^2$ J mol⁻¹ K⁻¹ ($T$ < 3080 K); $2.022 + 1.315\times10^{-2}\,T$ (3080–3695 K) | [24] | 300–3080 K and 3080–3695 K | evaluated as written |
| | $c_p$, liquid | 51.3 J mol⁻¹ K⁻¹ | [24] | liquid data to 5400 K | constant |
| | $k$, solid | $149.441 - 45.466\times10^{-3}\,T + 13.193\times10^{-6}\,T^2 - 1.484\times10^{-9}\,T^3 + 3.866\times10^{6}/T^2$ W m⁻¹ K⁻¹ | [24] | 300–3695 K (data 2–3000 K) | evaluated as written |
| | $k$, liquid | $66.6212 + 0.02086\,(T - T_m) - 3.7585\times10^{-6}\,(T - T_m)^2$ W m⁻¹ K⁻¹ | [24] | 3695–6000 K | evaluated as written |
| | $p_{sat}$ | $10^{(12.74 - 44485/T)}$ Pa | [1], Table 2 | not stated | evaluated as written |
| | $L_v$ | 4009 kJ kg⁻¹ | [1], Table 2 | – | sensitivity: 4630 kJ kg⁻¹, the Clausius–Clapeyron slope of the same $p_{sat}$ law |
| | $M$ | 183.84 g mol⁻¹ | [42] | – | – |
| Li, liquid | $T_m$ | 453.7 K | [25] | – | the film never solidifies (runs start at ≥ 500 K) |
| | $\rho$ | $562 - 0.100\,T$ kg m⁻³ | [25] | correlations extrapolated to about 1600 K by the source | evaluated as written |
| | $c_p$ | 4169 J kg⁻¹ K⁻¹ | [25] | constant | constant |
| | $k$ | $21.874 + 0.056255\,T - 1.8325\times10^{-5}\,T^2$ W m⁻¹ K⁻¹ | [25] | within 3 % of the source correlation at 800 K | evaluated as written |
| | $p_{sat}$ | $\log_{10} p_{sat} = 10.015 - 8064.5/T$, $p_{sat}$ in Pa | [25] | data between about 800 and 1800 K | extrapolated; the Demo surface reaches 2218 K behind thick films, 2233 K behind 10 µm and 2874 K with $\alpha = 0.1$ |
| | $L_v$ | 22.25 MJ kg⁻¹ (154 kJ mol⁻¹) | ESTIMATED: Clausius–Clapeyron slope of the $p_{sat}$ correlation | – | NIST: 149.4 ± 3.4 kJ mol⁻¹ (ideal gas) at the 1617 K boiling point [43] |
| | $M$ | 6.94 g mol⁻¹ | [42] | – | – |
| Steel (substitute for ODS-RAFM) | $\rho$, $c_p$, $k$ | 7800 kg m⁻³; $500 + 0.6\,T$(°C) J kg⁻¹ K⁻¹; 33 W m⁻¹ K⁻¹ | [7], App. A | not stated | table limit 1800 K; a run that exceeds it stops (Demo Li films ≤ 3.2 µm) |
| | photon attenuation | pure Fe | ASSUMPTION | – | – |

**Material properties (Section 2.6).** Tungsten follows the recommended analytical expressions of Tolias [24] for the density, specific heat and conductivity of the solid and the liquid, with $T_m = 3695$ K and $L_f = 284.5$ kJ kg$^{-1}$. Its vapor pressure and heat of vaporization (4009 kJ kg$^{-1}$) are those tabulated by Raffray [1]. Lithium is liquid throughout, with the properties compiled by Davison [25]. Its vapor-pressure correlation rests on data between about 800 and 1800 K and is extrapolated above, and its heat of vaporization, 22.25 MJ kg$^{-1}$, is estimated from the slope of that correlation. For comparison, the NIST evaluation gives 149.4 ± 3.4 kJ mol$^{-1}$ (21.5 MJ kg$^{-1}$) for the ideal gas at the 1617 K normal boiling point, and 127.6 ± 6 kJ mol$^{-1}$ with the compressibility of the real vapor [43]. The steel has $\rho = 7800$ kg m$^{-3}$, $c_p = 500 + 0.6\,T$(°C) J kg$^{-1}$ K$^{-1}$ and $k =$ 33 W m$^{-1}$ K$^{-1}$ [7], with a table limit of 1800 K.

**Table B.2.** Tungsten-grade properties used in Section 4.6 (temperatures in K; recrystallization onsets from one-hour isochronal anneals), with the source of each value.

| Grade | Conductivity factor on pure W | Recrystallization onset | DBTT values used (test) |
|---|---|---|---|
| Pure W | 1 | 1523 (hot-rolled plate); 1673 and 1973 for swaged rods of high and low deformation [30] | tensile 423 [30]; bending 423 ± 25 [44]; fracture toughness 673 [44]; Charpy 823 (plate) [30] and 1073 ± 50 (rod) [45]; recrystallized, small punch 673 [46]; neutron-irradiated bending 823 (1 dpa at 1373 K) and > 873 (1 dpa at 873 K) [44] |
| W–K (about 30–90 ppm K [30, 35]) | 1.0: rolled plate within 1 % of pure W [35]; lower for a porous swaged rod [30] | 1623 (plate); 1873 and 1973 for rods of high and low deformation [30] | tensile 323 [30]; bending 348 ± 25 (7 mm plate) [44]; Charpy 623 (plate) [30] and about 1273 (Plansee WVM rod) [45]; neutron-irradiated bending 598 ± 50 (7 mm plate, about 1 dpa) [44] |
| ODS W–1%$La_2O_3$ | not sourced: proxy 1.0, 0.9, 0.8 | 1823, measured on a W–3%Re–1%$La_2O_3$ plate [30] | tensile 323 and Charpy 823 (the same plate) [30]; Charpy about 1223 ± 50 for WL10, estimated, in a review that also reports a test in which WL10 had a lower DBTT than pure W [45] |

The activation energy of the recrystallization indicator is 322–350 kJ mol$^{-1}$, mid value 336, fitted to pure-tungsten monoblocks under cyclic heat loads [47] and applied to all grades for want of alloy values. The indicator assumes a single activation energy and no incubation, and extrapolates one-hour anneal data at 1523–1823 K to microsecond excursions above 3000 K. It measures thermal exposure, not a change of hardness or ductility.

## Appendix C. Related work

**Related work (Section 1).** Table C.1 compares the studies closest to this one: dry tungsten walls, liquid films, and the few studies that treated both wall types with one code. González-Arrabal et al. [3] reviewed the limitations of tungsten in the HiPER scenarios. None of them computes the thermal response, melting and evaporation of a tungsten armor and of a lithium film in one shared calculation, and none analyzes a liquid wall in the HiPER scenarios. For the studies marked (a) this rests on their abstracts only.

**Table C.1.** Related work compared with this study. –: not reported, or not in the parts read; (a): only the abstract was read; (s): the abstract and parts of the text.

| Study | Target | Chamber | Wall | Source spectrum | Solver | Vapor | Repeated shots | Validation |
|---|---|---|---|---|---|---|---|---|
| Raffray [1] | NRL 154 MJ direct drive | 6.5 m, evacuated; other radii and Xe densities | W and C armor | NRL target spectra | RACLETTE, 1-D thermal with phase change; BUCKY with gas | Xe attenuation from BUCKY | shot histories at 10 Hz | – |
| Blanchard and Martin [7] | 150 MJ, fluence scaled with yield | 7.0 m and other radii | W-coated steel, 250 µm of W and other thicknesses | x-ray and ion loads | closed form; ANSYS 2-D finite elements, thermal and stress | – | 10 and 50 implosions at 5 Hz | – |
| Garoz et al. [8] | ARIES 154 MJ direct drive, rescaled to 20 and 50 MJ | 5 and 6.5 m (HiPER) | 1 mm W on 1 cm ODS-RAFM steel | ARIES spectra | Code_Aster finite elements: temperature, plastic strain, crack growth | – | to a steady state at 1 and 10 Hz | – |

**Table C.1 (continued).**

| Study | Target | Chamber | Wall | Source spectrum | Solver | Vapor | Repeated shots | Validation |
|---|---|---|---|---|---|---|---|---|
| Prometheus (a) [9] | KrF laser, 4 MJ, gain 124, 5.65 Hz; heavy ions, 7 MJ, gain 103, 3.54 Hz | – | SiC with a thin Pb film | – | design study | film evaporated every shot, recondensed between shots | – | – |
| ARIES-IFE [10]; (a) [6] | NRL 154 MJ direct drive (also a 400 MJ version); LLNL/LBL 458 MJ heavy-ion indirect drive | 6.5 m with and without Xe buffer gas; 4.5 m and 3 m for the liquid films | 3 mm W and C armor; 1 mm Pb and FLiBe films, injected through a porous wall or by tangential flow | ARIES-IFE reference target spectra | BUCKY 1-D radiation hydrodynamics (Xe-protected C armor, Pb film); operating windows; film fluid dynamics; aerosol | Pb evaporated per shot (9.2 kg direct, 21.4 kg indirect drive [10]); FLiBe aerosol formation and transport | cyclic operation; chamber clearing between shots | fluid-dynamics experiments |
| Zaghloul and Raffray (a) [11] | indirect drive | – | liquid-wall candidates | prompt x-rays | volumetric deposition with explosive boiling and spall | ablated material | – | – |
| Furukawa et al. (s) [12] | high-gain direct drive, 3.0 MJ laser | – | liquid Pb | x-rays, $\alpha$ and other charged particles | ACORE, 1-D ablation | screening of later ions by the ablated Pb plasma | – | – |
| Medin et al. [13] | heavy ions, 741 MJ | 5 m | 2 mm Pb film | x-ray prepulse and main pulse, debris, neutrons | DEIRA-4 target; 1-D spherical film and fireball | the prepulse vapor shields the film, and its re-radiation revaporizes it | – | – |
| Peterson et al. [48] | KrF direct drive, 425 MJ; heavy-ion indirect drive, 365 MJ | 6.5 m with 0.5 torr Xe; 3.5 m with low-density gas | graphite; FLiBe film | each target's output | CONRAD, 1-D Lagrangian radiation hydrodynamics | vaporization, condensation and gas hydrodynamics | – | – |
| Hassanein and Morozov [16] | NRL direct drive | – | bare W and C composite; for a Li/Pb film on C composite, x-ray deposition only (the preprint adds ion ranges in a 1 mm Li film [17]); no thermal response of a film | NRL target spectra | HEIGHTS-IFE: deposition, conduction with melting and vaporization fronts | chamber-gas hydrodynamics | – | HEIGHTS-MFE models tested against simulation devices and tokamaks |
| Dagousset et al. [49] | 10 kJ direct-drive laser, D–D and D–T | 0.6 m test chamber | dry wall; 3 mm flowing PbLi on F82H, 316L, W or ODS FeCrAl | fusion products only, no x-rays or debris | PHITS Monte Carlo: deposition, damage, activation | – | continuous operation | – |

**Table C.1 (continued).**

| Study | Target | Chamber | Wall | Source spectrum | Solver | Vapor | Repeated shots | Validation |
|---|---|---|---|---|---|---|---|---|
| Habtemariam et al. [14] | 2520 MJ at 1 Hz, the power of HYLIFE-II (315 MJ at 8 Hz) | first wall at 3 or 6 m; HYLIFE-II geometry | 0–100 cm of FLiBe in front of a 316 steel first wall | neutrons and gammas; x-rays and debris left out | Serpent 2; GeN-Foam and OFFBEAT, steady conjugate heat transfer | – | steady state | Serpent benchmarks against MCNP and OpenMC, by others |
| Habtemariam et al. [15] | 100–1000 MJ, indirect drive | HYLIFE-II, in 1-D, 2-D and 3-D | FLiBe jets | x-ray deposition profile of earlier studies | ray tracing; OpenFOAM Euler solver with vacuum tracking | ablation by flash boiling; gas dynamics of the ablated vapor | single event | numerical benchmarks |
| This work | NRL 154 MJ direct drive scaled to 20, 50 and 154 MJ (HiPER) | 5 and 6.5 m, evacuated | 1 mm W on 1 cm steel; Li film on 1 cm steel | ARIES x-ray and ion spectra, number and energy of every species kept | 1-D enthalpy conduction with melting and evaporation; XCOM/Henke and SRIM deposition | parameterized, time-coupled screening model | to a periodic state, up to 600 shots (1090 with neutron heating) | analytical tests; Garoz et al. [8], Blanchard and Martin [7], Raffray [1], NIST [29], TRIM [21] |

# Supplementary material

Supplementary material for the paper

**Comparative analysis of tungsten armor and a liquid-lithium film as first walls for the HiPER laser-fusion chamber: thermal response, vapor shielding and helium transmission to the steel**

Shohruxbek Bozorov (ORCID: https://orcid.org/0009-0005-5165-093X),
Independent researcher, Chust, Uzbekistan

Contents: S1 thermal model, numerical scheme, verification and energy conservation; S2 source term, deposition and vapor-shielding details; S3 benchmark detail, further lithium-film results and helium in the steel; S4 literature inputs and quantities computed in this work; S5 limitations; S6 tungsten armor: further results, damage indicators, coolant sensitivity and grades; S7 helium-retention model; S8 reproduction.

Sections, tables, figures and equations numbered with the prefix S are those of this supplement; numbers without the prefix, and Appendices A–C, refer to the main paper.

## S1. Thermal model, numerical scheme, verification and energy conservation

**Thermal model (Section 2.2).** The plasma-facing surface sits at $\mu = M_e(t)$, the areal mass evaporated so far, which grows as $dM_e/dt = j$. Because the material does not move relative to $\mu$, Eq. (1) needs no advection term for the receding surface and no mesh motion for thermal expansion. The enthalpy includes the latent heat of fusion $L_f$, released linearly over a melting interval of $\Delta T_m = 1$ K around the melting point $T_m$, so that melting needs no front tracking.

The evaporation coefficient $\alpha = 1$ is an upper bound. Measured rates agree with it when a condensing surface lies within one collision length of the evaporating one. For a distant condenser, as in a chamber, the flux at low pressure is 0.844 of that maximum [1]. The evaporated mass carries away its own enthalpy plus $L_v$. The flux of Eq. (2) is evaluated at the temperature $T_0$ of the surface cell (1 nm thick for W, 10 nm for Li). At the Demo peak, $T_0$ exceeds the extrapolated surface temperature by less than $10^{-7}$ K for W and by 1.7 K for Li. Evaluating $j$ at the surface instead would lower the Li evaporated mass by about 0.36 % (a first-order estimate).

The slab is divided into cells of fixed areal mass, graded geometrically from 1 nm at the tungsten surface (10 nm for lithium) with a growth factor of 1.05 in the armor and the film and 1.15 in the steel. The tungsten case has 308 cells. Each time step advances the cell energy balances by backward Euler and solves them by Newton iteration on the cell enthalpies, with a tridiagonal linear solve per iteration. The step is limited to keep the largest cell-temperature change near 5 K, to 2 % of the time since the last change of a source, and to fractions of the x-ray pulse and of the ion arrival window. It never crosses the switch-on or switch-off of a source, and it grows from $10^{-13}$ s at the x-ray arrival to about 2 ms late in a 0.1 s shot period. When the surface cell has lost half its mass to evaporation, the front layer is remapped onto its original mass grid, conserving mass and enthalpy.

**Grids.** The 1 mm of tungsten has 261 cells, starting at 1 nm at the surface and growing by $\times 1.05$, and the grading is mirrored from a 2 µm cell at the W/steel interface. Its largest cell is 34 µm. Where the two gradings meet, 500 µm below the surface, the cell size jumps by a factor of 2.56. The 1 cm of steel has 47 cells, from 2 µm growing by $\times 1.15$ to 1.75 mm. Its last cell takes the remainder of the layer and is 1.63 times its neighbor. In the lithium cases the first cell is min(10 nm, $\delta/100$), growing by $\times 1.05$ and refining to min($\delta/50$, 2 µm) at the Li/steel interface. The steel starts at min($\delta/50$, 1 µm) and grows by $\times 1.15$.

**Discrete equations.** With backward Euler, cell $i$ obeys $m_i^{n+1} (H_i^{n+1} - H_i^n)/\Delta t = G_{i-½}(T_{i-1} - T_i) - G_{i+½}(T_i - T_{i+1}) + S_i - [i = 0]\, j\, L_v$, with all temperatures and $j$ at the new time level. The surface cell's new mass is $m_0^{n+1} = m_0^n - j\Delta t$, and the conductances $G_{i+½} = 1/(\Delta x_i/(2k_i) + \Delta x_{i+1}/(2k_{i+1}))$ use the cell masses at the start of the step, with $\Delta x_i = m_i/\rho(T_i)$ and $k_i$ at the new temperature. The conductance is zero at the plasma-facing surface and $1/(\Delta x_N/(2k_N) + 1/h)$ from the last cell $N$ to the coolant. $S_i$ is the exact integral of each source's arrival profile over the step, divided by $\Delta t$. The depth profiles of the sources are recomputed when the evaporated mass has changed by more than 1 % of the surface-cell mass or the cells have been remapped. The vapor column of Section 2.5 is taken from the mass evaporated when the depth profiles were last recomputed, which happens whenever the evaporated mass has changed by more than 1 % of the initial surface-cell mass or the cell list changes. The enthalpy is tabulated on a 0.5 K grid, with the melting interval resolved by 21 points 0.05 K apart, and $T(H)$ and $dT/dH$ are read from the same table. Each Newton iteration solves a tridiagonal system with the temperature dependence of the conductances lagged in the Jacobian. Iteration stops when the largest estimated cell-temperature update is below $10^{-6}$ K after at least two iterations. An update whose largest estimated temperature change exceeds 200 K is scaled down to 200 K. A step is rejected and retried with $\Delta t/4$ if Newton fails within 60 iterations, an iterate leaves the property table, a cell changes by more than 10 K, or evaporation would remove the surface cell. If the step falls below $10^{-18}$ s, the run stops with an error. The remap after evaporation is piecewise constant in enthalpy (first order) and conserves the layer's mass and enthalpy to rounding.

**Step control.** Each accepted step is the smallest of four limits: a largest cell-temperature change of about 5 K, after which $\Delta t$ is multiplied by min(1.3, max(0.5, 0.9 × 5 K/$\Delta T_{max}$)); 2 % of the time since the last source event or the start of the integration part, but not less than the initial step of the current integration part; 0.1 of the x-ray pulse duration while it is on; and 1/200 of the ion arrival window while ions arrive, about 420 ns at 6.5 m. A step never crosses the start or end of the x-ray pulse or the first or last ion arrival, and at each such event it is reset to its initial value. Each shot is integrated to 50 µs with an initial step of $10^{-13}$ s. A second part, started with a $10^{-9}$ s step, then runs to the next shot (W, Prototype and Demo) or to 5 ms (single lithium shots; repeated lithium shots run to the next shot) and is skipped for W, Experimental. The quasi-steady Demo cycle takes 4,441 accepted steps, 3,273 of them in the first 50 µs. The largest steps fall near the end of the 0.1 s period.

A further set of closed-form tests extends these: the full erfc profile under a constant surface flux, an instantaneous plane source, the one-phase Neumann problem at the default 1 K melting interval, energy conservation and the exact mixing temperature in closed slabs, and the invariance of a uniform state under zero flux for $10^4$ steps over 1 s. Their time- and mesh-refinement studies give observed orders of 0.93–1.05 in time and 2.00 in space for conduction on the code's own graded meshes. For the moving melt front, whose crossing of cell boundaries adds a non-smooth component, the order in time is 0.7–1.2. The tests use constant properties and therefore check the numerics only.

**Table S1.** Verification tests gating the production runs (constant-property test materials with tungsten-like magnitudes).

| Test | Problem | Reference solution | Result |
|---|---|---|---|
| 1 | Semi-infinite slab, constant surface flux 1 GW m$^{-2}$, 10 ns–10 µs | $T_{surf} - T_{init} = (2q_0/k)\sqrt{(at/\pi)}$, with the flux $q_0$, the initial temperature $T_{init}$ and the diffusivity $a = k/(\rho c_p)$ | max error 0.065 % |
| 2 | Instantaneous exponential source of e-folding length 0.5 µm, insulated surface, 0.1 ns–0.1 µs | even extension and Gaussian kernel | surface and profile error ≤ 0.28 % |
| 3 | Two-phase Stefan (Neumann) problem, different solid and liquid properties | Neumann similarity solution | melt-front error ≤ 0.22 % |
| 4 | Melting, evaporation (synthetic vapor law), 26 rezonings, two layers and a coolant | energy conservation | relative residual $1.1\times10^{-14}$ in magnitude |
| 5 | The melting and evaporating slab of test 4, run to 2 µs with a 2 nm surface cell, a 10 K temperature target, a pulse-step fraction of 0.1 and the 2 % step rule, against all four halved | self-convergence | peak temperature changes by 0.02 %, ablated depth by 0.06 % |
| 6 | Beer-law photons through two layers; 1 MeV alphas in thick and thin slabs | Beer–Lambert transmission; NIST ASTAR CSDA range [2] | transmission error $4\times10^{-16}$; end point 1.849 µm against 1.835 µm [2] (within one cell) |

**Convergence of the production cases.** Table S2 repeats four production cases with one numerical setting changed at a time: the surface cell halved and doubled, the armor or film growth factor halved to 1.025 and doubled to 1.10 (the steel to 1.075 and 1.30, so that $g - 1$ is halved and doubled in both), and every time-step limit (the 5 K temperature target, the 2 % rule and the x-ray and ion-window fractions) halved and doubled. The quasi-steady tungsten cycle is one 0.1 s period started from the state after 600 shots. For the grid settings, that state is interpolated in depth onto the new cells, which shifts the pre-shot and interface temperatures of every grid variant by 0.03–0.06 K.

The time-step changes converge at first order. The growth-factor changes converge at about second order for the tungsten surface and at about first order behind the lithium film. At the W/steel interface, as for the pre-shot surface, the halved and the doubled growth factor both raise the temperature, by the 0.03–0.06 K that the interpolated starting state gives every grid variant, so no order is observed and the nominal second order is assumed. Halving or doubling the surface cell changes no quantity by more than 0.07 K or 0.03 %. The last column extrapolates the baseline to converged steps and cells by Richardson's rule with the observed orders (the growth and time-step errors added). By this rule the tungsten peaks are about 1 K too high, so that 3027 K and 3317 K become 3026 K and 3316 K. The peak steel temperature behind a 31.6 µm film, 907 K, and the single-shot minimum film, 30.1 µm, do not change at their quoted precision. Of the other quantities of Table S2, only the evaporated depth of the tungsten single shot changes at its quoted precision ($3.381\times10^{-7}$ and $3.346\times10^{-7}$ nm). The thick-film lithium evaporation was not part of the convergence study. Behind the 31.6 µm film, the evaporated mass extrapolates 0.3 % higher.

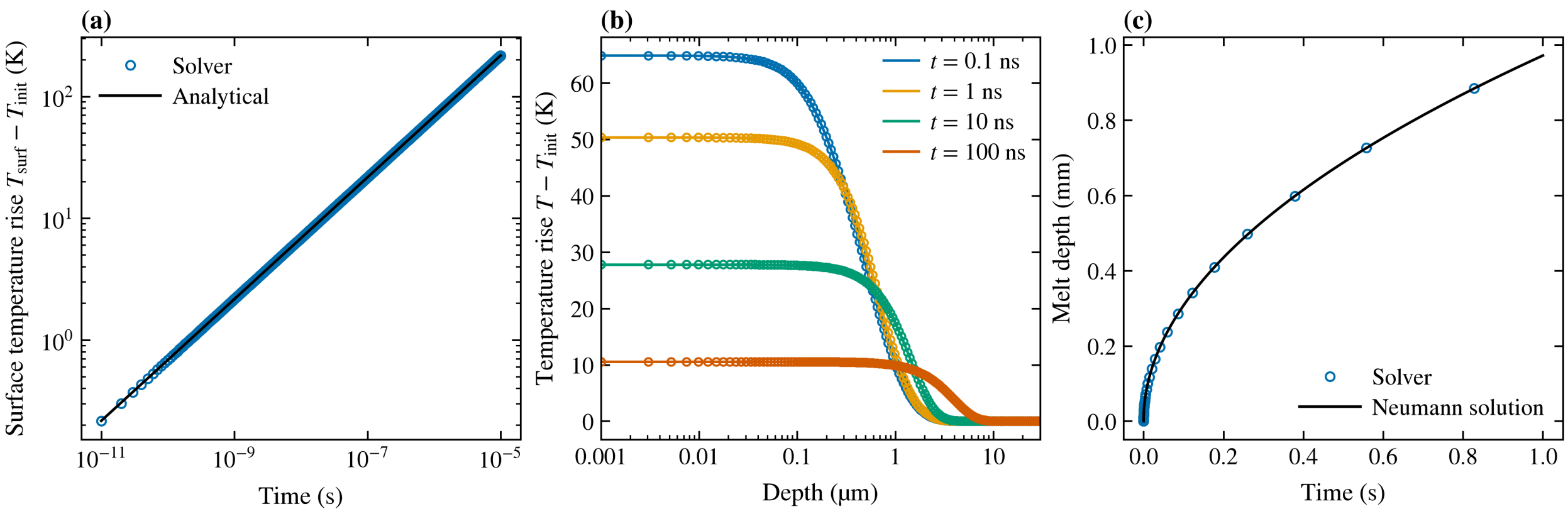

**Figure S1.** Verification tests 1–3, solver (circles) against the analytical solutions (lines): surface temperature rise under a constant flux (a), temperature profiles after an instantaneous exponential source (b), and the melt front of the two-phase Stefan problem (c).

**Energy ledger.** The solver accumulates, separately from the solve, the energy deposited by the sources, the energy removed by the coolant and by evaporation (latent plus sensible), and the change of stored enthalpy. The relative residual is $1.8\times10^{-12}$ (Demo), $2.3\times10^{-12}$ (Prototype) and $2.2\times10^{-14}$ (Experimental) for the single tungsten shots to 50 µs and at most $1.0\times10^{-11}$ in the verification tests. It is $-5.3\times10^{-8}$ and $-1.7\times10^{-7}$ over the 600-shot Demo and 100-shot Prototype runs and $-3.6\times10^{-6}$ to $-6.7\times10^{-6}$ over the four tungsten runs with neutron heating. In every lithium run it is below $10^{-11}$ with or without vapor shielding, and $8\times10^{-15}$ to $7\times10^{-13}$ in magnitude in the repeated-shot lithium runs, with or without neutron heating. The residuals of the repeated tungsten runs come from rounding in double precision. The solver spreads the source energy of a step over the difference of the clock readings at its ends but applies it over the requested step length, and the two differ for short steps at large clock times. The 5 ns neutron pulse is integrated from a step restarted at $10^{-13}$ s, at clock times of up to 109 s. A replay of that step sequence accounts for 53–62 % of the four residuals, with their sign. At $1.8–3.5\times10^{-5}$ of the neutron energy of a shot, the residuals are far too small to change any quoted temperature. The lithium runs restart the clock at every shot. In the Demo tungsten shot, 82,354 J m$^{-2}$ is incident and 398 J m$^{-2}$ is not absorbed (backscattered ions and x-rays that pass the domain).

**Table S2.** Grid and time-step convergence of the production cases: change from the baseline when one setting is halved (½) or doubled (2), observed order of the growth-factor (g) and time-step (t) refinements (2 assumed where the halved and the doubled growth factor move the quantity the same way), and the baseline extrapolated to converged steps and cells. The 17.8 µm row, which with the 31.6 µm rows brackets the minimum film, enters only through it and was not extrapolated.

| Quantity | Baseline | Surface cell ½ / 2 | Growth 1.025 / 1.10 | Time-step limits ½ / 2 | Order g; t | Extrapolated |
|---|---|---|---|---|---|---|
| W Demo single shot: peak surface | 3026.86 K | −0.00 / −0.01 K | −0.17 / +0.69 K | −0.49 / +0.94 K | 2.0; 0.93 | 3025.6 K |
| W Demo single shot: evaporated depth | $3.381\times10^{-7}$ nm | +0.02 / −0.01 % | −0.18 / +0.80 % | −0.36 / +0.68 % | 2.1; 0.89 | $3.346\times10^{-7}$ nm |
| W Demo quasi-steady cycle: next pre-shot surface | 920.93 K | +0.06 / +0.06 K | +0.06 / +0.06 K | −0.02 / +0.04 K | 2 (assumed); 0.95 | 921.0 K |
| W Demo quasi-steady cycle: peak surface | 3317.49 K | +0.01 / −0.00 K | −0.17 / +0.70 K | −0.48 / +0.93 K | 2.0; 0.97 | 3316.3 K |
| W Demo quasi-steady cycle: peak interface | 933.86 K | +0.03 / +0.03 K | +0.03 / +0.05 K | +0.02 / −0.04 K | 2 (assumed); 0.97 | 933.9 K |
| Li Demo, 31.6 µm single shot: peak steel | 907.08 K | +0.03 / −0.04 K | −0.28 / +0.66 K | +0.17 / −0.31 K | 1.2; 0.85 | 907.0 K |
| Li Demo, 31.6 µm single shot: evaporated mass | $7.452\times10^{-4}$ kg m$^{-2}$ | −0.01 / +0.02 % | +0.07 / −0.14 % | +0.06 / −0.11 % | 0.99; 0.86 | $7.472\times10^{-4}$ kg m$^{-2}$ |
| Li Demo, 17.8 µm single shot: peak steel | 1115.65 K | +0.06 / −0.06 K | −0.49 / +1.21 K | +0.15 / −0.26 K | – | – |
| Li Demo single-shot minimum film | 30.097 µm | +0.003 / −0.003 µm | −0.025 / +0.059 µm | +0.014 / −0.026 µm | 1.2; 0.85 | 30.085 µm |

**Verification (Section 3).** A constant surface flux of 1 GW m$^{-2}$ on a semi-infinite slab is reproduced within 0.065 % from 10 ns to 10 µs, an instantaneous exponential source within 0.28 %, and the melt front of the two-phase Stefan problem within 0.22 % (Figure S1). A synthetic case with melting, evaporation, 26 rezonings, two layers and a coolant conserves energy to $1\times10^{-14}$. Halving the surface cell, the temperature target and the step fractions together changes its peak temperature by 0.02 % and its ablated depth by 0.06 %. Further closed-form tests, with refinement in time and space, show the expected first-order convergence in time and second-order convergence in space on the code's own graded meshes (above, after Table S1). In the production runs the energy ledger closes to below $10^{-11}$ in single shots and to about $10^{-7}$ over the 600-shot Demo run. Halving and doubling the surface cell, the growth factor of the cells and every time-step limit of the production cases changes no result by more than 1.2 K or 0.8 %, with first-order convergence in time and first- to second-order convergence in the grading (Table S2). Extrapolated to converged steps and cells, the tungsten peaks are about 1 K lower, and the lithium results change at most in their last quoted digit.

The vapor-shielding option is checked by tests of its limiting cases, its coupling and its energy bookkeeping. With no column it returns the unshielded deposition exactly; a column held at the whole per-shot mass reproduces the static bound of Section 6 to $10^{-12}$; a 10 kg m$^{-2}$ column stops all of the ion energy and, with $f_{rr}$ = 1, returns all of it; and a column that grows linearly in time ahead of monoenergetic ions of constant stopping power stops the analytically integrated energy to within 0.2 %, the error of taking the column at the start of each step. The delayed return follows the exponential solution of Eq. (3) to $10^{-9}$. The column used in each step of a coupled lithium run differs from the mass evaporated before that step by at most 1 % of the surface-cell mass, and a coupled run closes its energy ledger to $10^{-10}$. They do not test the stopping power of the vapor itself. The stopping could be tested against the attenuation of the same target's ions by xenon fill gas that Raffray reports [3]. That benchmark is left for future work. It needs SRIM stopping tables for xenon gas, which could not be generated here (their input files and the script that makes the comparison are in the repository). It would also test the stopping only, not the re-radiation.

## S2. Source term, deposition and vapor-shielding details

**Source term (Section 2.3).** The source term is that of the NRL 154 MJ direct-drive target of the ARIES-IFE study: 2.1 MJ of x-rays, 19.5 MJ of burn ions and 22.1 MJ of debris ions. The 109 MJ of neutrons deposit most of their energy deep in the blanket [3, 4]. Their heating of the armor and the steel, which Garoz et al. neglected as small against that of the ions and x-rays [5], is left out of the baseline and added as a separate case (Sections 4.2 and 5.3). The 14.1 MeV neutrons arrive 127 ns after ignition at 6.5 m. They heat tungsten volumetrically by about 800 MW $cm^{-3}$ for 5 ns, the value Sharafat et al. give for the same target and chamber [6], scaled by $Y/154$ MJ and $(6.5\ m/R)^2$. As no published value for the steel was found, it is assigned 0.5 or 1 times the tungsten value per unit volume as a bounding assumption, and the lithium film receives none. At Demo conditions this is 4 J $cm^{-3}$ per shot in tungsten, 40 MW $m^{-3}$ averaged at 10 Hz, for a neutron wall load of 2.05 MW $m^{-2}$ (0.067 MW $m^{-2}$ in Prototype). The ion spectra are the ARIES-IFE tables for burn p, D, T, $^3He$ and $^4He$ and debris p, D, T, $^3He$, $^4He$, $^{12}C$ and Au [7]. Their bin edges are not published and are reconstructed from the tabulated energies (below), which gives 125 non-empty bins. Every species keeps its number of ions and its energy in the ARIES accounting, and the raw tables are run as a sensitivity case. The x-ray spectrum was digitized from Raffray's Fig. 2 [3] (1119 points, $\pm 10$ % trace width) and normalized to 2.1 MJ. The fluence at the wall is $E/(4\pi R^2)$.

The x-rays arrive at $R/c$, where $c$ is the speed of light, as a square pulse of 20 ns. Sensitivity cases use 0.1, 10 and 50 ns for tungsten, and 0.1 and 50 ns for lithium. Every ion component (species and energy bin) is emitted from a point at $t = 0$ and flies freely to the wall, so that its power on the wall follows from its spectrum and its time of flight. At 6.5 m the first ions arrive at 0.105 µs and the last at 83.5 µs. The energy-median arrival is 0.50 µs for the burn ions and 1.78 µs for the debris ions in Demo. Table 2 gives the resulting energy and power densities on the wall.

**Energy deposition (Section 2.4).** Counting incoherent scattering as absorbed slightly overestimates the deposit. Fluorescence and photoelectron transport are neglected, and the steel is treated as iron.

Ion ranges come from tables of projected range for every ion in W, Li and Fe computed with SRIM-2012 (The Stopping and Range of Ions in Matter) [8], used per unit areal mass. The energy is placed at depth through an effective projected range $R_{p,eff}(E) = f_{TRIM}(E)\ R_p(E)$. An ion that enters with energy $E_0$ keeps, at depth $x$, the energy $E$ for which $R_{p,eff}(E) = R_{p,eff}(E_0) - \xi(x)$, where $\xi$ is the areal mass above that depth (Table A.1). The deposit between two depths is the energy lost between them, so that the cell deposits sum to the incident energy. The range factor $f_{TRIM}$ is the ratio of the mean range from Monte Carlo runs of TRIM (the transport code of SRIM), made for this study, to the table projected range. It differs from one only for keV hydrogen isotopes and helium in tungsten, where it reaches about 2.1 at 1 keV. In tungsten the deposit is reduced by the TRIM backscattered-energy fraction. The SRIM electronic and nuclear stopping powers enter only a continuous-slowing-down-approximation (CSDA) mode, which uses the path length as depth and is run as a sensitivity case (Figure S2).

**Vapor-shielding model (Section 2.5).** Vapor produced during a shot stands between the wall and the ions that arrive later. The x-rays are not affected: they arrive within 42 ns, before there is vapor, and even the whole per-shot lithium vapor would absorb only 0.16 % of their energy. The ions arrive over 0.1–84 µs. The lithium surface evaporates while the debris ions heat it, so the vapor made by the first debris ions shields the later ones. Lithium vapor leaving a 2218 K surface has a sound speed of about 2.1 km $s^{-1}$ (monatomic ideal gas) and moves by centimeters during the ion pulse, which is small against the 6.5 m radius. In planar geometry, an expanding column keeps its areal mass. The column crossed by an ion arriving at time $t$ is therefore the mass evaporated so far in the shot, $\sigma_V(t) = M_e(t)$, whatever the details of the expansion.

The model represents this column as one cell of the wall's own front material, of areal mass $s\ \sigma_V$, placed in front of the mesh, and deposits every ion component on the extended mesh with the production depth model of Section 2.4. The share left in the extra cell is stopped in the vapor; the rest goes to the wall. Because the vapor and the wall are the same element, the wall then receives the tail of the deposition profile that the ion would have left from the front of the column. The factor $s$ represents ionization of the vapor, which raises its stopping power: $s = 1$ is cold, neutral vapor with the stopping of the condensed material per unit mass, and $s = 2$ is an assumed enhancement. No measured or computed factor for hydrogen or helium ions in lithium vapor was found. For heavy ions in hydrogen plasma, stopping is enhanced by free electrons and higher projectile charge states [9], by up to a factor of 35 for slow krypton ions in fully ionized hydrogen [10]. Calculations for heavy ions show non-equilibrium charge states shortening the ranges, in some cases by factors of 2–7 [11].

Immediate return is conservative for the surface temperature, as it provides an upper bound at a given $f_{rr}$. In a uniformly loaded spherical chamber, radiation emitted towards the center lands on the opposite wall within $2R/c \approx 43$ ns, so every wall element receives all the radiation the vapor emits—not half of it. $f_{rr}$ is therefore the fraction of the stopped energy that is radiated during the shot. Two reference points from other regimes span its range. In HEIGHTS simulations of 1 ms ITER transients, a lithium vapor plasma absorbs 82–88 % of the incident energy but re-radiates only 0.36–1.26 % of the incident energy, 0.4–1.4 % of what it absorbed, and keeps the rest as heat [12]. Xenon fill gas in IFE chambers, by contrast, re-radiates much of the stopped energy to the wall over a time long compared with the deposition [13], 200–700 μs in the calculations reported by Raffray [3, 14]. The calculations use $f_{rr} = 0$, 0.5 and 1 with immediate return and $s = 1$ and 2, and one xenon-like case with $f_{rr} = 1$ and $\tau_{rr} = 300$ μs. The cases $s = 5$ and 10 test larger stopping on the thick Demo film, and the repeated-shot runs use $f_{rr} = 0$ and $s = 1$.

The model neglects the hydrodynamics and energy balance of the vapor, its ionization balance and radiation transport, the return flux of vapor onto the surface, lateral loss, explosive boiling and droplet ejection, and residual vapor from earlier shots. These are the content of radiation-hydrodynamics codes [15, 16]. It is applied to single shots and, for lithium, to repeated shots in which the vapor of each shot is assumed to have cleared before the next and the column restarts from zero.

**Table S3.** Energy partition of the NRL 154 MJ direct-drive target [3].

| Component | Energy (MJ) [3] | Share [3] |
|---|---|---|
| X-rays | 2.1 | 1.4 % |
| Neutrons | 109 | 71 % |
| Gamma rays | 0.009 | 0.006 % |
| Burn (fast) ions | 19.5 | 13 % |
| Debris ions | 22.1 | 14 % |
| Residual energy | 1.29 | – |
| Total | 154 | |

**Ion spectra.** The ARIES tables give $dN/dE$ at up to 15 mid-bin energies per species. Rows holding the padding value 1.0 per keV are dropped, which leaves 125 non-empty bins. The debris energies are spaced logarithmically (at a constant ratio for p, D, T and helium), and their bin edges are the logarithmic midpoints, as are those of the five-point burn tables of p, D, T and $^3$He. The first eight burn-alpha energies, 0.23–3.54 MeV, lie one in each 0.5 MeV interval from 0 to 4 MeV, so these bins are 0.5 MeV wide. Logarithmic midpoints would make the 3.54 MeV bin 1.5 MeV wide (3.31–4.83 MeV) and add 5 % to the burn alphas and 14 % to their energy. With these edges the tables integrate to 0.8–1.5 times the per-species energies of the ARIES accounting for every species except $^3$He, which integrates to 2–6 times. An alternative construction scales the number of each species to match its accounting energy (option `ion_spectra=normalized`). It gives 1.20 alphas per fusion, raises the debris alphas by 24 % and lowers the debris deuteron and triton counts by 15 % and 19 %, to below the unburned fuel inventory of the target. The production spectra instead keep both the number and the energy of every species. The number is the table's for every species except $^4$He. Its burn and debris parts, 1.15 alphas per fusion in the tables with these edges, are scaled together to one alpha per fusion in their tabulated ratio. The energy is then set to the accounting. Burn ions, born at fixed nuclear energies, keep the tabulated energies. Their bin weights are tilted by a power of the energy, so that no ion is moved above the birth and knock-on energies that bound the table. The spectrum of each debris species (ions of the expanding target plasma) is stretched in energy by factors of 0.17 ($^3$He), 0.66–0.96 (Au, C, T, D, p) and 1.42 ($^4$He). The ion energy is 41.6 MJ in total in both constructions. The raw tables of the sensitivity case (Fig. 3 of the main paper) are the tables as read, with logarithmic-midpoint edges for every species and no scaling. They hold 1.18 alphas ($^4$He) per fusion instead of one (1.19 helium ions with $^3$He) and 47.2 MJ of ions, 13 % more than the 41.6 MJ of the accounting: 20.8 MJ of burn ions against 19.5 MJ, and 26.3 MJ of debris ions against 22.1 MJ (+19 %).

The tables can also be checked against the target itself. Against what the target of Raffray's Fig. 1 can produce [3], with the solid D–T density of the Peterson et al. target [13] (which gives Raffray's target 4.19 mg of D–T, 10.9 % of it burned; computed here), the deuterons and tritons of the tables are within 3.5 % of the fuel left after burn. Carbon and hydrogen are about 60 % of those in the CH foam and coating, and gold is within 8 % of the 30 nm layer (densities of polystyrene and bulk gold assumed for the coatings). With the alternative construction, only helium exceeds its source.

Within a bin the spectrum follows a log–log interpolation of the species table multiplied by a tilt $\exp(\beta \ln(E/E_{mid}))$, with amplitude and tilt solved so that the bin carries the number and energy of the construction above. The power arriving at the wall is $(dN/dE)\, E\, |dE/dt|$, evaluated at 24 energies per bin. Every component deposits its energy with the depth profile of its mid-bin energy $E_{mid}$: the tabulated energy for burn ions, and the tabulated energy times the stretch factor for debris species. At 6.5 m the arrivals span 0.105–83.5 μs. Only the first and last are treated as events.

**X-rays and scaling.** The digitized spectrum (1119 points between $1.9\times10^{-3}$ and 107 keV) integrates to 2.18 MJ, 3.9 % above the partition, and is normalized to 2.1 MJ $\times$ $Y/154$ MJ and rebinned onto 100 logarithmic bins. The scaling of the ion fluence, 0.27 $Y/(4\pi R^2)$, reproduces the fluences of González-Arrabal et al. [4] to within 2 %, which checks the arithmetic of the scaling but not the spectral shapes.

**Depth model.** For ions without their own TRIM runs, p and t use the d runs and $^3$He the $^4$He runs at equal velocity. Below 1 keV the TRIM quantities are held at their 1 keV values, and above the highest TRIM energy they relax to $f_{\mathrm{TRIM}} = 1$ and $\eta_E = 0$ over one decade. The stopping below the SRIM table is taken proportional to $\sqrt{E}$. For C in W and every ion in Fe, $f_{\mathrm{TRIM}} = 1$ and $\eta_E = 0$. Lithium backscatter (at most 0.14 %) is neglected, and straggling is not represented.

**The TRIM-profile path.** A second deposition path for keV hydrogen isotopes and helium in tungsten uses interpolated TRIM depth profiles and stretches the table model for hydrogen isotopes above 100 keV by an estimated factor (1.31 at 50 keV per nucleon, 1.04 at 500 keV per nucleon and 1 above 5 MeV per nucleon). It applies to a single tungsten layer, so it was used in the Raffray benchmark (one 3 mm slab) but in no production run, all of which have tungsten on steel. Enabling it for tungsten on steel lowers the single-shot peaks from 868, 1371 and 3027 K to 865, 1366 and 3010 K, by 0.3–0.6 %, and the quasi-steady Demo peak from 3317 to 3300 K. The production values are kept because the bias is below 1 % and conservative.

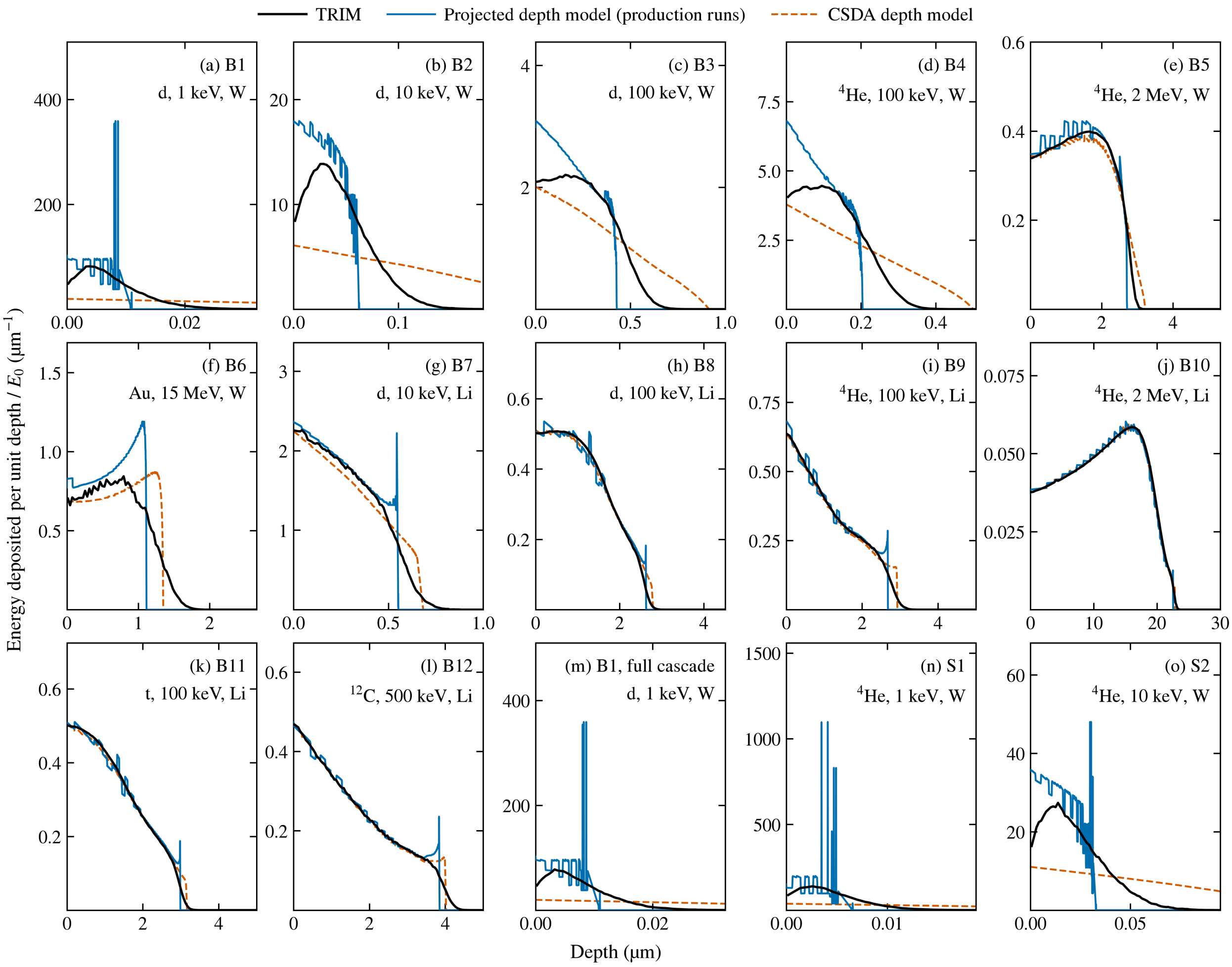

**Figure S2.** Deposited energy per unit depth, normalized to the ion energy, for 15 single-ion cases: TRIM [8] (black), the projected depth model of the production runs (blue) and the CSDA model (orange-red, dashed). For lithium targets and MeV ions in tungsten (B5–B12) the projected model follows TRIM, except for 15 MeV gold in tungsten (B6), whose projected profile is shallower and more peaked (50 % and 90 %-energy depths 0.93 and 0.85 of TRIM's); for d and He at 1–100 keV in tungsten (B1–B4, S1, S2) it is too shallow, with 50 %-energy depths 0.69–0.76 of TRIM's. Panel (m) repeats B1 with a full-cascade TRIM run. The narrow spikes of the blue curves are artifacts of the SRIM range tables, which print the projected range in whole ångströms at low energy; their height scales with the cell width.

**Mean ion energies.** The debris-ion mean energies agree with those of Garoz et al. [5] within rounding only for the debris protons (0.092 against 0.09 MeV). The debris D and T are 12 % and 20 % lower here and the debris helium 43 % higher. The burn-ion means do not agree either: D and T are 23 % and 25 % higher here, and burn $^4$He is 2.09 MeV (2.42 MeV energy-weighted) against their 3.81 MeV. The mean depth that Garoz et al. give for burn $^4$He, 1.8 μm, fits the lower energy: the depth model gives 1.73 μm for the ARIES burn-$^4$He spectrum and 2.99 μm for monoenergetic 3.81 MeV alphas. Garoz et al. define neither column and give no bin edges, so this does not establish an error in their table. No input of this work uses their mean energies.

**Table S4.** Demo lithium film with the reduced vapor-shielding model (single shots, $T_{\mathrm{Li}} = 600$ K, $\alpha = 1$). $f_{\mathrm{rr}}$: fraction of the stopped energy returned to the surface (at once unless a time constant is given); $s$: stopping of the vapor relative to cold vapor. Thick-film values are medians over $\delta \geq 100$ μm; the $s = 5$ and 10 cases were run for the 1 mm film only, and their ratios refer to the unshielded 1 mm film.

| Case | Ion energy stopped in the vapor | Li evaporated per shot, thick film (ratio to no shielding) | Peak Li surface, thick film | Peak steel, 31.6 μm film | Minimum film, single shot | Stopped energy per vapor atom |
|---|---|---|---|---|---|---|
| No shielding | 0 | $7.43\times10^{-4}$ kg m$^{-2}$ (1) | 2218 K | 907 K | 30.1 μm | – |
| $f_{\mathrm{rr}} = 0$, $s = 1$ | 7.5 % | $5.87\times10^{-4}$ (0.79) | 2197 K | 895 K | 29.1 μm | 0.72 eV |
| $f_{\mathrm{rr}} = 0.5$, $s = 1$ | 8.1 % | $6.74\times10^{-4}$ (0.91) | 2227 K | 898 K | 29.3 μm | 0.68 eV |
| $f_{\mathrm{rr}} = 1$, $s = 1$ | 8.8 % | $7.78\times10^{-4}$ (1.05) | 2262 K | 901 K | 29.5 μm | 0.64 eV |
| $f_{\mathrm{rr}} = 0$, $s = 2$ | 12.3 % | $4.89\times10^{-4}$ (0.66) | 2180 K | 888 K | 28.4 μm | 1.41 eV |
| $f_{\mathrm{rr}} = 0.5$, $s = 2$ | 13.6 % | $6.23\times10^{-4}$ (0.84) | 2233 K | 893 K | 28.8 μm | 1.23 eV |
| $f_{\mathrm{rr}} = 1$, $s = 2$ | 15.0 % | $7.99\times10^{-4}$ (1.08) | 2300 K | 896 K | 29.2 μm | 1.06 eV |
| $f_{\mathrm{rr}} = 1$, $s = 1$, $\tau_{\mathrm{rr}} = 300$ μs | 7.5 % | $5.89\times10^{-4}$ (0.79) | 2197 K | 896 K | 29.1 μm | 0.72 eV |
| $f_{\mathrm{rr}} = 0$, $s = 5$ (1 mm film) | 19.6 % | $3.39\times10^{-4}$ (0.46) | 2160 K | – | – | 3.25 eV |
| $f_{\mathrm{rr}} = 1$, $s = 5$ (1 mm film) | 23.0 % | $8.24\times10^{-4}$ (1.11) | 2356 K | – | – | 1.57 eV |
| $f_{\mathrm{rr}} = 0$, $s = 10$ (1 mm film) | 24.5 % | $2.43\times10^{-4}$ (0.33) | 2135 K | – | – | 5.69 eV |
| $f_{\mathrm{rr}} = 1$, $s = 10$ (1 mm film) | 27.6 % | $8.41\times10^{-4}$ (1.13) | 2410 K | – | – | 1.85 eV |

**Vapor shielding of tungsten (Section 6).** The tungsten armor evaporates $7\times10^{-12}$ kg m$^{-2}$ during the first Demo shot, from 600 K, a column that would stop $4\times10^{-10}$ of the ion energy even if it stood in front of the wall from the start. In the coupled tungsten runs this column stays below the change of evaporated mass at which the solver recomputes the deposition, $1.9\times10^{-7}$ kg m$^{-2}$, and is never applied. When the deposition is recomputed after every step, so that the column follows the evaporated mass, the vapor stops $1\times10^{-5}$ to $3\times10^{-5}$ J m$^{-2}$ of the $7.8\times10^{4}$ J m$^{-2}$ of ion energy and changes the Demo peak by at most $4.5\times10^{-7}$ K. Even a strongly degraded case, the halved conductivity with a 5 μm surface layer at $\varphi = 0.25$ (Table S12), evaporates only $6\times10^{-8}$ kg m$^{-2}$ per shot in its quasi-steady state, a column that would stop 3–$7\times10^{-6}$ of the ion energy. In the first shot the lithium film evaporates about a hundred million times more mass, and in Demo that vapor stands between the wall and the later ions.

Coupled in time, the vapor stops much less than the static bound suggests. A column holding the whole per-shot mass from the start would stop 24 % of the ion energy (the limiting case of the model, reproduced to $10^{-12}$). The column, however, grows during the shot, and most of the debris energy arrives while it is still thin. Cold vapor stops 7.5–8.8 % of the ion energy, and vapor with twice that stopping power 12–15 %. Shielding also feeds back on itself, because less heating of the surface means less vapor: with $f_{\mathrm{rr}} = 0$ and $s = 1$ the column at the end of the shot is 21 % smaller than without shielding.

**Return of the stopped energy and ionization (Section 6).** The fate of the stopped energy decides the sign of the effect on evaporation. If none of it returns to the wall ($f_{\mathrm{rr}} = 0$), the thick-film evaporation falls by 21 % with cold vapor and by 34 % with $s = 2$. If all of it returns at once ($f_{\mathrm{rr}} = 1$), evaporation rises by 5 % and 8 %, and the surface peaks up to 82 K hotter. The energy that the ions would have spread over their micrometer ranges is then returned to the very surface, where it drives evaporation most efficiently. A xenon-like return spread over 300 μs acts like $f_{\mathrm{rr}} = 0$ for evaporation ($-21$ %): by the time the energy comes back, the surface has cooled and evaporation has practically stopped. The 300 μs lies within the 200–700 μs over which the xenon fill gas re-radiates its energy in the BUCKY calculations reported by Raffray [3]. The literature on lithium vapor points to the low end of $f_{\mathrm{rr}}$ [12].

The energy stopped per evaporated atom, 0.6–1.4 eV, is far below the 5.39 eV needed to ionize lithium. An equilibrium Saha estimate, with the column spread over the 4 mm to 8 cm that the vapor expands at its sound speed in 2–40 μs ($5\times10^{23}$–$1.6\times10^{25}$ atoms m$^{-3}$) and the stopped energy shared between heating and ionizing the vapor, gives ionized fractions of 0.8–3 % for $s = 1$ and 4–12 % for $s = 2$. Only if all of it heated the vapor without paying for ionization would the $s = 2$ cases reach 20–95 %. $s = 1$ is therefore the closer of the two stopping values.

Larger stopping factors move the evaporation further in both directions (Table S4, 1 mm film). With $s = 5$ and 10 the thick-film evaporation falls to 0.46 and 0.33 of its unshielded value if none of the stopped energy returns, and rises to 1.11 and 1.13 times it if all of it returns at once. The vapor then stops 20–28 % of the ion energy, 3.3 and 5.7 eV per vapor atom without return. These cases test the sensitivity of the model rather than a likely state of the vapor: stopping several times that of cold vapor needs strong ionization, and the ionized fractions estimated above are small.

The steel hardly notices the vapor. The column stops mainly debris ions, which deposit their energy in the first micrometers of the film; the steel is heated by the burn ions that pass the film and by conduction from it. Shielding lowers the peak steel temperature behind a 31.6 μm film by 6–19 K and the single-shot minimum film from 30.1 μm to 28.4–29.5 μm, 2–6 % less. In repeated operation, with $f_{rr} = 0$ and $s = 1$ and the column restarting every shot, the steel behind 31.6 and 56.2 μm films settles at peaks of 992 K and 863 K instead of 1006 K and 871 K. The Demo minimum film, with the direct runs of both cases, falls by 2 μm, or 4.5 %, from 44 to 42 μm. For the 873 K compatibility limit, the film falls by 2 μm, from 56 to 54 μm.

## S3. Benchmark detail, further lithium-film results and helium in the steel

**Benchmarks (Section 3).** The energy fluences reaching the wall agree with those of Garoz et al. [5] to within their printed rounding for 31 of 33 per-species, group and x-ray values. The other two are the debris totals of Prototype and Demo, 1.353 and 4.167 J cm$^{-2}$ against 1.3 and 4.0 J cm$^{-2}$, 4.1 % and 4.2 % higher. The agreement is expected by construction: both works normalize to the same ARIES accounting energies, so it tests the bookkeeping, not the spectral shapes. The number-weighted mean photon energy of the digitized x-ray spectrum, 7.09 keV, matches the 7 keV given by Garoz et al. At every energy of the NIST PSTAR and ASTAR tables [2] from 0.01 to 10 MeV, the SRIM projected ranges of protons and alphas in tungsten agree with NIST to within 7.3 %, and the electronic stopping powers to within 5.1 % from 0.1 to 10 MeV. The largest differences are those of the proton range at 50 keV (0.927), the proton stopping at 225 keV (0.958) and the alpha stopping at 0.2 MeV (1.051). At 0.01 MeV the electronic stopping of alphas is 11 % below the NIST value. The depth model follows TRIM for lithium targets and for MeV ions in tungsten, except for the 90 %-energy depth of 15 MeV gold (0.85 of TRIM's), but places the energy of keV deuterons and helium in tungsten too close to the surface, with 50 %-energy depths 0.69–0.76 of TRIM's (Fig. S2). Every production tungsten run deposits these ions with this model. In lithium the ranges of this work agree to within 15 % with those that Hassanein and Morozov computed with HEIGHTS-IFE for the same target ions in a 1 mm lithium film on carbon composite, in the preprint of their paper [17]. The journal version shows only the x-ray deposition in a Li/Pb film [18], and neither version computes the thermal response of a film. The debris ions at the highest energies of their figure stop 0–14 % shallower here, and fast deuterons and tritons leave 1 mm of lithium with energies within about 0.1 MeV of theirs (below).

On the inputs of Blanchard and Martin [14] (250 μm of tungsten on 3 mm of steel, their properties and their uniform-depth loads, from 400 °C), the solver reproduces the rises of their finite-element peak temperatures to within 9.9 % (4.0 % for the peaks in kelvin). The surface rise is 4.4 % lower than theirs and the rises at 10 and 25 μm are 4.4 % and 9.9 % higher. Both calculations use the same prescribed deposition, so this pattern reflects the conduction calculation, not the deposition model. With this study's spectra and deposition, the peak rises of a 3 mm tungsten slab above its 500 °C initial temperature agree with Raffray's [3] to within 3 %. The single-shot and repeated-shot peaks of the HiPER scenarios agree with those of Garoz et al. [5] to within 3 %.

Raffray's values are read by eye from his Fig. 7, with an uncertainty of about ±25 °C; his text gives the surface maximum as about 2900 °C. The Raffray check uses the TRIM-profile deposition path, which applies to a single tungsten slab; the production runs use the projected model, whose tungsten peaks are 0.3–0.6 % higher (Section 2.4).

"Within rounding" means that the value falls inside the printed rounding interval of Garoz et al. Prototype and Demo both deviate by +1.8 %, but 2.545 J cm$^{-2}$ rounds to 2.5 and 7.840 does not round to 7.7. At 0.01 MeV the electronic stopping of alphas is 11 % below the NIST value and the nuclear stopping of protons 9 % above it. For the peaks themselves the Blanchard–Martin ratios are 0.963, 1.032 and 1.052 in °C and 0.967, 1.026 and 1.040 in kelvin, and the Raffray ratios 0.993 and 1.016 in °C and 0.994 and 1.013 in kelvin. The rises are the fairer comparison, because a ratio of Celsius temperatures depends on the zero of the scale. The Raffray values were read by eye from his Fig. 7 with an uncertainty of about ±25 °C. The digitized traces in the repository miss the peak between 1.4 and 2.3 μs. Their 100 μm trace starts at 502 °C, consistent with the 500 °C initial temperature. The temperatures of Garoz et al. are printed to two or three significant figures. The Garoz rows compare like with like. For Experimental they use the first pulse, and for Prototype and Demo the last shots of the 100- and 600-shot runs against their "pulse during steady state regime".

The HEIGHTS-IFE values of kind (f) were read by eye from Fig. 8 of the preprint of Hassanein and Morozov

[17] (to about $\pm 0.5$ µm and $\pm 0.3$ MeV), whose 1 mm lithium film lies on carbon composite. Only the curves inside the lithium are compared, and the starting energies are those of their figure. Their lithium density is not stated. The depths of this work are for liquid lithium at 600 K (502 kg m$^{-3}$), and would be 6 % shorter at the solid density of the SRIM tables. The fast protons of the figure start at 23 MeV, above the 20 MeV end of the SRIM tables of this work, and are not compared. The production runs deposit every ion at its mid-bin energy, the tabulated one for burn ions and the tabulated one times the stretch factor of Section S2 for debris species: at most 14.8 MeV for the hydrogen and helium isotopes and 26.4 MeV for the gold debris (39.8 MeV in the table).

**Table S5.** Benchmark checks. Kind: (a) source term against Garoz et al. [5]; (b) solver on the inputs of Blanchard and Martin [14]; (c) whole chain against Raffray [3] and Garoz et al.; (d) stopping data and depth model against NIST [2] and TRIM; (e) lithium evaporation against an energy-balance hand estimate; (f) ion ranges in lithium against HEIGHTS-IFE [17]. The ratios of the Blanchard–Martin and Raffray rows are ratios of the temperature rises above the initial temperature of each calculation, 400 °C and 500 °C, with the peaks in °C as in the original papers; fluences in J cm$^{-2}$, as in Garoz et al.

| Check | Kind | Reference | This work | Ratio / verdict |
|---|---|---|---|---|
| Total ion fluence, Experimental | a | 1.7 J cm$^{-2}$ [5] | 1.721 J cm$^{-2}$ | within rounding |
| Total ion fluence, Prototype | a | 2.5 J cm$^{-2}$ [5] | 2.545 J cm$^{-2}$ | within rounding |
| Total ion fluence, Demo | a | 7.7 J cm$^{-2}$ [5] | 7.840 J cm$^{-2}$ | +1.8 % |
| Per-species, group and x-ray fluences | a | 33 values [5] | – | 31 of 33 within rounding |
| X-ray mean photon energy | a | 7 keV [5] | 7.09 keV | within rounding |
| Peak W surface temperature, Experimental (first shot) | c | 875 K [5] | 868 K | 0.992 |
| Peak W surface temperature, Prototype (last of 100 shots) | c | 1410 K [5] | 1381 K | 0.979 |
| Peak W surface temperature, Demo (last of 600 shots) | c | 3400 K [5] | 3317 K | 0.976 |
| Blanchard–Martin benchmark, peak surface | b | 2554 °C [14] | 2460 °C | 0.956 |
| Blanchard–Martin benchmark, peak at 10 µm | b | 1393 °C [14] | 1437 °C | 1.044 |
| Blanchard–Martin benchmark, peak at 25 µm | b | 849 °C [14] | 893 °C | 1.099 |
| Raffray, 3 mm W slab, surface peak | c | ~2915 °C [3] | 2895 °C | 0.992 |
| Raffray, 10 µm peak | c | ~1525 °C [3] | 1549 °C | 1.023 |
| SRIM vs NIST in W, electronic stopping (0.1–10 MeV) | d | NIST [2] | SRIM-2012 [8] | 0.958–1.051 |
| SRIM vs NIST in W, projected range (0.01–10 MeV) | d | NIST [2] | SRIM-2012 [8] | 0.927–1.041 |
| Depth model vs TRIM (Li targets; MeV ions in W), 50 %-energy depth | d | TRIM [8] | model | 0.935–1.005 |
| Depth model vs TRIM (d and $^4$He, 1–100 keV in W), 50 %-energy depth | d | TRIM [8] | model | 0.69–0.76 |
| Li thick film, Demo: peak surface temperature | e | below 2318–2467 K (evaporation-limited, no conduction) | 2218 K | consistent |
| Li thick film, Demo: evaporated mass | e | 6.0×10$^{-4}$ to 1.4×10$^{-3}$ kg m$^{-2}$ | 7.4×10$^{-4}$ kg m$^{-2}$ | inside the bracket |
| Debris ions in Li at 3.42 (C), 1.10 ($^4$He), 0.83 ($^3$He), 0.83 (T), 0.55 (D) and 0.27 MeV (H): end of the path | f | 13.3, 14.5, 10.0, 28.5, 19.2, 9.5 µm [17] | 12.9, 13.2, 10.0, 25.3, 16.7, 8.2 µm | 0.86–1.00 |
| Fast D (16.3 MeV) and T (16.7 MeV): energy after 1 mm of Li | f | 14.0, 13.4 MeV [17] | 13.9, 13.3 MeV | 0.99 / 1.00 |
| Fast $^4$He (18.4 MeV) in Li: end of the path | f | 750 µm [17] | 723 µm | 0.96 |

The lithium hand estimate treats the $4.17\times10^4$ J m$^{-2}$ of debris-ion energy of a Demo shot as a surface heat flux. The depth model deposits half of it within 1.8 µm and 90 % within 4.9 µm of the surface, comparable with the 7.2 µm that heat diffuses during the arrival. The flux acts over the 10–90 % debris arrival window (1.28–2.66 µs; mean flux $2.4\times10^{10}$ W m$^{-2}$, peak $3.9\times10^{10}$ W m$^{-2}$) on semi-infinite liquid lithium with the property correlations of the solver. The burn ions and x-rays, which deposit deep or not at all, are neglected. Without evaporation the surface would reach 3620 K. With evaporation alone removing the flux, the surface would be held at 2318 K for the mean and 2467 K for the peak flux. Conduction, neglected in these caps, lowers the surface temperature further, consistent with the computed 2218 K. If the surface conducted heat at the capped temperature for the whole window, $6.0\times10^{-4}$ kg m$^{-2}$ would evaporate; if it conducted none, $1.4\times10^{-3}$ kg m$^{-2}$.

The computed $7.4\times10^{-4}$ kg m$^{-2}$ lies inside this bracket, near its lower end, as expected for a surface that conducts most of the time.

In the Blanchard–Martin benchmark the rise of the surface peak is 4.4 % low while the rises at 10 and 25 μm are 4.4 % and 9.9 % high. The shared prescribed deposition is uniform over 2 μm for the x-rays, 10 μm for the burn ions and 1 μm for the debris ions, so the pattern means that slightly less heat stays at the surface and slightly more reaches 10–25 μm in this model than in theirs. Possible causes, none tested, are how the uniform loads are applied (here per unit mass over the stated depth at the initial density, from the current surface), the near-surface resolution of the benchmark mesh (5 nm cells growing by ×1.06), and the evaluation of the temperature-dependent properties.

**Derived lithium quantities (Section 2.7).** The thick-film evaporated mass is the median over $\delta \geq 100$ μm, where it no longer depends on $\delta$. Whole-chamber totals are obtained by multiplying by $4\pi R^2$, an upper bound because the beam openings are not subtracted. The flow speed needed to replace the film once per shot assumes that it travels half a great circle, $\pi R$, per shot period; the chamber height 2R gives a lower bound.

**Lithium film, single shot (Section 5.2).** Films of 3.2 μm and less let the steel pass the 1800 K limit of its property table within the first 2 μs. Across the sensitivity cases the Demo minimum ranges from 24 to 79 μm, mostly through the film-and-coolant temperature, because that case moves the film, the starting steel and the coolant together. At 800 K the steel begins every shot only 125 K below its limit. The evaporation coefficient $\alpha = 0.1$ gives 36 μm.

For films thicker than a few tens of micrometers the evaporated mass no longer depends on the thickness. A Demo shot then evaporates $7.4\times10^{-4}$ kg m$^{-2}$ of lithium, 0.39 kg over the whole chamber if the film covers the whole sphere, which would correspond to a mean vapor number density of $3.0\times10^{22}$ m$^{-3}$ if it stayed in the chamber ($4.2\times10^{19}$ and $3.6\times10^{20}$ m$^{-3}$ for the thick films of Experimental and Prototype). Three inputs control this mass. First, the thick-film surface reaches 2218 K, above the 800–1800 K range of the vapor-pressure data, so the pressure is extrapolated. Second, the heat of vaporization is estimated from the slope of the same correlation. Third, the evaporation coefficient $\alpha = 0.1$ lowers the mass to $4.2\times10^{-4}$ kg m$^{-2}$. The mass is that of surface evaporation only. The surface layer is superheated by about 600 K above the 1617 K normal boiling point [19], and the volumetric or explosive boiling that Zaghloul and Raffray found to dominate the response of a liquid wall to prompt x-rays [20] is outside the model. An energy-balance estimate made independently of the solver is consistent with both results (above; Table S5). It treats the $4.2\times10^{4}$ J m$^{-2}$ of the debris ions, which deposit 90 % of their energy within 5 μm, as a surface heat flux, caps the surface at 2318–2467 K against the computed 2218 K, and brackets the evaporated mass between $6.0\times10^{-4}$ kg m$^{-2}$ and $1.4\times10^{-3}$ kg m$^{-2}$. Of the inputs varied one at a time, $\alpha$ has the largest effect on the evaporated mass and the film-and-coolant temperature the largest effect on the minimum thickness, while the x-ray pulse length changes neither (Section S6).

**Surface pressure and superheat (Section 5.2).** The numbers are taken from one fully recorded shot of the converged repeated-shot runs at 56.2, 100 and 1000 μm, which agree to within 0.1 K in the peak temperatures and 1.3 % in the impulse. They use the vapor-pressure law of Table B.1 at and above the surface peak. Anisimov's kinetic treatment of a metal evaporating into vacuum gives, at the sonic point of the vapor, $T_1 = 0.67\ T_0$ and $n_1 = 0.31\ n_0$ for a monatomic vapor, with about 18 % of the evaporated atoms condensing back [21]. Here $T_0$ is the surface temperature and $n_0 = p_{\mathrm{sat}}(T_0)/(k_{\mathrm{B}}T_0)$ the saturated vapor density, with $k_{\mathrm{B}}$ the Boltzmann constant. The momentum flux there, $(1 + \gamma)\ n_1 k_{\mathrm{B}} T_1$ with $\gamma = 5/3$, is $0.554\ p_{\mathrm{sat}}(T_0)$, the recoil pressure used in Section 5.2. The model itself evaporates at the full Hertz–Knudsen–Langmuir rate ($\alpha = 1$, no back flux, Section 2.2). At the 2218 K surface peak the saturation pressure is 2.4 MPa and the recoil 1.33 MPa. Its impulse is 1.14 Pa s over the 1.28–2.66 μs debris window and 1.6 Pa s over the shot. The hottest liquid, 2483 K, lies 2.3 μm below the surface 2.1 μs after ignition, when the surface is at 2216 K and evaporation cools it. At the recoil pressure the boiling point, from the same vapor-pressure law, is 2072 K. The critical-temperature estimates that Zaghloul and Hassanein tabulate range from 3223 to 4176 K, and their own model gives 3495 K, corrected to 3084 K in the version read [22]. The low-pressure spinodal of the van der Waals equation, 27/32 of these, is 2602–3524 K, 119 K above the hottest liquid at the low end. The recoil pressure, below 0.05 of any of the critical pressures they list, does not move it appreciably. The same source quotes a tensile strength below 15.2 MPa for solid lithium from a handbook, which is not a cavitation threshold of the liquid.

**Lithium film, repeated shots (Section 5.3).** The repeated-shot runs replace the film at full thickness and 600 K before every shot and carry the steel's temperature profile forward, with the coolant at 600 K as in the tungsten case. Each run continued until the pre-shot temperature of the steel front face rose by less than 0.05 K per shot, which took 100–390 shots. The direct runs at 45, 56 and 183 μm took 340, 320 and 200 shots and settled at steel peaks of 917, 872 and 754 K. Without these runs, interpolation between 31.6 and 56.2 μm for both limits gives 44.7 and 55.9 μm. With the direct points the bracketing pairs become 31.6–45 μm and 45–56 μm and the minima are 43.6 and 55.8 μm.

Raffray gives a separate compatibility limit of about 600 °C (873 K) for the interface between ferritic steel and lithium coolant [3]. Between Demo shots the steel front face sits at 728–795 K behind 17.8–100 μm films in repeated operation, below that limit, and it exceeds it only during the peak of each shot. Applied to the peak, the limit would raise the minimum film to 37 μm in a single shot and to 56 μm in repeated operation (14.8 and 15.1 μm in Prototype). Whether these excursions during the peaks matter for compatibility is not known. Table S6 gives the steel front face over one fully recorded cycle of each converged run: its time mean and the time it spends above 925 and 873 K, together with the energy partition of the shot. Judged by the time mean, every film computed would pass both limits.

Neutron heating of the steel (Section 2.3 and Section S2; the film itself receives none) raises the repeated-shot minimum. With half the tungsten heating per unit volume, the steel behind 31.6, 56.2 and 100 μm films settles at peaks of 1026, 891 and 820 K, and the Demo minimum becomes 48 μm, 4 μm more than without neutron heating. With the full tungsten value the peaks are 1045, 910 and 844 K and the minimum becomes 53 μm, 9 μm more. The film for the 873 K compatibility limit grows from 56 μm to 63 and 83 μm. All these minima include direct runs at the thicknesses interpolated from the runs above, with and without neutron heating.

At the thicknesses that matter the steel is heated mainly by ions—not by x-rays. Averaged over the run, per Demo shot behind a 31.6 μm film, the x-rays deposit $3.9\times10^{3}$ J m$^{-2}$ directly in the steel, the burn ions that pass the film $1.8\times10^{4}$ J m$^{-2}$, and the film conducts a further $3.0\times10^{4}$ J m$^{-2}$ into it. The x-rays are 8 % of the load, but they set the sharp first peak of the steel front face because they arrive within 20 ns. The film replacement itself acts as a heat sink on the steel front face, because the replaced film leaves at the steel temperature and the new one arrives at 600 K. Relative to a stagnant film this is non-conservative for the steel temperature; relative to a real flowing film, which also carries heat along the wall during the period, its sign is unknown.

**Table S6.** Steel front face and energy partition over one fully recorded shot of each converged repeated-shot run ($T_{\mathrm{Li}}$ = 600 K, no shielding or neutron heating): pre-shot, peak and time-mean temperature of the steel (peak: hottest steel cell), time per shot above 925 and 873 K, and the shares of the energy absorbed per shot ($8.2\times10^{4}$ J m$^{-2}$ in Demo, $2.7\times10^{4}$ J m$^{-2}$ in Prototype) that leave with the vapor (latent plus sensible heat), with the replaced film (its enthalpy above 600 K) and to the coolant; the rest, 1.6–2.6 % in Demo and below 0.1 % in Prototype, is the slow warming that remains in the converged state. Heat flux: time-mean heat flux through the steel to the coolant. Last two columns (pulsed-load indicators; no design rule for steel was found): the per-shot swing of the steel front face, and the heat-flux factor of the energy entering the steel, in its worst window (the 20 ns x-ray flash that the film lets through) and as the square pulse that gives the same swing, $F_{\mathrm{HF}} = \Delta T\,(\pi\,k\,\rho\,c_p)^{1/2}/2$ with the steel properties at the pre-shot temperature.

| Scenario, film | Pre-shot front (K) | Peak steel (K) | Time-mean front (K) | Above 925 [5] / 873 K [3] (μs per shot) | Vapor (%) | Replaced film (%) | Coolant (%) | Heat flux (MW m$^{-2}$) | Face swing (K) | $F_{\mathrm{HF}}$, worst window / square pulse (MW m$^{-2}$ s$^{1/2}$) |
|---|---|---|---|---|---|---|---|---|---|---|
| Demo, 17.8 μm | 795 | 1212 | 805 | 168.2 / 552.9 | 29.3 | 7.4 | 61.0 | 0.502 | 416 | 32.6 / 5.34 |
| Demo, 31.6 μm | 783 | 1006 | 791 | 97.0 / 316.7 | 27.4 | 13.4 | 57.0 | 0.469 | 223 | 28.2 / 2.85 |
| Demo, 45 μm (direct) | 769 | 917 | 777 | 0 / 156.2 | 26.7 | 18.1 | 52.7 | 0.434 | 147 | 26.8 / 1.87 |
| Demo, 56 μm (direct) | 759 | 872 | 767 | 0 / 0 | 26.5 | 21.5 | 49.5 | 0.408 | 113 | 26.7 / 1.43 |
| Demo, 56.2 μm | 759 | 871 | 766 | 0 / 0 | 26.5 | 21.6 | 49.5 | 0.407 | 112 | 26.7 / 1.42 |
| Demo, 100 μm | 728 | 796 | 734 | 0 / 0 | 26.2 | 31.6 | 39.6 | 0.326 | 53 | 26.7 / 0.66 |
| Demo, 183 μm (direct) | 694 | 754 | 698 | 0 / 0 | 26.1 | 42.7 | 28.6 | 0.236 | 45 | 26.5 / 0.56 |
| Demo, 1000 μm | 626 | 667 | 626 | 0 / 0 | 26.1 | 65.1 | 7.3 | 0.060 | 41 | 25.3 / 0.50 |
| Prototype, 10 μm | 609 | 969 | 611 | 4.0 / 8.2 | 1.1 | 0.7 | 98.2 | 0.026 | 360 | 12.0 / 4.29 |
| Prototype, 17.8 μm | 609 | 834 | 611 | 0 / 0 | 1.0 | 1.2 | 97.8 | 0.026 | 225 | 10.8 / 2.68 |
| Prototype, 31.6 μm | 609 | 742 | 610 | 0 / 0 | 0.9 | 2.2 | 96.9 | 0.026 | 133 | 9.4 / 1.58 |
| Prototype, 56.2 μm | 609 | 685 | 610 | 0 / 0 | 0.9 | 3.8 | 95.3 | 0.025 | 77 | 8.8 / 0.91 |

The recorded shot is the next shot of each stored run, fired from its last checkpoint exactly as the run fires every shot, with the temperature of the first steel cell and of the film recorded at every solver step; its peaks reproduce the stored last-shot peaks to within 0.04 K. Behind thick films the hottest steel lies below the front face (781 K at the face behind 100 μm, 739 K behind 183 μm), where the fast burn ions that cross the film stop. The vapor carries a quarter of the Demo energy for every film, because the surface evaporates the same mass. The replaced film carries more and the coolant less as the film thickens. Tungsten passes all of its absorbed energy, 0.82 MW $m^{-2}$ in Demo and 0.027 MW $m^{-2}$ in Prototype, through the steel (Section 2.7).

**Latent heat returned to the film (Section 5.3).** As a bounding case for vapor that condenses on the wall, three runs return to the film surface, as a uniform flux applied from 50 μs to the end of the period, the latent heat of the lithium evaporated in the first 50 μs of each shot. The sensible heat is not returned, and the energy ledger closes to $3\times10^{-13}$. Behind 45, 56.2 and 100 μm the steel then settles at peaks of 944, 897 and 828 K (917, 871 and 796 K without), its front face before each shot at 816, 803 and 765 K (769, 759 and 728 K), and the film evaporates 6.5, 4.5 and 1.3 % more per shot. The minimum films become 49.2 μm for 925 K, between 45 and 56.2 μm, and 68.7 μm for 873 K, between 56.2 and 100 μm, against 43.6 and 55.8 μm.

**Helium and displacements in the steel (Section 5.3).** Every ion sub-bin of the helium source term of Section S7.4 (24 sub-bins per energy bin here) crosses the film with the production range–energy mapping of lithium and reaches the steel with the energy left at its projected range minus the areal mass of the film. In the steel, modelled as iron with the SRIM iron tables, every transmitted helium and hydrogen ion is counted as implanted, because no TRIM reflection data for iron were made. The displacements follow the NRT model of Section S7.4 with the iron threshold $E_\mathrm{d} = 40$ eV [23], along the projected-range mapping in iron and blurred by its straggling. The ion energy carried into the steel agrees with that of the production deposition to within 2 % for films up to 100 μm and 7 % at 1 mm. Because each sub-bin either reaches the steel or not, the helium fluence is a step function of the film thickness. The helium-exclusion films of Table 5 are those of the step at which it falls below $10^{12}$ or $10^{13}$ $m^{-2}$ per shot ($10^6$ shots to $10^{18}$ or $10^{19}$ $m^{-2}$), rounded up to the next whole micrometer. The 3.5 MeV burn-alpha population, up to the 4.0 MeV top of its bin, stops within 60 μm. The helium of higher energy, a tail of the burn spectrum whose top sub-bins reach about 16 MeV, needs 0.69 mm, set by burn $^3$He at 15.8 MeV. Behind the 44 μm minimum the steel also receives $2.3\times10^{16}$ hydrogen isotopes $m^{-2}$ per shot. Its ion dose, $6\times10^{-7}$ dpa per shot in the first 5 μm, is $2.0\times10^{2}$ dpa per full-power year at 10 Hz, and still $1.0\times10^{-7}$ dpa per shot behind the 184 μm helium-exclusion film, from the fast hydrogen isotopes. The neutron dose of the steel is the same behind either wall and is not computed here. For a full-power year at 10 Hz, $3.2\times10^{8}$ shots, the Demo steel stays below $10^{18}$ and $10^{19}$ $m^{-2}$ of helium only behind 0.24 and 0.21 mm of lithium. The steel threshold is taken over from tungsten, for lack of one for steel under pulsed helium. Figure S3 shows the helium per shot against the film thickness. Each helium-exclusion film is set by a single sub-bin of the burn spectrum near the top of a table bin. These are burn alphas of 8.25 MeV for $10^6$ shots to $10^{18}$ $m^{-2}$ in Demo, and burn $^3$He of 8.56 MeV and alphas of 8.83 MeV for one full-power year to $10^{18}$ and $10^{19}$ $m^{-2}$. The films therefore depend on how the energy of the spectra is matched to the accounting (Section S2). Stretching the burn spectra in energy instead of tilting their weights gives 213 μm ($10^6$ shots) and 239 μm (one year) in Demo, against 184 and 236 μm. The alternative, energy-normalized construction of Section S2 gives 182 and 662 μm.

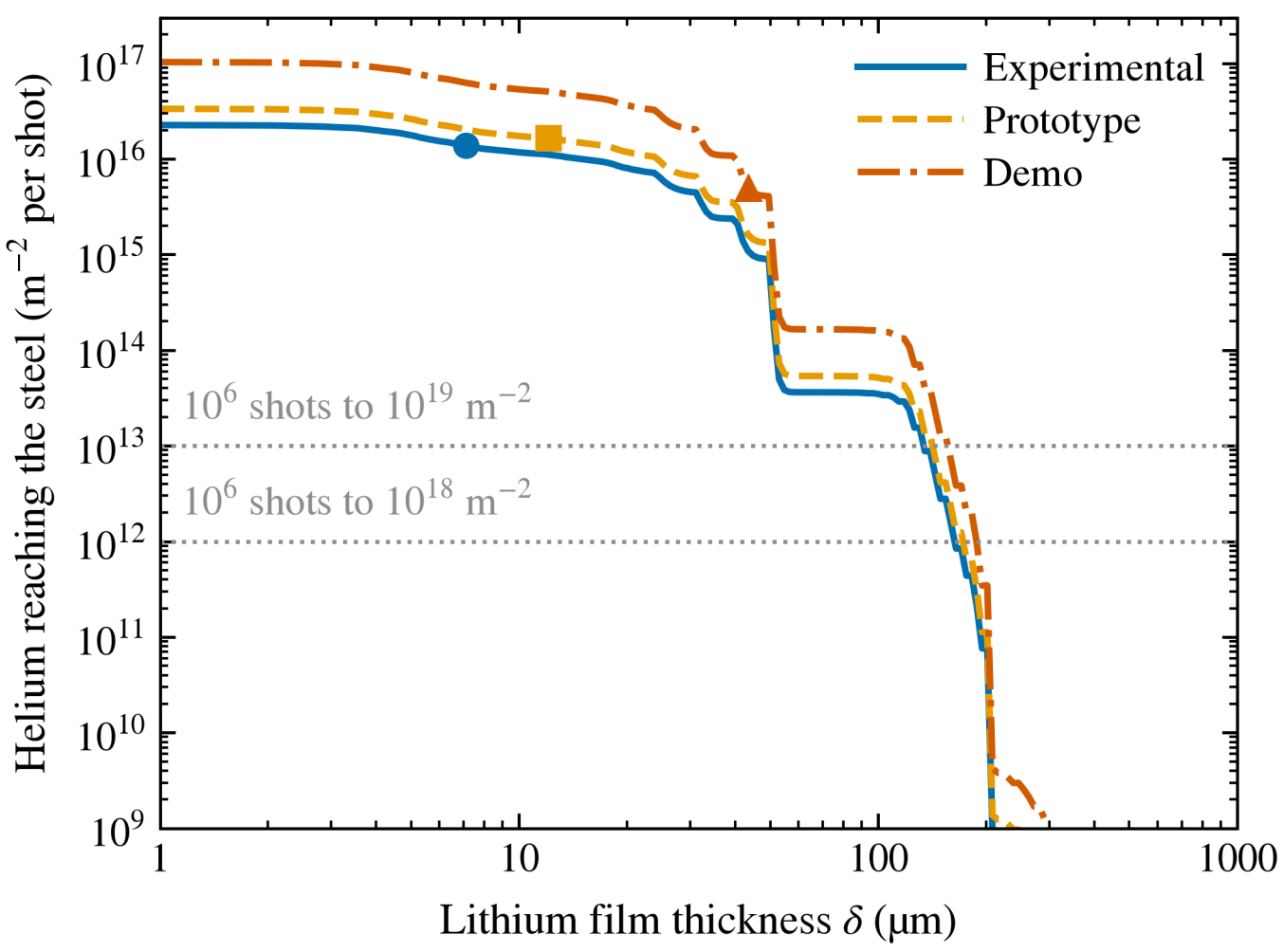

**Figure S3.** Helium reaching the steel behind a lithium film per shot against the film thickness, with all of it counted (fine grid of 241 films; the steps are those of the 24 sub-bins per energy bin). Dotted: the per-shot levels at which $10^6$ shots reach $10^{18}$ and $10^{19}$ $m^{-2}$ [4]. Markers: the 925 K minimum films of Table 5 (repeated shots; Experimental single shot).

**Partial renewal (Section 5.4).** The runs with partial renewal (`run_li_multi.py renew=<f>`) replace, at the start of every shot after the first, the fraction $f_{ren}$ of the lithium left after the previous shot by fresh lithium at 600 K and make up the evaporated mass with fresh lithium. The retained lithium keeps its enthalpy, and the film is rebuilt on its original mass grid at the uniform mixed enthalpy. The thin film is close to isothermal at the end of the period ($\delta^2/a \approx 40$ µs for 31.6 µm, with a the thermal diffusivity, against 0.1 s), which justifies this assumption. $f_{ren} = 1$ is the full replacement of the production runs, reproduced exactly. In Demo the film left after renewal is at 720–731 K for $f_{ren} = 0.3$ and 786–791 K for $f_{ren} = 0.03$ at the quasi-steady state. The steel peaks at 1050, 930 and 852 K behind 31.6, 56.2 and 100 µm for $f_{ren} = 0.3$ and at 1069, 957 and 892 K for $f_{ren} = 0.03$, against 1006, 871 and 796 K with full renewal, and the runs converge after 350–400 shots. The minimum films, interpolated linearly in temperature and logarithmically in thickness, are 58.4 µm (925 K) and 85.5 µm (873 K) for $f_{ren} = 0.3$ and 74.6 µm (925 K) for $f_{ren} = 0.03$, whose 873 K film is thicker than 100 µm. With full renewal the same interpolation gives 44.7 and 55.9 µm. Intermediate fractions and Prototype were not computed.

**Film replacement (Section 5.4).** Replacing the whole film once per shot requires the lithium to sweep the wall within one period. If it flows from pole to pole, half a great circle per period, the Demo film must move at $\pi R f = 204$ m s$^{-1}$, and at 20 m s$^{-1}$ in Prototype; the chamber height, $2R f = 130$ m s$^{-1}$, is a lower bound. Both are criteria of this study, not design values. The demonstrated free-surface lithium flows are an order of magnitude slower than the Demo requirement and close to the Prototype one. The IFMIF/EVEDA lithium test loop ran a 25 mm thick free-surface flow at 15 m s$^{-1}$ for 1300 h [24], in a loop built to validate a target at up to 20 m s$^{-1}$ [25]. The films here are not 25 mm but only tens of micrometers thick, and viscous drag decides how far they can travel. Once the momentum has diffused across a film, after about $\delta^2/\nu$, a free film on a no-slip wall with no driving force has the half-parabolic profile and a wall shear stress of $3\mu U/\delta$. Its mean speed $U$ then decays with the time constant $\delta^2/(3\nu)$, so that it travels about $U\delta^2/(3\nu)$ in all. With the density of Table B.1 and Davison's viscosity correlation [26], $\mu = 5.1\times10^{-4}$ Pa s and $\nu = 1.0\times10^{-6}$ m$^2$ s$^{-1}$ at 600 K. In a 44 µm film, momentum diffuses across the film within 1.9 ms, after which the flow decays viscously, and the film stops within 0.9 cm at 15 m s$^{-1}$ and within 13 cm at 204 m s$^{-1}$. A 56 µm film stops within 1.5 cm and 21 cm at the same speeds, and a 184 µm film within 17 cm and 2.3 m. The Prototype film of 12.1 µm stops within 1 mm. Only distributed injection, with inlets a few centimeters apart or a porous wall, can renew such films over the whole wall.

**Vapor shielding in context (Section 7.2).** If the re-radiated fraction is as small as in the lithium plasma of Sizyuk and Hassanein [12], the reduced model gives an evaporation 21–34 % lower than without shielding, about 5–6$\times10^{-4}$ kg m$^{-2}$ per shot. Even then the chamber must clear about 0.3 kg of vapor per shot, so none of the conclusions of Section 7.1 changes, and in all seven shielded cases the minimum film moves by only 2–6 %. A reduced model cannot do better than this bracket. The bracket is narrow where it matters, since the effect of shielding on the steel and on the film thickness is small for every value of the two parameters. Fixing the parameters themselves needs radiation hydrodynamics with lithium opacities, as in HEIGHTS or BUCKY [15, 16]. The only benchmark found with the same target spectra, the attenuation of the ions by xenon fill gas that Raffray reports from BUCKY calculations [3], of the kind that Peterson et al. describe for a xenon-filled 6.5 m chamber [13], is left for future work. It needs SRIM stopping tables for xenon gas that could not be generated here, and it would test the stopping only.

Liquid-metal experiments and simulations give a rough check of the size and direction of the effect, under conditions that differ from those here in every respect: plasma loading of steady devices or 1 ms tokamak transients instead of microsecond ion pulses, magnetized plasma that deposits its energy in the vapor instead of passing through it, and vapor clouds that build up over milliseconds to seconds. In Pilot-PSI, liquid tin exposed to hydrogen and helium plasmas at up to 22 MW m$^{-2}$ formed a vapor cloud whose shielding oscillated and kept the surface temperature nearly constant while the heat flux rose [27]; lithium targets in Magnum-PSI loaded with up to 9 MW m$^{-2}$ reached a temperature plateau that a model reproduced only if a significant fraction of the power was dissipated in the lithium vapor [28]; and in HEIGHTS simulations of ITER transients a lithium vapor plasma absorbed 82–88 % of the incident energy and re-radiated only 0.4–1.4 % of what it absorbed [12]. All three show the direction found here, a vapor that lowers the energy reaching the surface and limits its temperature, and the last supports a small re-radiated fraction. The effect is far larger there because the vapor has time to build up before most of the energy arrives. In an IFE shot the column grows during the few microseconds in which the debris ions arrive, which is why it stops only 8–15 % of the ion energy here. No quantitative comparison is possible.

## S4. Literature inputs and quantities computed in this work

**Scenarios and walls (Section 2.1).** The tungsten wall is 1 mm of W on 1 cm of steel, the configuration of Garoz et al. [5]. Their substrate is an oxide-dispersion-strengthened reduced-activation ferritic-martensitic (ODS-RAFM) steel, whose property set was not available here. The steel is therefore modeled with the ferritic-steel properties of Blanchard and Martin [14] as a substitute. It is judged against the 925 K operating limit used for ODS-RAFM by Garoz et al. In the Prototype and Demo scenarios, lead–lithium cools the back face of the steel at $T_c$ = 600 K with a heat-transfer coefficient $h$ = 10 kW m$^{-2}$ K$^{-1}$ [5]. The 600 K is the coolant temperature, not a prescribed wall temperature. The wall starts at 600 K and its temperature is computed from then on. The heat absorbed at the surface leaves only by conduction through the 1 cm of steel and convection into the coolant. The published Experimental scenario fires a few shots per bunch. Here it is computed as single shots, each from a wall at 300 K with an adiabatic back face (the author's choice). The published configuration removes the heat through the concrete shielding without stating the thermal condition there.

The lithium wall is a film of thickness $\delta$ on the same 1 cm of steel. Thirteen thicknesses from 1 μm to 1 mm, spaced logarithmically, are computed in every scenario. In the single-shot calculations, each shot hits a fresh film at $T_{\mathrm{Li}}$ = 600 K on steel already in equilibrium with it, and the coolant, where there is one, is also at $T_{\mathrm{Li}}$. This represents a film fully replaced between shots, on steel that keeps no heat from earlier shots. The repeated-shot calculations of Section 5.3 drop the second assumption. There the film is replaced at full thickness and 600 K before every shot, while the steel carries its temperature profile from shot to shot and is cooled as in the tungsten case. The baseline film temperature, 600 K, equals the coolant and initial wall temperature of the Prototype and Demo tungsten cases and lies about 150 K above the melting point of lithium. Sensitivity cases are run at 500 and 800 K. Table S7 gives the status of every input.

The quantities computed in this work are the surface, pre-shot, interface and steel temperatures and melt depths; the evaporated masses and depths; the absorbed fractions in lithium; the minimum film thicknesses and whole-chamber totals; the helium shot counts (with a published threshold as input); the sputtering erosion (with published fits as input); the mean interface temperature from the heat balance; the flow speeds (arithmetic on $R$ and $f$); the vapor-shielding results; and the validation ratios. The design-rule cycle counts at the printed heat-flux factors are arithmetic on two literature inputs and contain no output of the model. Those at the factors of this work use the computed ion power history.

**Table S7.** Inputs from the literature or author choices (status: SOURCED, read from the cited page; SOURCED (secondary), printed in the cited paper from another work; DIGITIZED; ESTIMATED; ASSUMPTION; COMPUTED, made for this study with the cited code) and the quantities computed in this work.

| Input | Value | Source and status |
|---|---|---|
| Yields, radii, repetition rates | Table 1 | [4, 5] (SOURCED) |
| Energy partition of the NRL 154 MJ target | Table S3 | [3] (SOURCED) |
| Ion spectra and accounting energies | tabulated | [7] (SOURCED; bin edges reconstructed) |
| X-ray spectrum | 1119 points | [3], Fig. 2 (DIGITIZED, ±10 %) |
| Scaling to 20 and 50 MJ | × $Y$/154 MJ | practice of [3]; method not stated in [4] (ESTIMATED) |
| X-ray pulse | 20 ns | consistent with the "tens of nanoseconds" stated in [4] (ASSUMPTION) |
| Photon attenuation | $\kappa(E)$ | [29, 30] (SOURCED) |
| Ion stopping, ranges, TRIM factors, backscatter | tables | SRIM-2012 and TRIM runs made for this study [8] (COMPUTED) |
| W, Li, steel properties | Table B.1 | [3, 14, 26, 31] (SOURCED or secondary; Li $L_v$ ESTIMATED) |
| Steel limit 925 K; wall and coolant | – | [5] (SOURCED, secondary) |
| W recrystallization range | 1300–1800 K | [4] (SOURCED, secondary) |
| Tungsten-grade properties | Table B.2 | [32–36] (SOURCED); ODS-W conductivity (ESTIMATED band) |
| Pulsed He damage fluence | $10^{18}$–$10^{19}$ m$^{-2}$ | [4], from experiments including [37] (SOURCED, secondary) |
| Heat-flux factors | 9.7, 14.4, 44.4 MW m$^{-2}$ s$^{1/2}$ | [4] (SOURCED) |
| Surface-load design rule | $N = \exp(100/F_{HF})$ | [38] (SOURCED; surface loads only) |
| Sputtering fit parameters | – | [39, 40] (SOURCED; C and Au proxies ESTIMATED) |
| Evaporation coefficient | $\alpha = 1$ (0.1) | the Hertz–Knudsen–Langmuir flux with a coefficient of unity applies when a condensing surface is within one collision length; for a distant condenser the flux is 0.844 of that maximum [1] (SOURCED; $\alpha = 1$ is therefore an upper bound); 0.1 ESTIMATED |
| Li film temperature | 600 K (500, 800 K) | ASSUMPTION |
| Experimental back boundary | adiabatic | ASSUMPTION |
| Full-replacement flow criterion | $\pi R f$ ($2R f$ lower bound) | ASSUMPTION |
| Steady-state criterion | 0.05 K per shot | ASSUMPTION |
| Neutron heating (separate case) | tungsten: 800 MW cm$^{-3}$ for 5 ns at the Demo conditions, scaled by $Y$/154 MJ and $(6.5\text{ m}/R)^2$; steel: 0.5 or 1 times the tungsten value per unit volume; lithium film: none | [6] for tungsten (SOURCED); steel and lithium ASSUMPTION |
| Vapor stopping factor | $s = 1$ (2; 5 and 10 on the 1 mm Demo film) | ASSUMPTION; no value for H or He ions in Li found |
| Re-radiated fraction and time | $f_{rr} = 0, 0.5, 1$; $\tau_{rr} = 0$ (300 µs) | ASSUMPTION, bracketed by [3, 12, 13] |

## S5. Limitations

**Limitations in the Discussion (Section 7.4).** The model computes temperature fields, not stress or strain. The tungsten damage indicators of Section 4.4, which mark the onset of surface modification, therefore rest on a published helium threshold and an empirical design rule, not on a computed damage state. No life is computed, and the plastic region of about 100 µm and the fatigue life of about $10^6$ cycles that Garoz et al. [5] computed for Demo are neither reproduced nor tested. Hassanein and Morozov point out that a liquid-metal surface under intense pulses is expected to lose mass as splashed droplets, through the boiling and bursting of bubbles, the momentum of the debris and hydrodynamic instabilities. In disruption-simulation experiments, they add, the losses of molten metals can be much larger than their surface vaporization [18]. The evaporated masses of Sections 5 and 6 are therefore surface-evaporation estimates, not the whole loss of a lithium film per shot. The same holds for the melt layer of a damaged tungsten surface (Section 4.5).

Several simplifications act in a known direction (Table S8). The slightly too shallow deposition of keV ions in tungsten overestimates the surface temperatures, which is conservative. So does the neglect of surface radiation if the chamber interior is not isothermal; under the isothermal-chamber argument of Section 2.2 it has no effect. Leaving neutron heating out of the baseline underestimates the surface, interface and steel temperatures, which is not conservative. The separate case of Section 4.2 raises the mean Demo interface temperature by 67 K or 117 K, depending on the assumed heating of the steel, and cuts the melting margin of the quasi-steady Demo surface from 378 K to 319 K or 275 K. The substitute steel properties, the fixed coolant heat-transfer coefficient and the one-dimensional, normal-incidence geometry act in an unknown direction. The source term is the ARIES spectrum scaled by yield, not the HiPER group's own. The raw ion tables, with 13 % more ion energy and 18 % more helium than the production spectra, raise the Demo single-shot tungsten peak by 526 K, so how the tables are made into spectra is a first-order input. The lithium evaporation rests on a vapor-pressure law extrapolated above 1800 K.

**Table S8.** What is excluded or simplified, which results it affects, and in which direction it acts.

| # | Excluded or simplified | Results affected | Direction |
|---|---|---|---|
| 1 | Thermomechanics: stress, plasticity, residual stress, ductile-to-brittle transition, crack growth, W/steel expansion mismatch | every statement about tungsten life | not computed; the life indicators are not lives |
| 2 | Vapor shielding only in the reduced form of Section 2.5: no vapor hydrodynamics, ionization balance, radiation transport or Knudsen-layer return flux; $s$ and $f_{rr}$ as ranges ($s$ = 5 and 10 as a sensitivity test on the thick film); repeated shots for $f_{rr}$ = 0, $s$ = 1 only, with the vapor cleared between shots | Li evaporation and surface temperature; Li minimum thickness | bracketed: −34 % to +8 % on the Demo evaporation for $s$ = 1–2 (−67 % to +13 % for $s$ up to 10); < 6 % on the single-shot minimum film and 4.5 % on the repeated-shot minimum; none on W |
| 3 | Radiation from the hot surface | surface temperatures | conservative if the chamber interior is not isothermal; none under the isothermal-chamber argument [14] |
| 4 | Neutron heating of the W and the steel, left out of the baseline; a separate case uses the Sharafat et al. value [6] for W and an assumed steel heating of 0.5 or 1 times the W value per unit volume, none in the Li film | W surface peak (melting margin 378 → 275–319 K); interface and steel temperatures; Li minimum thickness; Li evaporation (+7–13 %) | non-conservative in the baseline: the separate case raises the mean Demo W/steel interface by 67 or 117 K and the repeated-shot Li minimum film from 44 to 48 or 53 µm, all three with direct runs at the interpolated thicknesses, and cuts the Demo melting margin from 378 K to 319 or 275 K (Sections 4.2, 5.3) |
| 5 | 1D, normal incidence, isotropic point source, single spherical wall; no lateral conduction, edges or beam ports | all results | not assessed |
| 6 | TRIM depth profiles not applied with W on steel (Section S2) | W temperatures | conservative: W peaks 0.3–0.6 % too high |
| 7 | Other depth-model simplifications: no straggling, equal-velocity scaling for p, t, $^3$He, no TRIM data for C in W or for ions in Fe | deposition depth; W and Li temperatures | small; the CSDA alternative changes the Demo peak by −40 K |
| 8 | X-ray deposition: incoherent scattering counted as absorbed; no fluorescence or photoelectron transport; steel as Fe | near-surface x-ray deposit | slightly conservative |
| 9 | Source term: ARIES NRL 154 MJ spectra [7], not HiPER's; yield scaling ESTIMATED; number and energy of every species kept (one alpha per fusion, the ARIES accounting energies); reconstructed bin edges; digitized x-ray spectrum (±10 %) | all results | construction of the spectra: the raw tables (13 % more ion energy, 18 % more helium) raise the Demo single-shot peak by 526 K; stretching instead of tilting the burn spectra moves the helium-exclusion film from 184 to 213 µm (Section S3); scaling: unknown |
| 10 | Ion emission from a point at $t$ = 0, free streaming | timing of the heating | not assessed |
| 11 | Steel: ferritic substitute for ODS-RAFM [5], constant $k$ = 33 W m$^{-1}$ K$^{-1}$ [14] | W surface peak; interface and steel temperatures; Li minimum thickness | unknown sign; a 2.2 % higher $k$ brings the mean Demo interface to 925 K, and 29 W m$^{-1}$ K$^{-1}$ raises it to 965 K and the quasi-steady surface peak by 29 K (Section S6) |

**Table S8 (continued).**

| # | Excluded or simplified | Results affected | Direction |
|---|---|---|---|
| 12 | Li above 1800 K: vapor pressure extrapolated; $L_v$ ESTIMATED; $\alpha$ uncertain | Li evaporation and surface temperature | $p_{sat}$, $L_v$: unknown; $\alpha = 1$ is an upper bound on evaporation ($\alpha = 0.1$ lowers the evaporated mass by factors of 1.8 (Demo) to 9 (Prototype)) |
| 13 | Conductivity of undamaged bulk W, also in the near-surface layer | W temperatures, melting margin | non-conservative for damaged or imperfect surfaces: a halved conductivity melts the Demo surface (Section 4.5) |
| 14 | Coolant: fixed $h$ and $T_c$; no LiPb flow model | interface temperature | mean Demo interface 889–1012 K for $h$ = 20–5 kW m$^{-2}$ K$^{-1}$ |
| 15 | Li film flat and stable; no waves, droplets, splashing or dewetting; film static during the pulse; replacement once per shot | Li minimum thickness, evaporation and mass loss | splashed droplets and boiling of the superheated layer (Section 5.2) can remove more liquid than evaporation [18], so the evaporated mass is a surface-evaporation estimate, not the whole loss; the recoil pressure is computed but not applied to the film; the replacement acts as a heat sink on the steel: non-conservative relative to a stagnant film, unknown relative to a real flowing film |
| 16 | Chamber clearing, recondensation, target injection | feasibility of the Li wall | out of scope; no feasibility verdict is given |
| 16a | Steel behind the film: ion helium counted as fully retained against the tungsten threshold of $10^{18}$–$10^{19}$ m$^{-2}$ [4]; iron for the steel; no TRIM reflection in iron; neutron damage of the steel not computed | helium-exclusion film (Section 5.3) | the fully retained count is a lower bound on the shots; the threshold for steel is not known |
| 16b | Latent heat of the vapor lost from the wall budget; a bounding case returns it to the film between shots | Li minimum thickness, steel temperatures | non-conservative in the baseline: the bounding case raises the Demo minima from 44 to 50 µm (925 K) and from 56 to 69 µm (873 K) (Section 5.3) |
| 17 | Sputtering: extrapolated yields, C and Au proxies, gross erosion, no redeposition or helium-induced surface change, normal incidence | W erosion | order of magnitude only |
| 18 | Helium: threshold uncertain ($10^{18}$–$4{\times}10^{19}$ m$^{-2}$ [4, 37], and helium not confirmed as the cause of the pulsed-ion damage); threshold applied to shallow debris He as to burn He; temperature dependence of He effects; the retention model releases helium in a heating test where tungsten kept it (Section S7.10) | helium shot counts | the counts may be several times larger (Section 4.4); the full-retention counts are the main result, a lower bound |
| 19 | Design rule applied to volumetric µs heating, above its laboratory data and below the pulse length of its electron-beam data | design-rule cycle counts | unknown |
| 20 | No tritium or activation; displacement damage computed only in the helium model (ions, Section S7.4; neutrons at $2{\times}10^{-7}$ dpa per Demo shot, derived from the HAPL rate [6] and scaled with the neutron fluence in Prototype and Experimental, Table S15, 4–14 times the tungsten-specific rate of a magnetic-fusion DEMO [41]) and, for the ions, in the steel behind the film (Section 5.3), not fed back into the conductivity | W properties and life | non-conservative for the W conductivity: the ion damage reaches, within about 5 minutes of Demo operation (projected-depth count; sooner along the ion path), the 0.1 dpa from which the measured conductivity reduction saturates at about half (Section S6), which could melt the surface (item 13; Section 7.4); the neutrons alone would need 14 hours at this dose, 2.5–8 days at the tungsten-specific rate |
| 21 | Discretization: production grids and step limits (convergence study in Table S2); quasi-steady state judged by a 0.05 K per shot criterion | all results | tungsten peaks about 1 K too high, other results at most in their last quoted digit; the repeated Demo runs still rising by up to 5 K at their last shot (fitted; 938 K for the interface of the 600-shot run) |
| 22 | Experimental scenario as single shots on an adiabatic back face | Experimental results | no effect on the single-shot peak |
| 23 | Recrystallization: indicator only, extrapolated from one-hour anneals | W mechanical state | indicator only; no hardness or ductility change computed |
| 24 | Tungsten grades: only the conductivity changes; ODS-W conductivity a proxy band; pure-W properties otherwise; bubbles and particles not modeled | Section 4.6 | sensitivity band, not a prediction |

## S6. Tungsten armor: further results, damage indicators, coolant sensitivity and grades

**Indicators (Section 2.7).** Two damage indicators are reported for tungsten. Neither is a lifetime. The first is the number of shots after which the retained helium reaches the pulsed-damage fluence of $10^{18}$–$10^{19}$ $m^{-2}$ that González-Arrabal et al. [4] report from the experiments of Renk et al. [37] and of Dutta et al. (as cited there). The main count assumes that every implanted helium atom stays in the armor, which gives a lower bound on the number of shots. A reaction–diffusion model of helium, vacancies and self-interstitials, driven by the same source term and the computed temperatures (Section S7), gives a secondary estimate in which the heating spikes release part of the helium. Shot counts are interpolated between shots and rounded to the nearest shot. The second is the number of cycles allowed by the surface-load design rule of Mitteau et al. [38], $N = \exp(100/F_{\mathrm{HF}})$. It is evaluated at the heat-flux factors $F_{\mathrm{HF}}$ printed by González-Arrabal et al. [4], which contain no output of this model, and at the factors of the ion power history of this work (Design rule, below). Physical sputtering is computed from the Eckstein–Preuss fit formula with the parameters of Eckstein for H and He isotopes on tungsten [39, 40], extrapolated above 20 keV, where 95 % of the computed sputtering arises.

**Helium count.** The retained helium per shot is $n_{\mathrm{He}} = \Sigma_{\mathrm{bins}} N_{\mathrm{bin}} (1 - \eta_N)\, Y/(154\ \mathrm{MJ} \cdot 4\pi R^2)$, where $N_{\mathrm{bin}}$ is the number of helium ions of a bin emitted by the 154 MJ target and $\eta_N$ the TRIM backscattered-number fraction. At implantation, 99.6 % of the burn $^4$He and 98.0 % of all helium is retained. If all of it stayed, the Demo armor would reach $10^{18}$ $m^{-2}$ after 10 shots and $10^{19}$ $m^{-2}$ after 98 (Prototype: 30 and 303), counting burn and debris helium. Counting burn helium alone at $10^{19}$ $m^{-2}$ gives 181 shots. These are the main counts of Section 4.4. The reaction–diffusion model of Section S7 lets the retained fraction of newly implanted helium fall from 99 % at the first shot to 26 % after 1450 shots, as the heating spikes release it, which raises the counts to 11 and 125 shots (31 and 308; 11 and 131 with the projected-depth displacement count, Section S7.8). Because that model releases helium in heating tests in which tungsten kept it (Section S7.10), its counts are a secondary estimate.

Debris helium deposits its energy at a mean depth of about 0.3 μm and burn helium at about 1.7 μm. As González-Arrabal et al. [4] report the experiments behind the threshold, their 250–800 keV helium is comparable with the debris helium here (mean 0.33 MeV) and less energetic than the burn helium (mean 2.09 MeV). Whether shallow and deep helium count towards one threshold is not stated there. Renk et al. give TRIM ranges in tungsten of 0.43 μm at 250 keV and 1.11 μm at 800 keV, the maximum of their beam, against 6 μm at 4 MeV, and judge the beam closer in depth to debris ions than to fusion ions but a reasonable representation of the fusion helium [37]. Here the debris and burn helium stop at mean depths of 0.54 and 2.9 μm (Section S7.4).

**Design rule.** Garoz et al. [5] give $F_{\mathrm{HF}} = 10.4$, 15.4 and 47.3 MW $m^{-2}$ $s^{1/2}$ for the burn ions alone. With 47.3, the Demo count would be 8.3 instead of 9.5, so the choice does not change the conclusion. The factors of this work follow from the ion power history of the source term (Section 2.3), without the x-rays and with the energy backscattered by tungsten removed. The absorbed ion energy arrives between 0.27 and 2.31 μs (10–90 %; 2.04 μs) in Prototype and Demo and between 0.21 and 1.78 μs (1.57 μs) in Experimental. The absorbed energy within that window divided by the square root of its length gives 10.9, 14.2 and 43.7 MW $m^{-2}$ $s^{1/2}$ for Experimental, Prototype and Demo, from 1.5 % below to 13 % above the printed factors. These allow $9.3 \times 10^3$, 1144 and 9.8 cycles. The peak absorbed power, averaged over the intervals of Table 2, times the square root of the window gives 153, 199 and 612 MW $m^{-2}$ $s^{1/2}$ and 1.9, 1.7 and 1.2 cycles. The first form treats the load as a uniform pulse of the window length, the second as if the whole window were at the peak power, which bounds the factor from above. Mitteau et al. [38] describe fewer than $10^4$ cycles as a regime with significant unknowns.

**Conductivity reductions.** Self-ion damage at room temperature lowers the thermal diffusivity of tungsten by 40 % at 0.01 dpa and by about 55 %, roughly half, from 0.1 dpa up to 10 dpa, measured by transient grating spectroscopy [42]. In Demo the ions alone would bring the first micrometers to that dose within minutes. The displacement source of the helium model is $3.7 \times 10^{-5}$ dpa per shot averaged over the first 5 μm, with the displacements counted along the projected depth (Section S7.4; the path-length count behind the model helium counts of Section 4.4 gives 1.78 times more ion displacements). It gives 0.1 dpa after about $3 \times 10^3$ shots, 5 min at 10 Hz. The neutron damage of Table S15, $2 \times 10^{-7}$ dpa per Demo shot (derived here from the $10^{-6}$ dpa $s^{-1}$ of Sharafat et al., Section 3.2, at the 5 Hz HAPL rate of their Section 4.5; material not stated), would take about $5 \times 10^5$ shots, 14 hours. Gilbert et al. computed 4.5 dpa per full-power year with one nuclear-data library and 14.5 with another for the tungsten armor of a magnetic-fusion DEMO first wall [41]. At 10 Hz this is 1.4–$4.6 \times 10^{-8}$ dpa per shot, 4–14 times less than the HAPL rate, and would bring 0.1 dpa after 2.5–8 days. The Demo neutron wall load is 2.05 MW $m^{-2}$. That of their DEMO model is not given in that paper, so the rates are compared per full-power year rather than scaled. Whether damage made in microsecond pulses at these temperatures lowers the conductivity as in the measurements is not known. With 100 % retention the first 5 μm would gain 0.31 appm of helium per shot and reach 280 appm after about 900 shots. The retention model does not reach that level. Its local helium content peaks at 251 appm after 1450 shots.

The simulated peak interface temperatures of the Demo runs at $h = 5$ and 20 kW $m^{-2}$ $K^{-1}$ (811 and 804 K after 100 shots, against 807 K at shot 100 of the baseline) are far from steady state, because 1 cm of steel needs of order 10 s to heat through. Only Table S9 should be used for the coolant sensitivity. Figure S4 gives the one-at-a-time sensitivity of the Demo lithium film (Section 5.2). The Prototype runs had reached their quasi-steady

state by shot 100 (618, 615 and 614 K).

**Table S9.** Mean W/steel interface temperature at periodic steady state from the heat balance of Section 2.7 (exact in this model), against the coolant heat-transfer coefficient.

| Scenario | $h$ = 5 kW m⁻² K⁻¹ | $h$ = 10 kW m⁻² K⁻¹ (baseline [5]) | $h$ = 20 kW m⁻² K⁻¹ |
|---|---|---|---|
| Prototype | 613 K | 611 K | 609 K |
| Demo | 1012 K | 930 K | 889 K |

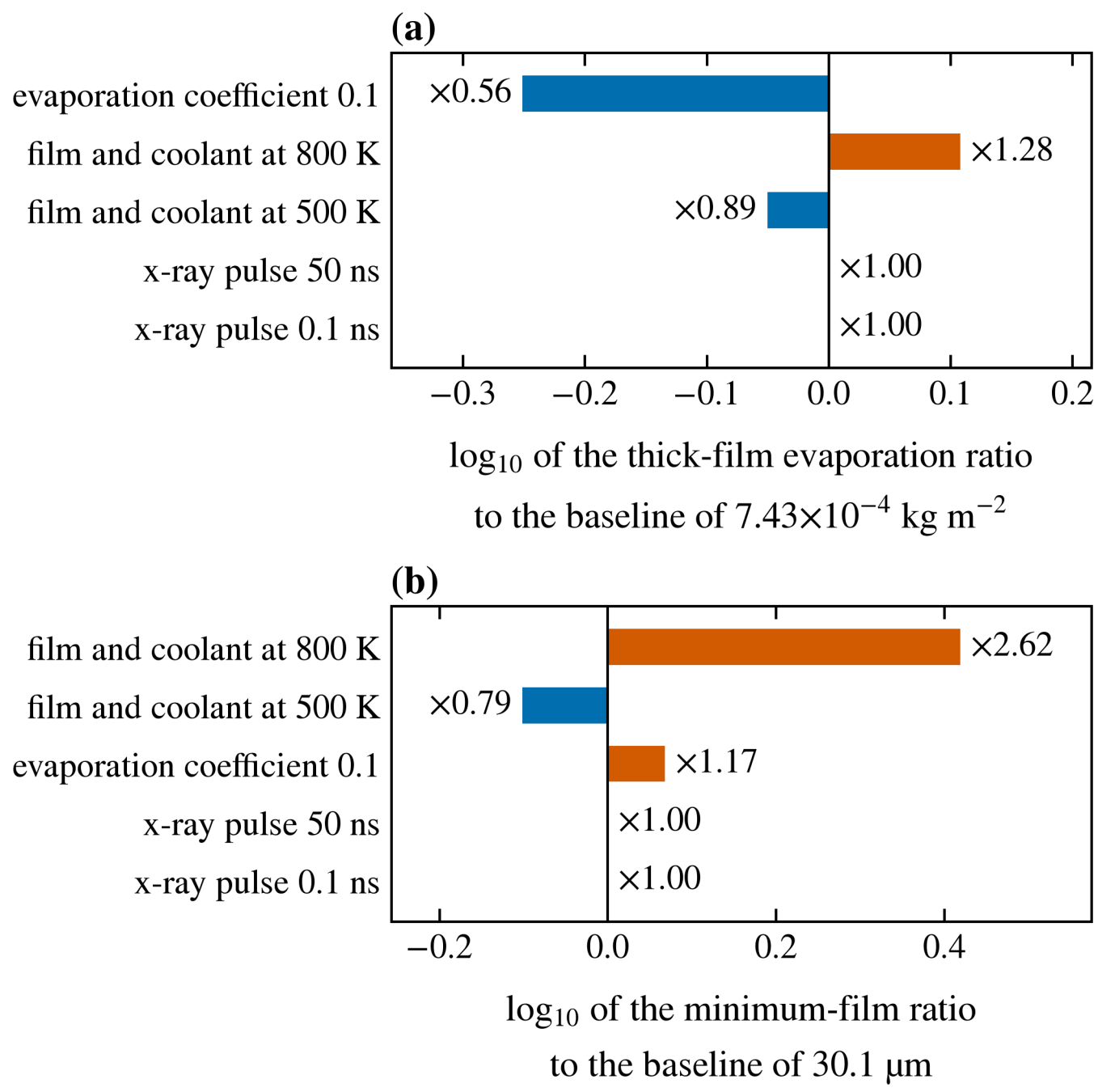

**Figure S4.** One-at-a-time sensitivity of the Demo lithium film (single shot): $\log_{10}$ of the ratio of the thick-film evaporation to the baseline $7.43 \times 10^{-4}$ kg m$^{-2}$ (a) and of the minimum film thickness to the baseline 30.1 μm (b). "Film and coolant at 500 K / 800 K" sets the film, the starting steel and the coolant to that temperature.

The proxy-0.8 grade peaks at 3558 K in a single shot from the pure-W quasi-steady state and at 3557 K in its own 550-shot run, 137–138 K below melting and well outside the 12–20 K bias of the model at this temperature. Nothing melted in the run. The W–K multi-shot run at 1 Hz is identical to the pure-W run in every per-shot value.

**Table S10.** Tungsten grades, Demo and Prototype: single shots from the quasi-steady inter-shot state of the pure-W runs (920.9 K and 609.5 K). Indicator: Arrhenius-equivalent time at the onset temperature $T_{rx}$ (Table B.2 of the main paper [32]; $Q = 336$ kJ mol$^{-1}$ [36]) accumulated over one cycle (0.1 s and 1 s), $t_{eq} = \int \exp[-(Q/R_g)(1/T - 1/T_{rx})]\,dt$; "shots to the one-hour exposure" is one hour divided by it, and the operating time is at 10 Hz (Demo) and 1 Hz (Prototype). Brackets: $Q = 322$–$350$ kJ mol$^{-1}$ [36], then $T_{rx}$ $-100$ K / $+100$ K. The depth is the deepest point of the recorder's logarithmic depth grid (1 nm–200 μm, 240 points, about 5 % apart) that reached $T_{rx}$.

| Scenario, grade (conductivity factor) | Peak surface $T$ | Evaporation per shot | $T_{rx}$ | Time above $T_{rx}$ per shot | Depth reached by $T_{rx}$ | Shots to the one-hour exposure | Operating time |
|---|---|---|---|---|---|---|---|
| Demo, pure W (1.0) | 3317 K | $6.9\times10^{-6}$ nm | 1523 K | 16.9 μs | 19.1 μm | $2.2\times10^{3}$ [1.2–$3.9\times10^{3}$; $3.4\times10^{2}$–$1.1\times10^{4}$] | 3.7 min |
| Demo, W–K (1.0) | 3317 K | $6.9\times10^{-6}$ nm | 1623 K | 12.5 μs | 16.4 μm | $1.1\times10^{4}$ [$6.8\times10^{3}$–$1.9\times10^{4}$; $2.2\times10^{3}$–$4.8\times10^{4}$] | 19 min |
| Demo, ODS W–$La_2O_3$, proxy 1.0 | 3317 K | $6.9\times10^{-6}$ nm | 1823 K | 7.8 μs | 12.1 μm | $1.7\times10^{5}$ [1.2–$2.6\times10^{5}$; $4.8\times10^{4}$–$5.5\times10^{5}$] | 4.8 h |
| Demo, ODS W–$La_2O_3$, proxy 0.9 | 3429 K | $1.8\times10^{-5}$ nm | 1823 K | 8.5 μs | 12.1 μm | $1.2\times10^{5}$ | 3.2 h |
| Demo, ODS W–$La_2O_3$, proxy 0.8 | 3558 K | $5.4\times10^{-5}$ nm | 1823 K | 9.5 μs | 12.1 μm | $7.4\times10^{4}$ [$4.9\times10^{4}$–$1.1\times10^{5}$; $2.1\times10^{4}$–$2.4\times10^{5}$] | 2.1 h |
| Prototype, pure W (1.0) | 1381 K | below $10^{-20}$ nm | 1523 K | 0 | 0 | $8.1\times10^{10}$ [7.1–$9.2\times10^{10}$; $1.3\times10^{10}$–$4.1\times10^{11}$] | $2.6\times10^{3}$ years |
| Prototype, W–K (1.0) | 1381 K | below $10^{-20}$ nm | 1623 K | 0 | 0 | $4.1\times10^{11}$ [3.4–$5.1\times10^{11}$; $8.1\times10^{10}$–$1.8\times10^{12}$] | $1.3\times10^{4}$ years |
| Prototype, ODS W–$La_2O_3$, proxy 1.0 | 1381 K | below $10^{-20}$ nm | 1823 K | 0 | 0 | $6.4\times10^{12}$ [4.6–$8.7\times10^{12}$; $1.8\times10^{12}$–$2.0\times10^{13}$] | $2.0\times10^{5}$ years |
| Prototype, ODS W–$La_2O_3$, proxy 0.8 | 1467 K | below $10^{-20}$ nm | 1823 K | 0 | 0 | $1.2\times10^{12}$ [$9.0\times10^{11}$–$1.5\times10^{12}$; $3.2\times10^{11}$–$3.6\times10^{12}$] | $3.6\times10^{4}$ years |

**Table S11.** Inter-shot W surface temperature against the sourced ductile-to-brittle transition temperatures (K) of Table B.2 of the main paper. Experimental, Prototype and Demo are the HiPER scenarios defined in [4, 5] (Table 1 of the main paper). "above"/"below": more than 50 K above or below; "near": within $\pm 50$ K. Pairs give pure W / W–K; the WL10 row gives its own value in the pure-W column.

| Transition (test, condition) | Pure W | W–K | Demo, 921 K | Prototype, 609 K | Experimental, 300 K |
|---|---|---|---|---|---|
| Tensile, as-rolled plate | 423 [32] | 323 [32] | above / above | above / above | below / near |
| Bending, bar / 7 mm plate | 423 [33] | 348 [33] | above / above | above / above | below / near |
| Charpy, plate | 823 [32] | 623 [32] | above / above | below / near | below / below |
| Charpy, rod | 1073 [35] | 1273 [35] | below / below | below / below | below / below |
| Fracture toughness | 673 [33] | – | above | below | below |
| Recrystallized, small punch | 673 [43] | – | above | below | below |
| Neutron-irradiated bending, about 1 dpa | 823 (irradiated at 1373 K); > 873 (irradiated at 873 K) [33] | 598 [33] | above (823) or near to below (> 873) / above | below / near | below / below |
| W–1%$La_2O_3$ (WL10), Charpy | 1223 [35] | – | below | below | below |

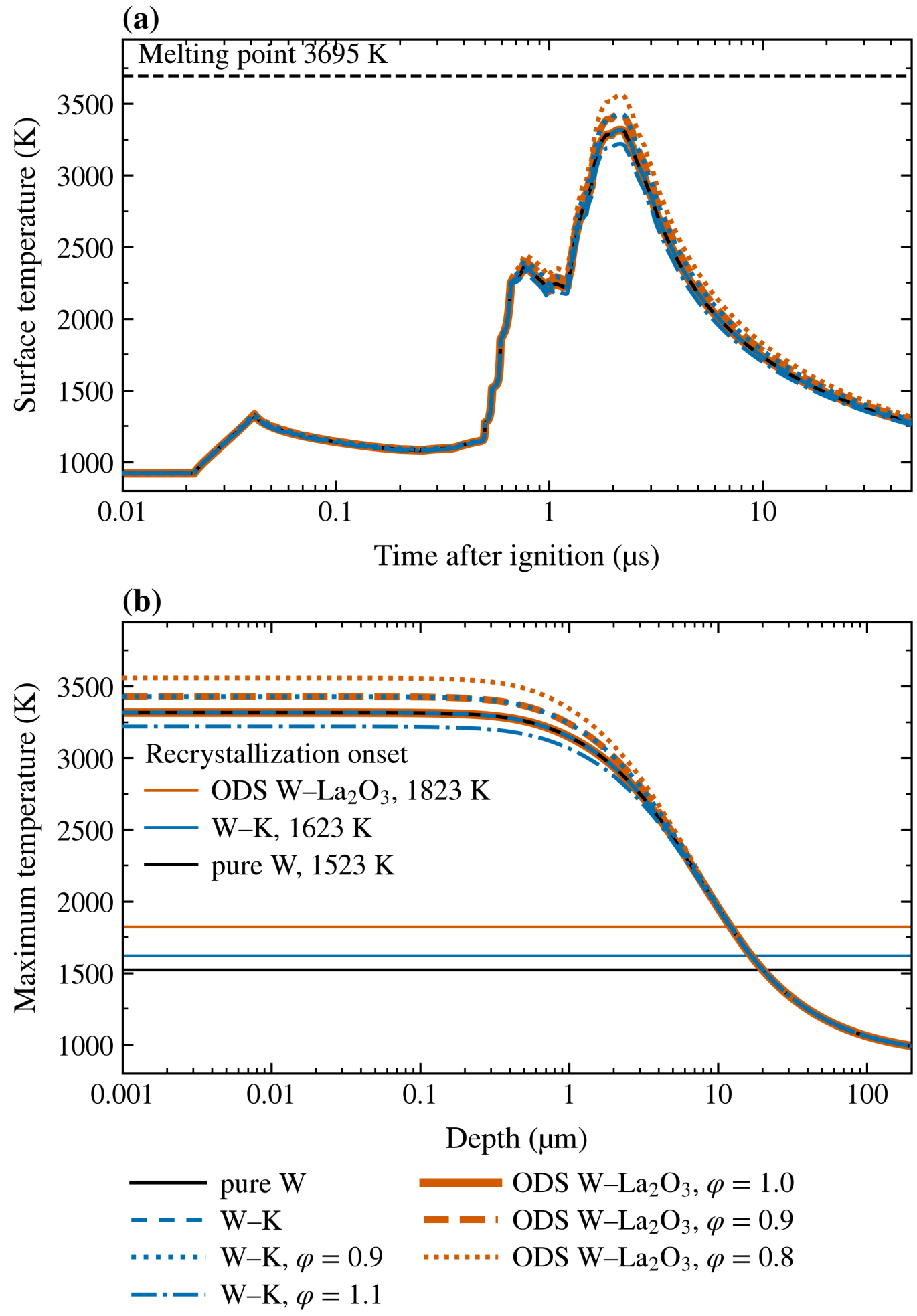

**Figure S5.** Demo, single shot from the quasi-steady 921 K state, by tungsten grade: surface temperature (a) and maximum temperature against depth with the onset temperatures of Table B.2 of the main paper [32] (b); dashed in (a): melting point, 3695 K [31]. W–K coincides with pure W; the dotted and dash-dotted W–K curves multiply its conductivity by 0.9 and 1.1, a ±10 % sensitivity test.

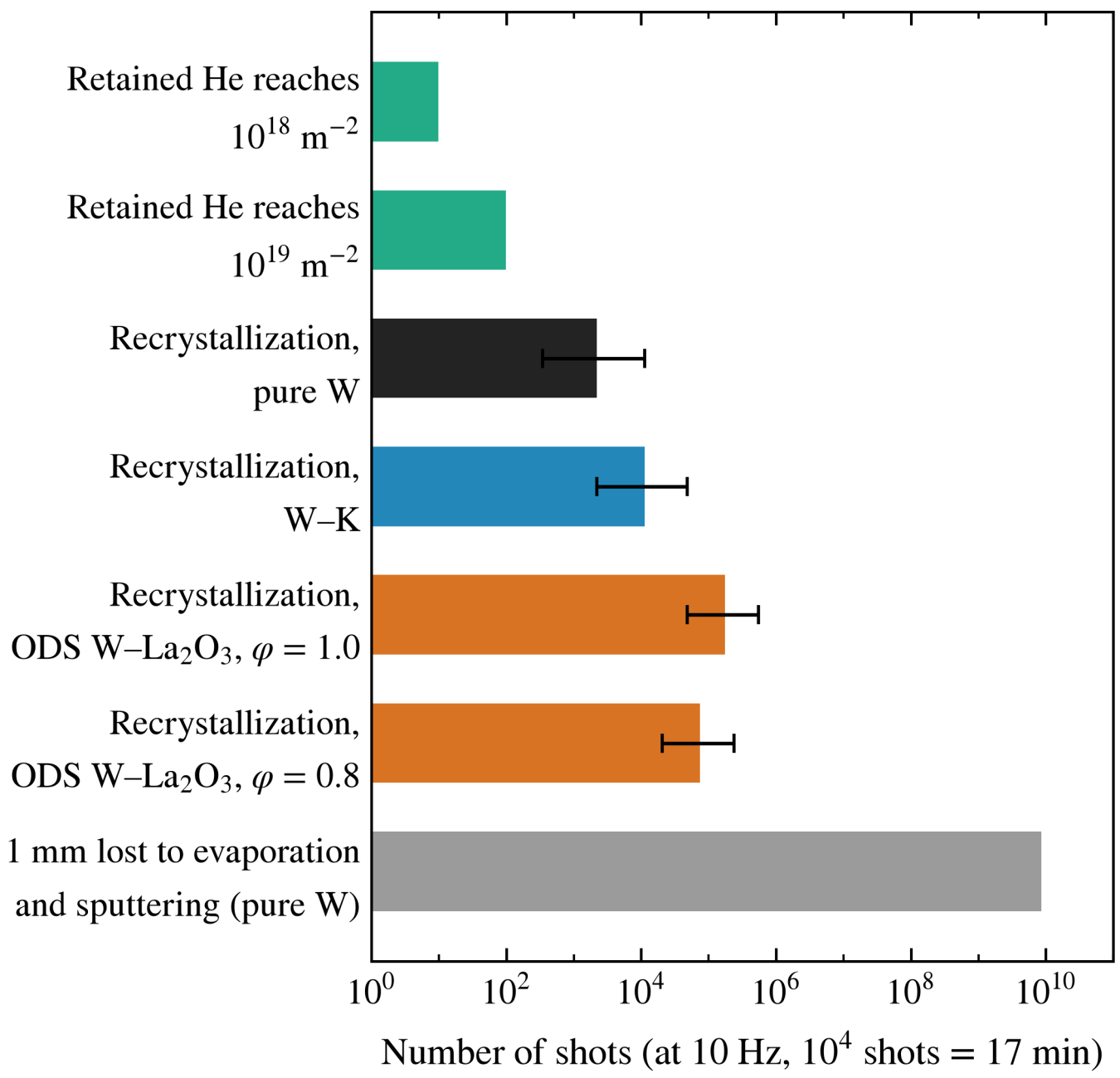

**Figure S6.** Demo shot-count indicators by grade: retained helium reaching $10^{18}$ and $10^{19}$ $m^{-2}$ [4] (100 % retention; the same for every grade), the one-hour recrystallization exposure at the surface (onsets of Table B.2 of the main paper [32]; error bars: activation energy 322–350 kJ $mol^{-1}$ [36] and onset ±100 K), and the consumption of 1 mm by evaporation and sputtering. None is a computed life.

**Reduced conductivity.** Table S12 gives the Demo runs behind Section 4.5 in which the conductivity of solid tungsten is multiplied by a uniform factor $\varphi$, while molten tungsten keeps its undamaged conductivity. Melting starts between $\varphi = 0.6$ and 0.5 from 600 K and between $\varphi = 0.75$ and 0.6 from the quasi-steady state. The first melting peaks, 3720 K at $\varphi = 0.5$ from 600 K and 3739 K at $\varphi = 0.6$ from 920.9 K, lie 25–44 K above the melting point, beyond the bias of the model (Section 2.4), so melting is established at these factors. The two 550-shot runs from 600 K melt on every shot and reach the quasi-steady criterion at shot 550. Both have the same pre-shot surface temperature, 922 K, and the same peak interface temperature, 930 K, against 932 K at shot 550 of the undamaged run. The reduced conductivity thus changes how the energy spreads through the first micrometers, not the heat that reaches the steel. Evaporation alone would consume 1 mm in 4.9 full-power years at $\varphi = 0.5$ and in 1.0 years with the 5 μm layer. Melt motion and splashing could remove more—neither is modeled. In Prototype a factor of 0.5 raises the single-shot peak to 1669 K from 600 K and to 1679 K from 609.5 K, without melting.

**Table S12.** Demo tungsten with the conductivity of the solid multiplied by $\varphi$: single shots from 600 K and from the quasi-steady 920.9 K state, and the last shot of two 550-shot runs from 600 K, the second with a 5 μm surface layer at a quarter of the undamaged conductivity.

| Case | Peak surface $T$ | Melt depth | Evaporation per shot |
|---|---|---|---|
| $\varphi$ = 1.0, from 600 / 920.9 K | 3027 / 3317 K | 0 / 0 | $3.4\times10^{-7}$ / $6.9\times10^{-6}$ nm |
| $\varphi$ = 0.85 | 3207 / 3491 K | 0 / 0 | $2.2\times10^{-6}$ / $3.1\times10^{-5}$ nm |
| $\varphi$ = 0.75 | 3353 / 3632 K | 0 / 0 | $8.7\times10^{-6}$ / $9.5\times10^{-5}$ nm |
| $\varphi$ = 0.6 | 3627 / 3739 K | 0 / 0.39 μm | $8.5\times10^{-5}$ / $3.4\times10^{-4}$ nm |
| $\varphi$ = 0.5 | 3720 / 3830 K | 0.31 / 0.77 μm | $2.8\times10^{-4}$ / $6.4\times10^{-4}$ nm |
| $\varphi$ = 0.4 | 3823 / 3931 K | 0.73 / 1.15 μm | $6.2\times10^{-4}$ / $1.2\times10^{-3}$ nm |
| 550 shots at $\varphi$ = 0.5, last shot | 3831 K | 0.77 μm | $6.4\times10^{-4}$ nm |
| 550 shots at $\varphi$ = 0.5 with a 5 μm layer at $\varphi$ = 0.25, last shot | 4085 K | 1.72 μm | $3.0\times10^{-3}$ nm |

**Single shot (Section 4.1).** The surface temperature follows the same history in all three scenarios (Fig. 1): the x-rays raise it within 20 ns of their arrival, the burn ions add from about 0.4 μs, and the debris ions set the peak at 1.65, 2.14 and 2.15 μs after ignition. The Prototype peak, 1371 K, falls inside the 1300–1800 K recrystallization range given by González-Arrabal et al. [4], but below the 1523 K one-hour onset of a pure-tungsten plate [32] (Table B.2). The Demo peak lies far above that range, and the Experimental peak below it.

**Quasi-steady state (Section 4.2).** The computed pre-shot temperatures match to within 13 K the "base temperature" of Garoz et al. [5], the temperature at the end of a pulse: 934 K for Demo and 610 K for Prototype.

**Interface with the steel (Section 4.2).** The steady-state heat balance gives a mean interface temperature of 930 K at $h = 10$ kW m$^{-2}$ K$^{-1}$, 889 K at 20 kW m$^{-2}$ K$^{-1}$ and 1012 K at 5 kW m$^{-2}$ K$^{-1}$. The excess is set by the cooling path, not by the armor. The 1 cm of steel has a thermal resistance of $3.0\times10^{-4}$ m$^2$ K W$^{-1}$, three times that of the coolant film at the baseline $h$. Its conductivity is the substitute value of Section 2.1. With 29 W m$^{-1}$ K$^{-1}$ instead of 33 W m$^{-1}$ K$^{-1}$ the heat balance gives 965 K instead of 930 K, and, with neutron heating, 1037 and 1091 K (for the 0.5× and 1× steel factors) instead of 997 and 1047 K (Prototype: about 1 K more). A 600-shot Demo run with 29 W m$^{-1}$ K$^{-1}$ meets the quasi-steady criterion at shot 600 with the surface at 953 K before each shot (921 K with 33 W m$^{-1}$ K$^{-1}$), a shot peak of 3346 K (3317 K), 349 K below melting, and an interface peak of 966 K (934 K). The 32 K rise of the interface matches the 35 K of the heat balance.

The values of Section 4.2 are those of the last shot of each run. Fitted like the interface of the 600-shot run, the surface peak would still rise by 4–5 K and the interface by 5 K in the neutron-heating runs, and by 3 K and 4 K in the 600-shot run. The heat balance extended to the volumetric heating gives mean interface temperatures of 997 K and 1047 K, in line with these runs. Bringing them down to 925 K would need $h = 31$ or 321 kW m$^{-2}$ K$^{-1}$, or a steel conductivity 33 % or 61 % above the substitute value. In Prototype, where the neutron heat per shot is a third of the Demo value and the repetition rate a tenth, the surface returns to 612–613 K before each shot, 3–4 K above the run without neutron heating, and peaks at 1383–1385 K, 2–4 K above it.

**Sensitivity of the Demo peak (Section 4.3).** Two inputs control the Demo single-shot peak (Fig. 3). Using the raw ion tables instead of the production spectra raises it by 526 K. The raw tables carry 47.2 instead of 41.6 MJ of ions (13 % more, the debris ions 19 % more) and 1.18 alphas ($^4$He) per fusion instead of one (Section S2), and no run separates the two effects. A 15 % lower tungsten conductivity raises the peak by 180 K (a 15 % higher one lowers it by 148 K). The depth model matters less ($-40$ K for the CSDA alternative), and the x-ray pulse length, the evaporation coefficient and the heat of vaporization change the peak by less than 1 K. Added to the quasi-steady 3317 K, the first two increments would bring the peak to about 3840 K and 3497 K. The conductivity increment (the second of the two) gives 3497 K, close to the 3491 K computed directly from the quasi-steady state (Table S12). The first exceeds the melting point. It is an extrapolation, because that case was not run from the quasi-steady state, but it shows that the Demo melting margin depends on how the ion spectra are constructed and on the conductivity.

**Erosion (Section 4.4).** With the conductivity of undamaged tungsten, erosion is negligible. Evaporation removes $3.4 \times 10^{-7}$ nm in the first Demo shot and $6.9 \times 10^{-6}$ nm per shot in the quasi-steady state, and physical sputtering $1.1 \times 10^{-4}$ nm per shot, or up to $1.7 \times 10^{-4}$ nm with proxies for the carbon and gold ions, whose yields on tungsten are not tabulated. At 10 Hz sputtering alone would consume 1 mm in 29 full-power years; adding the evaporation gives 27, the proxies 18, and both together 18 full-power years. With the evaporation included, the armor keeps more than the 100 μm that Garoz et al. found sufficient to protect the steel [5] for 16–24.5 full-power years (15.9 with the proxies, 24.5 without). With the degraded conductivities of Section 4.5, the halved conductivity and the halved conductivity with a 5 μm layer at $\varphi = 0.25$, evaporation and sputtering shorten this to 3.8 and 0.9 full-power years. The sputtering values rest on fit formulas extrapolated beyond 20 keV, include no redeposition and no helium-induced change of the surface, and are order-of-magnitude estimates. The conclusion that erosion does not limit an undamaged armor needs only their size.

**Helium threshold and retention (Section 4.4).** If every implanted helium atom stays, the retained helium in the Demo armor reaches $10^{18}$ m$^{-2}$ after 10 shots and $10^{19}$ m$^{-2}$ after 98 shots, and at 1 Hz in Prototype after 30 and 303 shots. These counts are lower bounds, because any release of helium can only delay the threshold. They also match how the threshold was obtained, as a fluence delivered in pulsed experiments with heat loads, in which any helium that was released is already accounted for. The retention model of Section S7 lets the heating spikes release part of the helium and gives 11 and 125 Demo shots (31 and 308 in Prototype). However, it releases nearly all of the helium in a 2000 °C flash anneal that left the measured retention of implanted tungsten unchanged (Section S7.10). Its counts are therefore a secondary, model-dependent estimate that probably overstates the release.

The threshold itself is uncertain. González-Arrabal et al. [4] take the $10^{18}$ m$^{-2}$ end from a single plasma-focus pulse whose damage decreased with the accompanying heat load, and read $10^{19}$ m$^{-2}$ from the pulsed-ion experiments of Renk et al. [37]. That paper places the onset of helium pores and bubbles at absorbed levels in the mid-$10^{15}$ He cm$^{-2}$ range. Its estimate, $4 \times 10^{15}$ He cm$^{-2}$ ($4 \times 10^{19}$ m$^{-2}$), is for single-crystal tungsten heated to 500 °C and given 400 helium pulses below 1 J cm$^{-2}$. Similar protrusions or bubbles lined the sides of stress cracks in oriented-grain tungsten after 800 pulses and in W–25Re and W–TiC, and open pores formed on W–TiC, W–25Re and single-crystal tungsten after 1200 pulses, none of them under nitrogen pulses. The other helium samples started at room temperature. A helium pulse of 1 J cm$^{-2}$ implanted about $1 \times 10^{13}$ He cm$^{-2}$ ($10^{17}$ m$^{-2}$), with a calculated peak surface temperature of about 1100 °C (a 1600-pulse series on oriented-grain tungsten averaged 1.2 J cm$^{-2}$ and a 1300 °C peak). That fluence is close to the $1.02 \times 10^{17}$ He m$^{-2}$ implanted by a Demo shot here (Section S7.4). The authors could not confirm that helium caused the pores and bubbles. If it did, the level lies far below the $10^{17}$–$10^{18}$ He cm$^{-2}$ reported for bubble nucleation under steady implantation at a fixed temperature; if it did not, the thermomechanical load of the helium pulses may damage the surface before that level is reached. They also note that helium implanted in cycles at elevated temperature may escape the surface more readily, so that bubbles may form differently [37]. The helium count could therefore be several times larger than the 10–98 shots above.

It is also unclear whether the threshold transfers to this temperature regime. Helium effects in tungsten are strongly temperature dependent, with the fibre-like nanostructure (fuzz) forming readily between about 1000 and 2000 K at helium ion energies above 20 eV [44]. The Demo surface sits at 921 K between shots, while the first 5 μm, which hold 95 % of the implanted helium, spend about 1 ms of each 100 ms cycle above 1000 K.

**Design-rule data (Section 4.4).** The rule is stated for surface loads only, whereas the ion energy here is deposited over micrometers in a few microseconds. The data closest to the Demo factor are one tokamak melting point, at about 40 MW m$^{-2}$ s$^{1/2}$ with large error bars, and one calculated point at 55 MW m$^{-2}$ s$^{1/2}$. Behind such correlations, the electron-beam data are no shorter than 0.48 ms for scanned loading and about 5 μs for a single spot (G. Pintsuk, personal communication, 2026), against a 10–90 % ion arrival window of 2.05 μs in Demo (incident ions, Table 2; 2.04 μs for the absorbed energy). The count is therefore an order-of-magnitude indicator of surface damage, not a fatigue life. That it is similar to the helium count is a coincidence of unrelated inputs, not two confirmations of a ten-shot life.

**Reduced conductivity and measured reductions (Section 4.5).** With the conductivity of solid tungsten multiplied by a factor $\varphi = 0.5$ and molten tungsten left undamaged, the Demo surface melts on every shot of a 550-shot run, 0.31 μm deep in the first shot and 0.77 μm in the last (Table S12). It then evaporates $6.4 \times 10^{-4}$ nm per shot, enough to consume 1 mm in about 5 full-power years if nothing else were lost. At $\varphi = 0.75$ a shot from the quasi-steady state peaks at 3632 K, 63 K below the melting point, more than the 0.3–0.6 % by which the production tungsten peaks are too high (Section 2.4), so that factor does not melt the surface.

Reductions of this size have been measured at room temperature. In the transient-grating measurements of Hofmann et al. [45], helium implanted at 573 K into a 2.6 μm layer at 280 and 3100 atomic parts per million (appm) lowers the diffusivity from about 0.68 cm$^2$ s$^{-1}$ for pure tungsten to about 0.53 and 0.36 cm$^2$ s$^{-1}$ (the last value printed in their text, the others read from their Fig. 3b), about 0.78 and 0.53 of the pristine value. Self-ion damage lowers it by about 55 %, saturating from 0.1 displacements per atom (dpa) [42]. After annealing at 1073 K, a residual reduction remains for doses of 0.03 dpa and above [46]. Whether such reductions occur in the HiPER wall is not established. González-Arrabal et al. [4], citing Linke et al., report no significant degradation of the conductivity at the HiPER operating temperatures after neutron irradiation to 0.6 dpa.

**Near-surface layer (Section 4.5).** The diffusion length in tungsten is 7–12 μm over the 2 μs in which the debris ions set the peak, comparable with the 1–10 μm in which most of the energy is deposited, so a degraded layer of that thickness governs the peak. Fig. 4 shows the peak for a surface layer of thickness $d = 1$, 3 or 10 μm whose solid conductivity is a hypothetical fraction $\varphi$ of the bulk value. From 600 K, a 1 μm layer at $\varphi = 0.25$ or a 3 μm layer at $\varphi = 0.5$ brings the Demo peak to within about 130 K of the melting point (3562 and 3564 K). A 3 μm layer at $\varphi = 0.25$ or a 10 μm layer at $\varphi \leq 0.5$ melts the surface 0.3–1.5 μm deep. From the quasi-steady state, a 10 μm layer at $\varphi = 0.75$ stays 73 K below the melting point (3622 K), and at $\varphi = 0.25$ the surface reaches 4102 K and melts 1.8 μm per shot. The 10 μm layer reproduces the uniform reductions to within about 10 K from the quasi-steady state and within 12 K from 600 K, so the top ten micrometers set the peak. The layer changes neither the interface nor the temperature before the next shot (610 K before the second Demo shot for every $d$ and $\varphi$), because the inter-shot state is set by the steel and the coolant. The Prototype peak crosses the 1523 K recrystallization onset of pure tungsten for $\varphi \leq 0.5$ at $d \geq 3$ μm and reaches it for $\varphi = 0.25$ at $d = 1$ μm (1522 K), but stays far from melting. A measured effusivity of the as-installed surface, for example by pulsed photothermal radiometry [47, 48], would decide between these cases.

**Tungsten grades (Section 4.6).** Potassium-doped tungsten (W–K) and $La_2O_3$-strengthened tungsten (ODS W–$La_2O_3$) enter this model only through their conductivity. For rolled W–K plate it equals that of pure tungsten within 1 % [34], although a swaged W–K rod with residual porosity conducted less [32]. Every temperature, melting margin and erosion rate above therefore applies unchanged to W–K plate (Figure S5). For ODS W–$La_2O_3$ no measured $k(T)$ was found, and a band of 1.0, 0.9 and 0.8 times the pure-tungsten curve was run as a sensitivity study. At 0.8 the quasi-steady Demo peak is 3557 K in a 550-shot run, 138 K below melting and well outside the bias of the model, while the inter-shot surface and interface temperatures stay within 1 K of the pure-tungsten run. A thermal-exposure indicator of recrystallization, the Arrhenius-equivalent time at the one-hour onset temperature with an activation energy of 336 kJ mol$^{-1}$ [36], reaches one hour after about $2.2 \times 10^3$ Demo shots (3.7 min) for pure tungsten with a 1523 K onset, $1.1 \times 10^4$ shots (19 min) for W–K with 1623 K, and $7.4 \times 10^4$ shots (2.1 h) for ODS W–$La_2O_3$ at 0.8 times the pure-tungsten conductivity if the 1823 K onset measured on a W–3%Re–1%$La_2O_3$ plate applies [32] (Figure S6). These times exclude neutron heating, whose higher peaks would shorten them. In Prototype, whose surface peaks at 1381 K, below all of these onsets, the same indicator needs about $8.1 \times 10^{10}$ shots for pure tungsten, more than a thousand years at 1 Hz, and $1.3 \times 10^{10}$ shots, 400 years, with the onset 100 K lower (Table S10). A $\pm 100$ K change of the onset changes the Demo counts by factors of 3.2–6.5, comparable with the gain of W–K over pure tungsten. The onset itself differs by 150–450 K between plates and rods of the same material (pure tungsten: 1523 K for a plate, 1673–1973 K for rods; W–K: 1623 K for a plate, 1873–1973 K for rods [32]).

**Tungsten grades in the Discussion (Section 7.3).** The interface temperature is set by the steel and the coolant, and the helium count by the ion spectrum. Neither depends on the grade in this model, and no grade-specific pulsed-helium threshold was found. The only quantity the grades move is the recrystallization indicator. It moves from minutes for pure tungsten to hours for the oxide grade if the onset temperature of a Re-containing plate applies, still far short of the $10^6$-shot thermomechanical life quoted for Demo [4]. The shift is comparable with the uncertainty of the onset temperature itself. If the conductivity of the oxide grade is 20 % below that of pure tungsten, its melting margin shrinks from 378 K to 138 K, which would make it worse than pure tungsten, not better. What the grades may offer is a wider window between the ductile-to-brittle transition (Table S11) and recrystallization, and better resistance to surface damage. The pulsed-ion experiments available for IFE conditions do not settle this. Oriented-grain tungsten performed better than polycrystalline tungsten but appeared to suffer added surface damage under helium [37]. Under nitrogen-ion pulses in the same program, W–25Re started with a higher threshold than powder-metallurgy tungsten (about 1.25 J cm$^{-2}$) but its cracking threshold fell from about 1.6 J cm$^{-2}$ after 400 pulses to 0.6 J cm$^{-2}$ after 1200 [37]. Assessing either needs a stress calculation, so a grade study in a purely thermal model can only bracket the conductivity, which is what Section 4.6 does.

**Coolant 3.5 mm behind the surface (Section 7.5).** With the coolant 3.5 mm behind the plasma-facing surface instead of 11 mm, 1 mm of tungsten on 2.5 mm of steel, the Demo wall reaches its quasi-steady state after 120 shots. The surface is then at 738 K before each shot and peaks at 3153 K, 542 K below melting, and the interface peaks at 752 K, well below the 925 K limit. The heat balance gives a mean interface temperature of 744 K without and 758–765 K with neutron heating. In Prototype the surface returns to 603 K and peaks at 1374 K. A lithium film on 3.5 mm of steel needs the same 30.1 μm in a single Demo shot, because the steel does not feel the coolant within one shot, but 37 μm instead of 44 μm in repeated operation, both with direct runs at the interpolated thicknesses (the method of Section S3). In Prototype it needs 11.9 μm in a single shot and 12.0 μm in repeated operation, against 12.1 μm in the production geometry.

**Conditions at Demo (Section 7.5).** For the tungsten armor three conditions must hold. First, the interface must stay below the steel limit. With the 82 kJ $m^{-2}$ absorbed per Demo shot, the heat balance of Section 2.7 brings the mean interface temperature down to 925 K for $h = 10.7$ kW $m^{-2}$ $K^{-1}$ instead of 10, or for a steel conductivity 2.2 % above the substitute value, so a modest improvement of the cooling path suffices without neutron heating. With neutron heating, the balance needs $h = 31$ or 321 kW $m^{-2}$ $K^{-1}$, or a steel conductivity 33 % or 61 % higher, so the coolant has to come closer to the surface, as in the check above. Second, the near-surface conductivity must stay close to that of undamaged bulk tungsten, since a conductivity halved over the top 3–10 μm melts the surface in a quasi-steady Demo shot. This needs a measured effusivity of the as-installed surface and of the irradiated surface. Third, the surface modification that pulsed-ion experiments at other temperatures associate with $10^{18}$–$10^{19}$ $m^{-2}$ of helium, which this model reaches after 10–98 shots in the first 5 μm, must not compromise the armor's function. This needs damage-evolution data for microsecond helium pulses at about 900 K, which the published thresholds do not settle.

The lithium film must be at least 44 μm thick at 10 Hz to keep the steel below 925 K, 48–53 μm with neutron heating, 56 μm for the 873 K compatibility limit and about 184 μm to keep the helium in the steel below $10^{18}$ $m^{-2}$ for $10^6$ shots. It must be renewed by injection a few centimeters apart (Section 5.4), and the chamber must clear 0.22–0.45 kg of lithium vapor between shots (full renewal). Whether a film of that thickness can be kept uniform and stable under the recoil pressure of Section 5.2 is outside this model, and the renewal and the clearing are system requirements that this study can quantify but not assess.

## S7. Helium-retention model

The secondary helium counts of Section 4.4 come from a one-dimensional reaction–diffusion model of helium, vacancies and self-interstitial atoms (SIAs) in the 1 mm tungsten armor. It was written for this study and is released with the code as the subpackage `ife_wall.he` (commands in Section S8). The model follows the helium implanted by the target ions and the Frenkel pairs they create. It uses the ion source term of Section 2.3 and the temperature history of the thermal model of Section 2.2, and does not feed back on the thermal calculation. The baseline holds helium only in single vacancies. An optional second level, used only as a sensitivity case, adds interstitial helium clusters and trap mutation (Section S7.7).

### *S7.1 Species and reactions*

In every depth cell the model follows mobile interstitial helium, He; free vacancies, V (also written $V_0$); SIAs, I; and vacancies holding $k$ helium atoms, $V_k$ (the $He_kV$ complexes), for $k = 1 \ldots K$ with $K = 7$. $K$ is the largest number of helium atoms a vacancy can hold; a helium atom that encounters $V_K$ is therefore not captured. All are concentrations in atomic parts per million (appm) of the tungsten atom density $N_W$. He, V and I diffuse; the $V_k$ do not move. A helium atom is trapped by a free or partly filled vacancy, $He + V_k \rightarrow V_{k+1}$ for $k = 0 \ldots K - 1$, and leaves it by thermal detrapping, $V_k \rightarrow V_{k-1} + He$. A helium atom that meets $V_K$ is not captured, so $K$ is also the largest number of helium atoms a vacancy can hold. It is the largest $k$ for which a binding energy was found (Table S15). It also lies at or above the size, five to seven atoms, at which an interstitial helium cluster pushes out a tungsten atom [49, 50]. A free vacancy and an SIA recombine, $V + I \rightarrow \varnothing$. An SIA that meets a helium-filled vacancy annihilates it and returns its $k$ helium atoms to the mobile population, $I + V_k \rightarrow k\,He$, an assumption of this work. Vacancies and SIAs are absorbed at dislocations. Helium has no sink in the bulk, because no binding energy of helium to dislocations or grain boundaries was found. The sources are the implantation of helium and the production of Frenkel pairs by the target ions and, uniformly and weakly, by the neutrons. All bimolecular reactions are diffusion limited. They use the rate constant $4\pi(r_A + r_B)(D_A + D_B)$ of Faney [51], where $r$ is a capture radius and $D$ a diffusion coefficient. Detrapping is of first order (Table S13).

### *S7.2 Equations, boundary and initial conditions*

With concentrations in appm, depth $x$ below the surface and time $t$, the model solves

$$\partial He/\partial t = \partial/\partial x\,(D_{He}\,\partial He/\partial x) + g_{He} - \Sigma_{k=0}^{K-1}\, a_k\, He\, V_k + \Sigma_{k=1}^{K}\, b_k\, V_k + e\, I\, \Sigma_{k=1}^{K}\, k\, V_k - D_{He}\, S_{He}\, He, \quad (S1)$$

$$\partial V/\partial t = \partial/\partial x\,(D_V\,\partial V/\partial x) + g_V - a_0\, He\, V + b_1\, V_1 - c\, V\, I - D_V\, S_V\, V, \quad (S2)$$

**Table S13.** Reactions of the baseline model. Concentrations in appm; rates per unit volume in appm $s^{-1}$; the coefficients are defined by Eqs. (S5)–(S8).

| Reaction | Rate | Coefficient |
|---|---|---|
| He + V → $V_1$ | $a_0$ He V | $a_0 = 4\pi\, r_c\, (D_{He} + D_V)\, N_C$ |
| He + $V_k$ → $V_{k+1}$, $1 \le k \le K-1$ | $a_k$ He $V_k$ | $a_k = 4\pi\, r_c\, D_{He}\, N_C$ |
| $V_k$ → $V_{k-1}$ + He, $1 \le k \le K$ | $b_k\, V_k$ | $b_k = v \exp[-(E_{b,k} + E_{m,He})/k_B T]$ |
| V + I → ∅ | $c$ V I | $c = 4\pi\, r_{iv}\, (D_I + D_V)\, N_C$ |
| I + $V_k$ → $k$ He, $1 \le k \le K$ | $e\, V_k$ I | $e = 4\pi\, r_{iv}\, D_I\, N_C$ |
| V → dislocation; I → dislocation | $D_V\, S_V$ V; $D_I\, S_I$ I | $S_V = Z_V\, \rho_d$, $S_I = Z_I\, \rho_d$; $S_{He} = 0$ |
| implantation; Frenkel pairs | $g_{He}$; $g_V = g_I$ | Section S7.4 |

$$\partial I/\partial t = \partial/\partial x\, (D_I\, \partial I/\partial x) + g_I - c\, V\, I - e\, I\, \Sigma_{k=1}{}^{K}\, V_k - D_I\, S_I\, I, \text{ (S3)}$$

$$\partial V_k/\partial t = a_{k-1}\, He\, V_{k-1} - a_k\, He\, V_k - b_k\, V_k + b_{k+1}\, V_{k+1} - e\, I\, V_k,\ k = 1 \ldots K,\ a_K \equiv 0,\ b_{K+1} \equiv 0, \text{ (S4)}$$

where $V_0 \equiv V$ and $g_I = g_V$, because every displacement makes one vacancy and one SIA. The coefficients are

$$D_s = D_{0,s} \exp(-E_{m,s}/k_B T),\ s = \text{He, V, I}, \text{ (S5)}$$

$$a_0 = 4\pi\, r_c\, (D_{He} + D_V)\, N_C,\ a_k = 4\pi\, r_c\, D_{He}\, N_C\ (1 \le k \le K - 1), \text{ (S6)}$$

$$b_k = \nu \exp[-(E_{b,k} + E_{m,He})/k_B T], \text{ (S7)}$$

$$c = 4\pi\, r_{iv}\, (D_I + D_V)\, N_C,\ e = 4\pi\, r_{iv}\, D_I\, N_C, \text{ (S8)}$$

with $N_C = 10^{-6}\, N_W$, the number density of one appm, so that $a_k$, $c$ and $e$ are in $appm^{-1}\ s^{-1}$ and $b_k$ in $s^{-1}$ (Table S14). $T = T(x, t)$ is the local temperature of the thermal model (Section S7.5) and $k_B$ the Boltzmann constant. Here $k$ is an index, not the thermal conductivity. $E_{b,k}$ is the binding energy of the $k$-th helium atom to $V_{k-1}$. The detrapping activation energy $E_{b,k} + E_{m,He}$ follows the convention of Faney [51]. Captures by the immobile $V_k$ ($k \ge 1$) contain only the diffusion coefficient of the mobile partner. The sink strengths are $S_V = Z_V\, \rho_d$ and $S_I = Z_I\, \rho_d$, with the dislocation density $\rho_d$ and the bias factors $Z_V$ and $Z_I$, and $S_{He} = 0$.

The three mobile species are absorbed at the plasma-facing surface, $x = 0$. The flux leaving the first cell, of width $\Delta x_0$, is $G_{surf}\, C_0$, with $G_{surf} = 2\, f_{surf}\, D_s/\Delta x_0$, $D_s$ evaluated in that cell and $C_0$ the concentration there. For $f_{surf} = 1$ this sets the concentration to zero at the surface; $f_{surf} = 0.1$ is a sensitivity case. The back face of the armor, $x = L = 1$ mm, has no flux, and the immobile $V_k$ need no boundary condition. The armor starts free of helium and defects, and no thermal vacancies are generated. The surface does not move. The recession by evaporation and sputtering, about $10^{-4}$ nm per shot (Section 4.4), is ignored.

**Table S14.** Symbols of the helium model and their names in the released package ife_wall.he. Concentrations in appm. The vacancy survival fraction is written $\xi_V$ because ξ is the areal mass in Table A.1.

| Symbol | Name in the code | Meaning | Unit |
|---|---|---|---|
| He, V, I | He, V, I (species 0, 1, 2) | mobile helium, free vacancies, SIAs | appm |
| $V_k$, $k = 1 \ldots K$ | HV_k (species 2 + k) | vacancy holding $k$ helium atoms ($He_k V$); $V_0 \equiv V$ | appm |
| $K$ | K | largest number of helium atoms per vacancy | – |
| $D_{0,s}$, $E_{m,s}$ | D0_He, Em_He, D0_V, Em_V, D0_I, Em_I | diffusion prefactor and migration energy of species s | $m^2\ s^{-1}$, eV |
| $E_{b,k}$ | Eb[k−1] | binding energy of the $k$-th helium atom to $V_{k-1}$ | eV |
| $v$ | nu | attempt frequency of detrapping | $s^{-1}$ |
| $r_c$, $r_{iv}$ | r_c, r_iv | He–V and V–I reaction radii (sums of capture radii) | m |
| $a_k$, $b_k$, $c$, $e$ | a, b, cc, ee | trapping, detrapping, recombination and annihilation coefficients | $appm^{-1}\ s^{-1}$; $s^{-1}$ |
| $S_{He}$, $S_V$, $S_I$ | sink_He, sink_V, sink_I | dislocation sink strengths | $m^{-2}$ |
| $\rho_d$, $Z_V$, $Z_I$ | RHO_D, Z_V, Z_I | dislocation density; bias factors | $m^{-2}$, – |
| $f_{surf}$ | surface_factor | surface conductance relative to a perfect sink | – |
| $g_{He}$, $g_V = g_I$ | GHe + bgHe, GV + bgV | helium and Frenkel-pair sources | appm $s^{-1}$ |
| $N_W$, $N_C$ | N_W, NC | tungsten atom density; number density of 1 appm | $m^{-3}$ |
| $\xi_V$ | xi | vacancy survival fraction applied to the NRT displacements | – |
| $G_{i+½}$, $G_{surf}$ | Gc, Gs | diffusive conductance between cells; to the surface | μm $s^{-1}$ |
| $s$ | – | species index (He, V, I) in $D_s$ and $E_{m,s}$; not the stopping factor of Table A.1 | – |
| $L$ | – | armor thickness, 1 mm | m |
| $j$ | – | index of the energies along an ion path (Section S7.4); not the mass flux of Table A.1 | – |

### *S7.3 Parameters*

Table S15 gives every physical parameter of the baseline with its source; values without a literature source are marked as assumptions. Where sources disagree, a low and a high value were run as sensitivity cases (Section S7.8).

**Table S15.** Baseline parameters. Status: P, primary source read; S, value printed in a secondary source that quotes the primary; D, derived here from numbers in the cited source; A, assumption or choice of this work.

| Quantity | Symbol | Baseline | Unit | Source | Status |
|---|---|---|---|---|---|
| Tungsten lattice parameter | $a_W$ | 0.31653 | nm | chosen so that $N_W = 2/a_W^3$; Mohamed et al. quote 0.316 nm [52] | D |
| Tungsten atom density | $N_W$ | $6.306\times10^{28}$ | $m^{-3}$ | $2/a_W^3$; $6.3\times10^{28}$ $m^{-3}$ in Delaporte-Mathurin et al. [53] | D |
| Helium migration | $D_{0,He}$, $E_{m,He}$ | $2.95\times10^{-8}$; 0.13 | $m^2$ $s^{-1}$; eV | Faney, Table 3.1, molecular-dynamics fit over 300–2000 K [51] | P |
| Vacancy migration | $D_{0,V}$, $E_{m,V}$ | $1.11\times10^{-6}$; 1.71 | $m^2$ $s^{-1}$; eV | He et al., Table 2 (preprint) [54] | P |
| SIA migration | $D_{0,I}$, $E_{m,I}$ | $8.77\times10^{-8}$; 0.009 | $m^2$ $s^{-1}$; eV | Faney, Table 3.2 [51]; the same values in He et al., Table 2 [54] | P |
| Binding energies, $k$ = 1–4 | $E_{b,1}$ ⋯ $E_{b,4}$ | 4.05, 3.11, 2.88, 2.41 | eV | thermal desorption of Kornelsen (1972), as tabulated by Faney, Tables 3.5 and 5.1 [51] | S |
| Binding energies, $k$ = 5–7 | $E_{b,5}$ ⋯ $E_{b,7}$ | 2.21, 2.27, 2.45 | eV | $E_{b,k} = E_f(He) + E_f(He_{k-1}V) - E_f(He_kV)$ from the formation energies in Table S3 of the preprint of Tirumala et al. [55] | D |
| Largest helium number per vacancy | $K$ | 7 | – | largest $k$ with a published binding energy | A |
| Helium capture radius | $r_{He1}$ | 0.30 | nm | Faney, Section 3.5 [51] | P |
| Vacancy capture radius | $r_{V1}$ | $(\sqrt{3}/4)\ a_W$ = 0.1371 | nm | Faney, Section 3.5 [51] | P |
| He–V reaction radius | $r_c$ | $r_{He1} + r_{V1}$ = 0.4371 | nm | sum of the capture radii, as in Faney, Eq. 2.6 [51] | D |
| V–I reaction radius | $r_{iv}$ | $r_{V1}$ + 1.15 $r_{V1}$ = 0.2947 | nm | derived from Faney, Eqs. 3.13–3.15 [51]; no primary value found | D |
| Attempt frequency | $\nu$ | $10^{13}$ | $s^{-1}$ | Dunand et al., Section 3.2 [50]; implied by Faney's rates [51] ($4\pi\ r_c\ D_{0,He}\ N_W = 1.02\times10^{13}$ $s^{-1}$) | P |
| Dislocation density | $\rho_d$ | $5\times10^{12}$ | $m^{-2}$ | Mohamed et al., Table 2, who take it from Shah et al. [52] | S |
| Dislocation bias | $Z_V$, $Z_I$ | 1.0; 1.2 | – | Mohamed et al., Table 2 [52] | S |
| Helium sink strength | $S_{He}$ | 0 | $m^{-2}$ | no binding energy of helium to dislocations or grain boundaries found | A |
| Surface conductance factor | $f_{surf}$ | 1 (perfect sink) | – | model assumption | A |
| SIA annihilation of $V_k$ | – | on | – | model assumption | A |
| Displacement threshold, NRT | $E_d$ | 90 | eV | ASTM E521 value, as quoted by He et al., Wielunska et al. and Mitsi et al. [54, 56, 57] | S |
| Vacancy survival fraction | $\xi_V$ | 1 | – | every NRT Frenkel pair survives, the high-vacancy choice | A |
| Thomas–Fermi scattering function | $\lambda$, $m_{TF}$, $q_{TF}$ | 1.309, 1/3, 2/3 | – | Winterbon–Sigmund–Sanders form; not checked against a source | A |
| Neutron displacement damage | – | $2\times10^{-7}$ | dpa per shot | Demo; derived here from the $10^{-6}$ dpa $s^{-1}$ of Sharafat et al., Section 3.2 (material not stated), at the 5 Hz HAPL rate of their Section 4.5 [6], and scaled by $Y$/154 MJ × (6.5 m/$R$)² in Prototype and Experimental; for tungsten first-wall armor of a magnetic-fusion DEMO, Gilbert et al. give 4.5–14.5 dpa per full-power year, 1.4–$4.6\times10^{-8}$ dpa per shot at 10 Hz [41] | D |
| Neutron helium | – | 0.3 | appm $dpa^{-1}$ | derived from Gilbert et al. and Stork et al. for the first wall of a magnetic-fusion DEMO [41, 58] | D |

### S7.4 Helium and damage sources

The sources are computed once per scenario from the ion spectra of Section 2.3, separately from the energy deposition of Section 2.4 but with the same range data. Each of the 125 energy bins is split into six sub-bins of equal logarithmic width (750 sub-bins, 270 of them helium). The number of ions in a sub-bin is the integral of the within-bin spectrum over it, and the representative energy assigned to the sub-bin is the number-weighted mean energy. The $^3$He and $^4$He ions of a sub-bin of energy $E$ stop with a gamma distribution in depth. Its mean is the depth $R_{p,eff}(E)/\rho_{SRIM}$, the effective projected range of Section S2 (an areal mass) divided by the SRIM density of tungsten, $\rho_{SRIM} = 19\ 350$ kg m$^{-3}$. Its standard deviation $\sigma_R(E)$ is the SRIM projected-range straggling multiplied by the same factor $f_{TRIM}(E)$ and converted to depth in the same way. The fraction stopping in each cell is the difference of the distribution function across the cell, multiplied by $1 - \eta_N(E)$. In Demo this implants $1.02\times10^{17}$ He m$^{-2}$ per shot of the $1.04\times10^{17}$ m$^{-2}$ incident. The burn helium has a mean stopping depth of 2.9 μm and the debris helium one of 0.54 μm (their energy is deposited shallower, Section S6). Prototype and Experimental receive $3.30\times10^{16}$ and $2.23\times10^{16}$ m$^{-2}$ per shot.

All ion species make Frenkel pairs. The displacement cross section $\sigma_d(E)$ of an ion of energy $E$ integrates the NRT number of displacements of each tungsten recoil over the Thomas–Fermi recoil spectrum. The NRT number is $N_d = 0.8\ E_{dam}/(2E_d)$ for $E_{dam} \geq 2E_d/0.8$, $N_d = 1$ for $E_d \leq E_{dam} < 2E_d/0.8$ and $N_d = 0$ below $E_d$ (the NRT model of Norgett et al., as printed by Nordlund et al. [23]). The recoil spectrum is $d\sigma = (\pi a_{TF}{}^2/2)\ f(u^{½})\ u^{-3/2}\ du$, with $u = \varepsilon^2\ E_r/E_{r,max}$, $E_{r,max} = 4A_1A_2E/(A_1 + A_2)^2$, $f(y) = \lambda y^{1/3}/[1 + (2\lambda y^{4/3})^{2/3}]^{3/2}$, the screening length $a_{TF} = 0.8853\ a_{Bohr}\ (Z_1{}^{2/3} + Z_2{}^{2/3})^{-½}$ and $\varepsilon = E/E_L$, $E_L = 30.724\ Z_1Z_2(Z_1{}^{2/3} + Z_2{}^{2/3})^{½}(A_1 + A_2)/A_2$ eV for an ion ($Z_1$, $A_1$) in tungsten ($Z_2 = 74$, $A_2 = 183.84$). At every ion energy the spectrum is rescaled so that its energy-weighted integral equals the SRIM nuclear stopping cross section. The shape only decides how much of that energy goes into recoils below the threshold. The damage energy of a recoil follows the Lindhard–Robinson partition for tungsten in tungsten, $E_{dam} = E_r/[1 + k_L\ g(\varepsilon_W)]$, $g(\varepsilon) = 3.4008\ \varepsilon^{1/6} + 0.40244\ \varepsilon^{3/4} + \varepsilon$, $k_L = 0.1337\ Z_2{}^{1/6}(Z_2/A_2)^{½} = 0.174$ and $\varepsilon_W = E_r/(1.998$ MeV), as printed by Chen et al. [59]. $\sigma_d$ is tabulated on 140 energies from 200 eV to 30 MeV with 900 recoil energies each. Along the path of a sub-bin ion, from its energy $E_0$ down to 300 eV on 70 logarithmic energies, the element between $E_j$ and $E_{j+1}$ is placed at the mean depth $\{R_{p,eff}(E_0) - ½[R_{p,eff}(E_j) + R_{p,eff}(E_{j+1})]\}/\rho_{SRIM}$ and makes $N_W\ \sigma_d(\bar{E})\ \Delta l$ displacements, $\bar{E} = (E_jE_{j+1})^{½}$. In the production runs the path element $\Delta l$ is the decrement $[R_{p,eff}(E_j) - R_{p,eff}(E_{j+1})]/\rho_{SRIM}$ of the projected range. For the Demo spectrum this gives 0.562 of the count with the path length $(E_j - E_{j+1})/[\rho_{SRIM}S_{SRIM}(\bar{E})]$, $S_{SRIM}$ being the SRIM stopping power per unit areal mass. The case "dpa_path" of Section S7.8 uses the path length. The displacements are spread over depth by a Gaussian whose variance is the straggling accumulated since the surface, $\sigma_R(E_0)^2 - ½[\sigma_R(E_j)^2 + \sigma_R(E_{j+1})^2]$ (with 0.1 nm added to the standard deviation to keep it positive), reflected at the surface. The total of the sub-bin is multiplied by $1 - \eta_E(E_0)$. In Demo the ions make $1.21\times10^{19}$ displacements m$^{-2}$ per shot ($2.15\times10^{19}$ m$^{-2}$ with the path length), 119 per implanted helium atom. The damage peaks at $3.6\times10^{-4}$ dpa in the first nanometer and averages $3.7\times10^{-5}$ dpa over the first 5 μm. The vacancy and SIA sources are $g_V = g_I = \xi_V \times 10^6 \times$ (displacement rate per atom), and $g_{He}$ is the helium stopping rate divided by $N_C$.

Each sub-bin reaches the wall between $R/v(E_{hi})$ and $R/v(E_{lo})$, the times of flight at its edge energies, at a constant rate within that window. The arrival interval, 0.105–83.5 μs at $R = 6.5$ m, is divided into 40 logarithmically spaced slices, which gives slice-average rates of implantation and displacement in every cell. The model uses the continuous piecewise-linear rate through the slice midpoints, zero at the first and last arrival, rescaled in every cell so that its time integral equals the slice sum. The neutrons add $2\times10^{-7}$ dpa and $6\times10^{-8}$ appm of helium per Demo shot (Table S15), scaled like the neutron heating by $Y/154$ MJ $\times$ (6.5 m/$R)^2$, which gives 0.325 times these in Prototype and 0.219 times in Experimental, uniformly in depth and over the whole cycle. Over the 1 mm this helium is $4\times10^{-5}$ of the ion helium of a Demo shot, but it is the only source deeper than about 50 μm.

### S7.5 Temperature history and repetition of shots

$T(x, t)$ is taken unchanged from the thermal model of Section 2.2. One shot period is run with the production solver. The temperature at the surface and at 240 logarithmically spaced depths from 1 nm to 1 mm is stored after every accepted step, and the helium model interpolates this record linearly in depth to its cell centers and linearly in time between the stored steps. The cycle starts from the quasi-steady pre-shot state of the repeated-shot runs of Section 4.2. In Demo it is taken as tungsten uniform at 920.9 K and the steel linear to 680.5 K at its back face (the true face value; the centre of the last steel cell, 1.75 mm thick, is at 701.9 K). Its period of 0.1 s takes 4441 steps, stored as 4442 states with the initial one. In Prototype the two values are 609.5 K and 602.7 K (603.4 K at the centre of the last steel cell), with a period of 1 s and 3859 stored states. The thermal runs end at 920.9 and 701.9 K, and 609.5 and 603.4 K, the second value of each pair at the centre of the last steel cell. Over one cycle the surface returns to within 1.7 K (Demo) and 0.47 K (Prototype) of its starting temperature. The Demo cycle peaks at 3317 K at the surface, 2442 K at 5 μm and 1957 K at 10 μm. It stays above 2000 K for 5.6 μs at the surface and 4.1 μs at 5 μm, and above 1000 K for about 1 ms at every depth of the implanted layer. The ions arrive during this spike.

Every shot repeats the same temperature cycle and the same sources. It starts from the helium and defect state at the end of the previous shot, so helium and defects accumulate while the temperature history is periodic from the first shot on. The real first shots start colder. Their pre-shot temperature rises from 600 to 921 K over the 600 shots of the run, 95 % of the rise within the first 370. A sensitivity case shifts the cycle of shot $n$ rigidly by the difference between the pre-shot surface temperature of shot $n$ of the 600-shot thermal run and 920.9 K. Two further cases shift the whole cycle by $\pm 50$ K. The Experimental scenario has no repetition rate and is computed as a single shot from 300 K, followed for 1 s. Repeating that cycle, every shot starting from 300 K, gives the 500-shot Experimental run behind the counts of Table 4.

### S7.6 Numerical method and verification

The equations are discretized by finite volumes on 186 cells over the 1 mm: 110 cells growing geometrically by a factor of 1.065 from 1 nm at the surface (102 of them in the first 10 μm), 35 cells of 1 μm from 15.7 to 50.7 μm, and 41 cells growing by a factor of 1.12 to the back face, the last 90 μm wide. The diffusive flux between cells $i$ and $i+1$ is $G_{i+½}(C_i - C_{i+1})$, with $G_{i+½} = 1/[\Delta x_i/(2D_i) + \Delta x_{i+1}/(2D_{i+1})]$ and $D$ evaluated at the temperature of each cell. Reactions, sinks and sources are evaluated at the cell centers. The resulting 1860 ordinary differential equations, together with six bookkeeping variables (the cumulative helium, vacancies and SIAs lost through the surface, and the cumulative helium implanted and vacancies and SIAs produced), are integrated with the stiff solver LSODA of SciPy. The solver uses the analytic Jacobian in banded form (half-bandwidth 10), a relative tolerance of $10^{-3}$ and an absolute tolerance of $10^{-9}$ appm. Each shot is integrated in three pieces, from ignition to the first ion arrival, over the ion arrival window and from the last arrival to the next shot, so that the solver cannot step over the ion pulse. The first step of each piece is $10^{-6}$ of its length. An independent implementation of the baseline equations, integrated with the Radau method, serves as the reference for cross-checks. With these variables the helium balance, implanted = mobile + trapped + released through the surface, is checked at every shot (Table S16).

**Table S16.** Verification of the helium model.

| Test | What is checked | Result | Status |
|---|---|---|---|
| 1 | helium pulse diffusing at 1000 K with an absorbing surface, against the method-of-images solution, on a uniform 25 nm grid and on the production grid | largest error 0.099 % and 0.23 % of the peak; mass error $3\times10^{-16}$ | pass (limit 1 %) |
| 2 | diffusion through immobile traps ($K$ = 1) against the Oriani effective diffusivity $D/[1 + (N_T/N_W)\exp(E_b/k_BT)]$, $N_T$ the trap density, at 1000 K, 0.90 eV, 50 appm and 1400 K, 1.20 eV, 200 appm | largest error 0.068 % | pass (limit 2 %) |
| 3 | helium balance at every shot | $2.1\times10^{-13}$ (Demo, 1450 shots), $1.4\times10^{-13}$ (Prototype, 450 shots) | pass (limit $10^{-10}$) |
| 3b | compiled kernels against the reference implementation; helium, vacancy and SIA balances with every channel active (60 cells, $K$ = 4, 1100 K, four pulses) | right-hand side $1.5\times10^{-16}$ and Jacobian $6.4\times10^{-17}$ relative; balances $9.9\times10^{-15}$ (He), $1.1\times10^{-15}$ (V), $4.9\times10^{-16}$ (I) | pass (limit $10^{-9}$) |
| 4 | helium per shot against a separate re-implementation from the raw spectrum table and against the 100 %-retention bookkeeping of Section 4.4; displacements per shot against a direct integral over each ion's slowing-down; displacements per ion against TRIM Kinchin–Pease runs at $E_d$ = 25 eV | helium below $10^{-15}$ against the re-implementation and $9.2\times10^{-4}$ against the bookkeeping of the burn $^4He$ alone (Section 4.4); displacements $4.2\times10^{-4}$ relative; TRIM against the path-length count: 147/141 (2 MeV $^4He$), 70/69.9 (100 keV $^4He$), 18/20.3 (10 keV $^4He$), 24/24.4 (100 keV D), 7/8.3 (10 keV D) | pass (limit 0.5 %); the TRIM comparison is informational |
| 5 | retained helium after 30 Demo shots with every cell halved, with tolerances $10^{-6}$ and $10^{-11}$, and with 12 sub-bins and 80 time slices | changes of $3.1\times10^{-5}$, $1.1\times10^{-5}$ and $4.8\times10^{-4}$; the helium in the 0–1 μm and 1–5 μm layers changes by 2.4 % and 1.4 % when the grid is halved | pass (limit 2 % on the retained total; the 0–1 μm layer exceeds it) |
| 6a | thermal desorption of a $He_4V$ layer at 40 K s$^{-1}$ | peaks at 991, 1180, 1253 and 1605 K; Redhead formula for $E_{b,k} + E_{m,He}$: 999, 1177, 1264, 1617 K | implementation check only |
| 7 | second-level terms against an independent implementation; a forced self-trapping case; the effect in the Demo baseline | $3.2\times10^{-16}$ per term; retained helium ×71 at 1000 K with no vacancies and 100 times the helium source; $+2.8\times10^{-4}$ after 400 Demo shots | pass |

Tests 1–3b check the discretization, the rate laws and the bookkeeping, test 4 the source calculation and test 5

the resolution. Test 6a reproduces the Redhead peak temperatures of the model's own energies within 12 K. They lie 30–60 K above Kornelsen's measured peaks (950, 1120, 1220 and 1560 K, as tabulated by Faney [51]) because $E_{\mathrm{m,He}}$ is added to binding energies that were obtained from those peaks and therefore already contain it.

### S7.7 Second level: helium self-trapping and trap mutation

The second level adds immobile interstitial helium clusters $\mathrm{He}_m$, $m = 2 \ldots 7$. They form by dimerization (He + He → $\mathrm{He_2}$, rate constant $4\pi(2r_{\mathrm{He1}})(2D_{\mathrm{He}})N_{\mathrm{C}}$), grow by helium capture and break up thermally. The binding energies $E_{\mathrm{bC},m} = 1.00$, 1.35, 1.69 and 1.84 eV for $m = 2 \ldots 5$ are derived from the formation energies of Tirumala et al. [55] and held at 1.84 eV for $m = 6$ and 7. The attempt frequency is $10^{13}$ s$^{-1}$ ($1.4\times10^{12}$–$1.3\times10^{13}$ s$^{-1}$ in Perez et al. [49]). They also capture free vacancies ($\mathrm{He}_m$ + V → $\mathrm{V}_m$) and undergo trap mutation, $\mathrm{He}_m \rightarrow \mathrm{V}_m$ + I for $m = 5$, 6 and 7 with the activation energies 1.20, 1.02 and 0.701 eV and prefactors $2.95\times10^{12}$, $1.28\times10^{13}$ and $6.64\times10^{12}$ s$^{-1}$ of Perez et al. [49]. The clusters do not diffuse in the model, although molecular dynamics gives mobile $\mathrm{He_2}$ to $\mathrm{He_6}$ [51]. In the Demo baseline the second level changes the retained helium by $+2.8\times10^{-4}$ after 400 shots. Clusters form during the spike, when detrapping fills the lattice with mobile helium, but dimerization is reversed almost completely.

### S7.8 Outputs and sensitivity

The retained helium is the areal amount of mobile and trapped helium, He + $\Sigma\, k\, \mathrm{V}_k$, in the whole 1 mm. The shot at which it reaches a threshold is interpolated linearly between shots. In Demo the armor keeps 98.9 % of the helium of the first shot, 93.2 % of all helium implanted up to shot 11 and 75.0 % up to shot 131. After 1450 shots it keeps only 26 % of each new shot, because the spike releases helium from the fuller $\mathrm{V}_k$. It reaches $10^{18}$ and $10^{19}$ m$^{-2}$ after 10.6 and 131.3 shots, against 9.9 and 98.5 shots if all implanted helium stayed. Prototype keeps 99.6 % of the first shot and 98.4 % of all helium after 308 shots, and reaches the thresholds after 30.7 and 308.2 shots (30.3 and 303.4 with full retention). The single Experimental shot keeps 99.6 %, and over the 500-shot Experimental run $10^{18}$ and $10^{19}$ m$^{-2}$ are reached after 45.3 and 452.3 shots (44.9 and 448.8 with full retention).

Counting the displacements along the ion path instead of the projected depth (Section S7.4) raises the displacements made by the ions by a factor of 1.78. The Demo armor then keeps 0.3 % more helium after 11 shots and 4.5 % more after 131, and reaches the thresholds after 10.5 and 124.9 shots. The Prototype counts stay at 30.7 and 308.4 (350 shots run). These path-length counts are the model counts quoted in Section 4.4 for Demo and Prototype; the Experimental counts are those of its projected-count run, the only one made. The local helium content stays below 0.03 at.% in every reported Demo and Prototype shot (Demo: 251 appm at 0.60 µm after 1450 shots; Prototype: 136 appm at 18 nm after 450 shots). It reaches 429 appm (0.043 at.%) at 0.5 nm after 500 Experimental shots, far below the peak concentration of about 0.7 at.% at which Sharafat et al. report surface blistering of polycrystalline tungsten [6]. Table S17 lists the Demo sensitivity cases, each run for 300–600 shots (150 for the path-length count).

**Table S17.** Sensitivity of the Demo helium counts. All counts are computed, not extrapolated.

| Case | Change from the baseline | Retained He at 300 shots relative to the baseline | Shots to $10^{18}$ m$^{-2}$ | Shots to $10^{19}$ m$^{-2}$ |
|---|---|---|---|---|
| baseline | – | 1 | 10.6 | 131 |
| $E_{\mathrm{m,He}}$ low / high | 0.06 eV, the ab initio value of Becquart and Domain, / 0.28 eV, the middle of the experimental range 0.24–0.32 eV, both as quoted by Faney [51]; $D_{0,\mathrm{He}}$ kept | 0.951 / 1.099 | 10.6 / 10.4 | 137 / 123 |
| $E_{\mathrm{b},k}$ low / high | 4.05, 2.89, 2.52, 2.41, 2.0, 2.0, 2.0 eV / 4.78, 3.17, 3.14, 3.19, 2.7, 2.7, 2.7 eV (for $k \le 4$ the lowest and highest of the measured and computed values tabulated by Faney [51] and the formation energies of Tirumala et al. [55]; for $k$ = 5–7 rounded bounds) | 0.907 / 1.160 | 10.6 / 10.5 | 142 / 122 |
| $\xi_{\mathrm{V}}$ low | 0.211, the arc-dpa efficiency at a damage energy of 10 keV [23] | 0.740 | 10.8 | 174 |
| arc-dpa | $E_{\mathrm{d}}$ = 70 eV, $b_{\mathrm{arc}}$ = –0.56, $c_{\mathrm{arc}}$ = 0.12 [23] | 0.970 | 10.6 | 135 |
| $f_{\mathrm{surf}}$ = 0.1 | surface conductance reduced to 10 % | 1.003 | 10.5 | 131 |
| $T \mp$ 50 K | whole cycle shifted by –50 / +50 K | 1.031 / 0.969 | 10.5 / 10.6 | 128 / 135 |
| rising $T$ | pre-shot temperature rising from 600 K as in the thermal run | 1.037 | 10.3 | 123 |
| second level | Section S7.7 | 1.0002 | 10.6 | 131 |
| displacements along the path | path-length count, 1.78 times the ion displacements (Section S7.4) | 1.050 at 150 shots | 10.5 | 125 |
| full retention | every implanted helium atom stays | – | 9.9 | 98.5 |

### S7.9 What the model leaves out

The model holds helium only in single vacancies with at most seven atoms. It has no multi-vacancy clusters, cavities or bubbles, and vacancies do not cluster with one another. It therefore describes neither bubble growth, coalescence and blistering nor the thermal stability that larger helium–vacancy clusters give. Helium is not trapped at dislocations, grain boundaries, impurities or the surface. The microstructure does not evolve: the dislocation density is constant, although the Demo surface passes through the recrystallization range on every shot. SIA clusters and dislocation loops are not formed, and small helium clusters are immobile in the second level and absent in the baseline. The hydrogen isotopes implanted with the helium, about 17 per helium ion in the ARIES spectra, compete for the same vacancies but are not modeled. No thermal vacancies are generated, although at the 3317 K surface peak the Boltzmann factor $\exp(-E_f/k_B T)$ for a vacancy formation energy of 3.5 eV [55] is $5\times10^{-6}$. Diffusion is Fickian, with no thermal or stress-driven drift in the steep gradients of the spike, and every NRT Frenkel pair survives, with no recombination or clustering inside the cascade. The temperature cycle is the same for every shot and is not changed by helium or damage. Mobile helium, including helium that has diffused deep into the armor, counts as retained because the back face is closed. After 1450 Demo shots it is $1.6\times10^{14}$ of the $5.8\times10^{19}$ m$^{-2}$ retained. Depths in the source calculation use the SRIM density, 0.5 % above the density implied by $N_W$. Among the inputs, the binding energies for $k$ = 5–7 come from one preprint, the vacancy and SIA prefactors from molecular-dynamics fits or a preprint, the V–I radius has no primary source, and the Thomas–Fermi parameters were not checked against a source.

### S7.10 Comparison with experiment

The model was compared with the helium experiments of the HAPL program in which tungsten was implanted at high temperature and flash-annealed. These experiments combine MeV helium, the damage it makes, high temperature and heating cycles, and reach the $10^{19}$ m$^{-2}$ of the upper damage threshold. Only their abstracts were read. Hashimoto et al. implanted single-crystal and polycrystalline tungsten with $^3$He at 850 °C up to $10^{19}$ m$^{-2}$ and measured the retained helium by nuclear reaction analysis [60]. As implanted, both retained helium similarly, and a single flash anneal at 2000 °C did not change the retention. When the same dose was given in 1000 cycles of $10^{16}$ m$^{-2}$, each followed by a flash to 2000 °C, the helium yield dropped by 95 % in the single crystal and by 70 % in the polycrystal, read here as relative to the single-step samples. The companion paper attributes the difference to the grain boundaries, at which cavities form and grow [61]. The abstracts of Gilliam et al. from the same group give 1.3 MeV $^3$He [62, 63], and Sharafat et al. give an 18 s heating cycle [6].

The unchanged model was run on the production grid with a closed back face and no neutrons. The source is 1.3 MeV $^3$He (mean stopping depth 1.72 μm) with its NRT Frenkel pairs, 13.3 per helium atom. The dislocation densities of $2\times10^{10}$ m$^{-2}$ for the single crystal and $5\times10^{12}$ m$^{-2}$ for the polycrystal are the only microstructural difference the model can express. The implantation rate and the flash profile are not in the abstracts and were assumed: a constant rate over 2000 s (200 s, $2\times10^4$ s, and 2000 s followed by 600 s at temperature as variants), a linear ramp from 850 to 2000 °C in 4 s and back to 850 °C 18 s after its start. In the cyclic case each $10^{16}$ m$^{-2}$ packet arrives in the first 2 s of an 18 s cycle, after which the flash ramps to 2000 °C by 6 s and returns to 850 °C at 18 s. The cyclic runs reach a periodic plateau within 20 cycles, so 60 cycles stand for the 1000 of the experiment. Because nuclear reaction analysis sees the helium near the surface, the helium within 5 μm of the surface is reported besides that in the whole millimeter (Table S18).

**Table S18.** Single-step and cyclic implantation of Hashimoto et al. [60] against the model.

| Quantity | Single crystal: experiment [60] | Single crystal: model | Polycrystal: experiment [60] | Polycrystal: model |
|---|---|---|---|---|
| Retained after $10^{19}$ m$^{-2}$ at 850 °C, fraction of implanted | not given; similar to the polycrystal | 8.3 % | not given | 19.5 % (4.3–54.7 % over the assumed rates) |
| Retained after one 2000 °C flash / before it | unchanged | $6.6\times10^{-4}$ ($3.0\times10^{-8}$ within 5 μm) | unchanged | $6.2\times10^{-4}$, $5.5$–$8.0\times10^{-4}$ over the variants ($1.7$–$2.5\times10^{-8}$ within 5 μm) |
| Retained after cycling / after the single step | 0.05 | $6.9\times10^{-5}$ ($2.4\times10^{-9}$ within 5 μm) | 0.30 | $3.0\times10^{-5}$ ($1.0\times10^{-9}$ within 5 μm) |

The model agrees with the experiment only in that heating cycles remove most of the helium. It disagrees in three respects. First, a single 2000 °C flash empties the model armor, whereas the measured retention did not change. At 2273 K even the HeV complex dissociates within about 0.2 ms ($1/b_1$, Eq. (S7)), and the model has no cavity or larger cluster that could hold the helium. The experiment formed cavities. Second, after cycling the polycrystal kept 30 % and the single crystal 5 % of the single-step yield, against $3\times10^{-5}$ and $7\times10^{-5}$ in the model, which has neither grain boundaries nor cavities. Third, the model retains 2.4 times less helium in the single crystal than in the polycrystal after implantation, where the experiment found them similar.

The experiment does not test the microsecond release that sets the Demo counts, because its anneals last seconds. It shows, however, that tungsten keeps helium through anneals that the $He_kV$ states of the model cannot survive once cavities have formed. Cavities and grain boundaries would both raise the retention of the polycrystalline Demo armor above the model's and move its $10^{19}$ $m^{-2}$ count from 125 towards the 98.5 shots of full retention, while the $10^{18}$ $m^{-2}$ count, 11 against 10, is insensitive. The full-retention counts are therefore the main result of Section 4.4, and the model counts a secondary estimate.

### *S7.11 Summary of the helium counts*

In Demo the armor keeps 98.9 % of the helium of the first shot but only 26 % of each new shot after 1450 shots, because the heating spikes release helium from the fuller vacancies. With the displacements counted along the ion path — the count quoted in Section 4.4 — the $10^{18}$ and $10^{19}$ $m^{-2}$ thresholds are reached after 11 and 125 shots (131 with the projected-depth count of the 1450-shot run), against 10 and 98 with full retention. Prototype reaches them after 31 and 308 shots (30 and 303), and Experimental, each shot from 300 K and with the projected-depth count, after 45 and 452 (45 and 449). The model releases nearly all of the helium in a 2000 °C flash anneal after which the measured retention of implanted tungsten was unchanged (Section S7.10), so it probably overstates the release and its counts are a secondary estimate. The ions make $1.21 \times 10^{19}$ displacements $m^{-2}$ per Demo shot, $3.7 \times 10^{-5}$ dpa per shot averaged over the first 5 µm with the projected-depth count (1.78 times more along the ion path). This is the source behind the dose estimate of Section 7.4.

## S8. Reproduction

The model is the Python package `ife_wall` with its run scripts, archived at Zenodo (https://doi.org/10.5281/zenodo.23577155) and developed at https://github.com/Shohruxbek-cyber/first-wall-comparison. The article corresponds to the release tagged v1.1.0. The production runs used Python 3.14.6 with NumPy 2.4.6, SciPy 1.18.1, Matplotlib 3.11.0 and numba 0.65.1 (for the helium model). The pinned list is in the repository. The SRIM-2012 stopping tables and TRIM outputs were generated for this study with SRIM-2012.01 and SRIM-2012.03. The reference PDFs and the page images used for digitizing are not distributed.

All commands run from the repository root (on Windows after setting `PYTHONUTF8=1`) and write into the repository's output folders, so they are best run in a copy. The batch drivers (`run_he_batch.py`, `run_alloy_study.py single`, `run_skin_sweep.py launch`) skip every case whose output already exists, so a case's stored output folder must be deleted in the copy before it is regenerated. The commands behind the published results, in the order they ran, are in `rerun_queue/queue.json`. The few run after it by hand (the two extended helium runs, the helium-depth diagnostic and the repeated helium-source check) are in the revision log. Table S19 gives the final form of each. The alternative, energy-normalized construction of the spectra (Section S2) is selected with the environment variable `IFE_WALL_ION_SPECTRA=normalized` (or the option `ion_spectra=normalized`).

**Table S19.** Commands that regenerate each table and figure.

| Output | Command |
| --- | --- |
| Table 1 (scenario inputs) | no command: the values are set in `src/ife_wall/scenarios.py` |
| Tables S3, S7 and S8 (energy partition, literature inputs, limitations) | no command: Table S3 and the inputs of Table S7 are literature values, cited in the tables; Table S8 summarizes the text |
| Tables S13–S15 (reactions, symbols and parameters of the helium model) | no command: they describe the equations and names of `src/ife_wall/he/`; the values of Table S15 are set in `src/ife_wall/he/constants.py` |
| Table S1, Fig. S1 and the analytical tests | `python -m pytest -q` (one helium test reads the Prototype temperature history that `python scripts/record_he_thermal.py Prototype converged --Tback=converged` writes), then `python scripts/plot_verification.py` |
| Table S2 (convergence) | `python scripts/run_w_multi.py Demo 600 --out=runs/w_multi/verify_2026-10-05` (the same run as the W multi-shot row below, whose outputs it is; its final state starts the cycle), then `python scripts/convergence_study.py` and `python scripts/convergence_extrapolate.py` |
| Rounding of short steps (energy ledger, Section S1) | `python scripts/diag_clock_rounding.py` |

**Table S19 (continued).**

| Output | Command |
|---|---|
| Table 2 (power density); neutron wall load (Section S2) | `python scripts/diag_power_density.py` |
| Figure S2; SRIM/NIST and depth rows of Table 3 and Table S5 | `python validation/srim_checks.py` |
| Fluence rows of Table 3 and Table S5 | `python scripts/check_source_vs_garoz.py` |
| Blanchard–Martin rows (Table 3 and Table S5) | `python validation/bm2005.py 10`; ratios of temperature rises: `python scripts/diag_benchmark_rises.py` |
| Raffray rows (Table 3 and Table S5) | `python validation/raffray2005.py baseline no_stretch csda xray_pe xray_0.1ns xray_1ns` |
| Lithium rows of Table S5 (hand estimate; ranges against HEIGHTS-IFE) | `python scripts/diag_li_hand_estimate.py`; `python scripts/benchmark_hm2002_li.py` |
| W single shots (Tables 3 and 4, Fig. 1) and the Fig. 3 variants | `python scripts/run_w_single.py`, e.g. `python scripts/run_w_single.py normalize_ions=false` |
| TRIM-profile bias (Section S2) | `python scripts/run_w_single.py trim_multilayer=true --out=results/revision/trim_multilayer_cold` |
| W multi-shot (Tables 3 and 4, Fig. 2, Table S9) | `python scripts/run_w_multi.py Demo 600`; `python scripts/run_w_multi.py Prototype 100`; `python scripts/run_w_multi.py Demo 100 h=5000` (and `h=20000`, and the same for Prototype); the heat-balance column of Table S9: `python scripts/table_h1_design_rule.py`; the drift left at the last shot and the 938 K fit: `python scripts/diag_quasi_steady_drift.py` |
| Helium counts, 100 % retention (Section 4.4, Table 4) | `python scripts/he_shot_count.py` |
| Design-rule cycle counts (Section 4.4, Table 4) | `python scripts/table_h1_design_rule.py`, at the heat-flux factors printed by González-Arrabal et al.; heat-flux factors of this work: `python scripts/diag_heat_flux_factor.py` |
| Experimental helium counts (Table 4) | `python scripts/diag_experimental_helium.py` (reads the 500-shot run of the retention model) |
| Neutron displacement rates (Section 7.4, Section S6) | `python scripts/diag_neutron_dpa.py` |
| Steel conductivity 29 W m⁻¹ K⁻¹ (Section 2.1, Section S6) | `python scripts/diag_steel_k.py`; `python scripts/run_w_multi.py Demo 600 steel_k=29 --out=results/revision/steel_k29/w_multi_Demo` |
| Helium counts, retention model (Section S7, with the full-retention reference counts of Section S7.8) | `python scripts/diag_converged_state.py` (the converged pre-shot state of the repeated-shot runs: Demo 920.9 K with the steel back face at 680.5 K, Prototype 609.5 K and 602.7 K), then `python scripts/record_he_thermal.py Demo converged --Tback=converged` (and Prototype), `python scripts/run_he_retention.py Demo --shots 1450 --case baseline` (Prototype: 450 shots), `python scripts/make_he_results.py` (thresholds and counts); path-length count: `python scripts/run_he_retention.py Demo --shots 150 --case dpa_path --out runs/he_retention/Demo_dpa_path` (Prototype: 350 shots), then `python scripts/he_case_thresholds.py runs/he_retention/Demo_dpa_path runs/he_retention/Prototype_dpa_path`; comparison with experiment (Table S18): `python scripts/he_validation_hashimoto.py` |

**Table S19 (continued).**

| Output | Command |
|---|---|
| Helium-model verification and sensitivity (Tables S16, S17; Section S7.8) | python scripts/run_he_batch.py --jobs 3 Demo:Em_He_low:600 Demo:Em_He_high:300 Demo:Eb_low:600 Demo:Eb_high:600 Demo:xi_low:300 Demo:dpa_arc:300 Demo:surface_0p1:300 Demo:T_m50:300 Demo:T_p50:300 Demo:level2:400; then python scripts/verify_he_1_diffusion.py, verify_he_2_trapping.py, verify_he_3_conservation.py, verify_he_3b_numpy_crosscheck.py, verify_he_4_sources.py, verify_he_5_convergence.py, verify_he_6_benchmarks.py 250, verify_he_7_level2.py (one after another; the last reads the level-2 run of the batch); python scripts/run_he_rising_T.py 300; Experimental: python scripts/record_he_thermal.py Experimental cold, then python scripts/run_he_retention.py Experimental --shots 500 --case baseline (its threshold counts are those of Table 4); then python scripts/make_he_results.py; mean stopping depths (Section S7.4): python scripts/diag_he_depths.py; time above the thresholds (Section S7.5): python scripts/he_time_above.py; hydrogen isotopes per helium ion (Section S7.9): python scripts/he_compare_sharafat.py (reads a text copy of the paper, not distributed) |
| Ion arrival times at the wall (Section 2.3, Section S2) | python scripts/plot_source_term.py |
| Sputtering and erosion lifetimes | python scripts/sputtering_erosion.py; python scripts/diag_erosion_lifetime.py |
| Neutron heating (Sections 4.2 and 5.3) | python scripts/diag_heat_balance_neutron.py; tungsten: copy the checkpoint of the 600-shot run of the Table S2 row into the output folder, then python scripts/run_w_multi.py Demo 1200 neutron_heating=true neutron_steel_factor=0.5 --converge --resume --out=results/neutron/w_multi_Demo_steel0.5 (and neutron_steel_factor=1, ..._steel1), python scripts/run_w_multi.py Prototype 100 neutron_heating=true neutron_steel_factor=0.5 --converge --out=results/neutron/w_multi_Prototype_steel0.5 (and 1); lithium: python scripts/run_li_multi.py Demo 600 delta_index=7 neutron_heating=true neutron_steel_factor=0.5 --converge --out=results/neutron/li_multi_Demo_d56.2um_steel0.5 (indices 6–8, factors 0.5 and 1, folders named the same way); then python scripts/diag_li_multi_variants.py (minimum films and the evaporation at them) |
| Coolant 3.5 mm behind the surface (Section 7.5) | python scripts/run_w_single.py steel_thickness=0.0025 --out=results/cooling_3p5mm/w_single; python scripts/run_w_multi.py Demo 600 steel_thickness=0.0025 --converge --out=results/cooling_3p5mm/w_multi_Demo (and Prototype 100, w_multi_Prototype); python scripts/run_li_scan.py Demo T_li=600 steel_thickness=0.0035 (and Prototype); python scripts/run_li_multi.py Demo 600 delta_index=6 steel_thickness=0.0035 --converge --out=results/cooling_3p5mm/li_multi_Demo_d31.6um (Demo indices 6–7, Prototype 100 shots, indices 4–5, folders named the same way); then python scripts/diag_cooling_summary.py |
| Reduced conductivity (Section 4.5, Table S12) | python scripts/run_damage_study.py --skip-multi --workers 6; python scripts/run_w_multi.py Demo 600 k_irr=0.5 --converge --out=runs/w_damage/multi/kirr0.5_conv; python scripts/run_w_multi.py Demo 600 k_irr=0.5 skin_um=5 skin_factor=0.25 --converge --out=runs/w_damage/multi/Heskin_conv; python scripts/plot_damage_study.py |
| Near-surface layer (Fig. 4) | python scripts/skin_sweep_diagnostic.py; python scripts/run_skin_sweep.py launch --jobs=4; python scripts/run_w_multi.py Demo 100 skin_um=10 skin_factor=0.25 --out=results/skin_sweep/multi_Demo_d10_f0.25; python scripts/run_skin_sweep.py summary |
| Tungsten grades (Section 4.6, Tables S10–S11, Figs. S5–S6) | python scripts/run_alloy_study.py single; python scripts/run_w_multi.py Demo 600 w_material=W_ODS_La2O3 alloy_k_proxy=0.8 --converge --out=results/alloys/multi/Demo_W_ODS_La2O3_p0.8, python scripts/run_w_multi.py Prototype 100 w_material=W_K --converge --out=results/alloys/multi/Prototype_W_K (and w_material=W_ODS_La2O3 alloy_k_proxy=0.8, Prototype_W_ODS_La2O3_p0.8); then python scripts/run_alloy_study.py summary |
| Li deposition (Fig. 5) | python scripts/deposition_profiles.py |
| Li single shots (Table 5, Fig. 6, Fig. S4) | python scripts/run_li_scan.py Demo T_li=600 and the variants in run_all.py; then python scripts/make_summary.py |
| Li repeated shots (Table 5, Fig. 7) | python scripts/run_li_multi.py Demo 600 delta_index=6 --converge --out=runs/li_multi/Demo_d31.6um (indices 5–8 and 12 for Demo; Prototype 100 shots, indices 4–7; folders runs/li_multi/<scenario>_d<thickness>um), then python scripts/li_multi_summary.py |
| Li/steel compatibility limit (Table 5) | python scripts/diag_li_compatibility.py |

**Table S19 (continued).**

| Output | Command |
|---|---|
| Direct repeated-shot runs and the minima of Table 5 | python scripts/run_li_direct.py baseline 925 (and baseline 873, he 925; the film, 45 and 56 µm interpolated from the finished runs and 183 µm the helium-exclusion film of steel_criteria.json, each rounded to the whole micrometer, and run with run_li_multi.py Demo 600 delta_um=<film> --converge); for the variants python scripts/run_li_direct.py neutron_steel0.5 925 (and 873; neutron_steel1, vapor_frr0_s1, steel_3.5mm); then python scripts/diag_li_minima_direct.py |
| Table S6 (one recorded shot of each converged run, with the steel-face swing and heat-flux factors) | python scripts/run_li_extra_cycles.py (runs scripts/li_extra_cycle.py for each converged production and direct run); python scripts/table_li_extra_cycles.py |
| Latent heat returned to the film (bounding case, Section 5.3) | python scripts/run_li_multi.py Demo 600 delta_index=7 --converge --return-latent --out=results/revision/li_return_latent/Demo_d56.2um (and delta_index=8); at the direct film python scripts/run_li_direct.py return_latent 925 |
| Helium and displacements in the steel (Table 5, Fig. S3) | python scripts/steel_behind_film.py; python scripts/diag_steel_criteria.py; python scripts/steel_at_films.py --auto (the minimum, direct-run and helium-exclusion films of the stored results, scripts/_films.py) |
| Particle count of the ion spectra (Section 2.3, Section S2) | python scripts/diag_particle_conservation.py |
| Partial film renewal (Section 5.4) | python scripts/run_li_multi.py Demo 600 delta_index=6 renew=0.03 --converge --out=results/revision/li_renewal/Demo_d31.6um_r0.03 (indices 6–8, renew=0.03 and 0.3, folders named the same way); python scripts/li_renewal_summary.py |
| Helium deposition depths (Sections S2 and S6) | python scripts/diag_he_energy_depth.py |
| Surface pressure and superheat (Section 5.2) | python scripts/diag_li_surface_pressure.py (after the Table S6 commands) |
| Viscous stopping distance (Section 5.4, Table 5) | python scripts/diag_film_travel.py |
| Vapor shielding (Table S4, Fig. 8) | python scripts/run_vapor_shielding.py --workers=6 |
| Static vapor bound (limiting case of Section 6); x-ray share and sound speed of the vapor (Section S2) | python scripts/vapour_shielding_bound.py |
| Vapor shielding: tungsten column, ionization estimate, larger stopping factors | python scripts/diag_w_vapor.py; python scripts/diag_saha.py; python scripts/diag_vapor_stopping.py |
| Vapor shielding over repeated shots (Section 6) | python scripts/run_li_multi.py Demo 600 delta_index=6 vapor_shielding=true vapor_f_rr=0.0 vapor_stopping_factor=1.0 --converge --out=results/vapor_shielding/multi/Demo_d31.6um_frr0_s1 (and delta_index=7, Demo_d56.2um_frr0_s1) |
| Xenon benchmark (future work: needs the SRIM xenon tables made from data/srim_xe_inputs) | python scripts/xe_benchmark.py |
| Alpha mean depth (Section S2) | python scripts/diag_alpha_depth.py |
| Surface-cell versus surface temperature (Section S1) | python scripts/diag_T0_vs_Ts.py |
| The figures as printed (Figs. 1–8 and S1–S6), drawn from the outputs above | python scripts/make_article_figures.py |
| The main paper and this supplement | python scripts/make_article.py (needs the paper source, which the release does not contain) |